\documentclass[11pt,a4paper]{article}   	
\usepackage{geometry}

\usepackage{graphicx}		
\usepackage{xcolor}

\usepackage{xspace}
\usepackage{subfig}

\usepackage{amsmath} 
\usepackage{ifthen}  
\usepackage{amssymb}
\usepackage{amsfonts}
\usepackage{upgreek} 

\usepackage{bm}

\usepackage{lineno}

\usepackage{hyperref}
\usepackage{cancel}
\newboolean{uprightparticles}
\setboolean{uprightparticles}{false}

\newcommand{\tev}{\ensuremath{\mathrm{\,Te\kern -0.1em V}}\xspace}
\newcommand{\gev}{\ensuremath{\mathrm{\,Ge\kern -0.1em V}}\xspace}
\newcommand{\mev}{\ensuremath{\mathrm{\,Me\kern -0.1em V}}\xspace}
\newcommand{\kev}{\ensuremath{\mathrm{\,ke\kern -0.1em V}}\xspace}
\newcommand{\ev}{\ensuremath{\mathrm{\,e\kern -0.1em V}}\xspace}
\newcommand{\gevc}{\ensuremath{{\mathrm{\,Ge\kern -0.1em V\!/}c}}\xspace}
\newcommand{\mevc}{\ensuremath{{\mathrm{\,Me\kern -0.1em V\!/}c}}\xspace}
\newcommand{\gevcc}{\ensuremath{{\mathrm{\,Ge\kern -0.1em V\!/}c^2}}\xspace}
\newcommand{\gevgevcccc}{\ensuremath{{\mathrm{\,Ge\kern -0.1em V^2\!/}c^4}}\xspace}
\newcommand{\mevcc}{\ensuremath{{\mathrm{\,Me\kern -0.1em V\!/}c^2}}\xspace}

\newcommand{\eg}{\mbox{\itshape e.g.}\xspace}
\newcommand{\ie}{\mbox{\itshape i.e.}}

\newcommand{\ps}{\ensuremath{{\cal P}_{\mathrm{s}}}}
\newcommand{\pb}{\ensuremath{{\cal P}_{\mathrm{b}}}}
\newcommand{\calpi}{\ensuremath{{\cal P}_{i}}}
\newcommand{\calpj}{\ensuremath{{\cal P}_{j}}}
\newcommand{\calpk}{\ensuremath{{\cal P}_{k}}}
\newcommand{\calpl}{\ensuremath{{\cal P}_{l}}}

\newcommand{\calp}{\ensuremath{{\cal P}}}

\newcommand{\vijinv}{\ensuremath{V_{ij}^{-1}}}
\newcommand{\viji}{\vijinv}
\newcommand{\vklinv}{\ensuremath{V_{kl}^{-1}}}
\newcommand{\vkli}{\vklinv}
\newcommand{\vijinvhat}{\ensuremath{\hat{V}_{ij}^{-1}}}
\newcommand{\vijihat}{\vijinvhat}

\newcommand{\vi}{\ensuremath{V^{-1}}}

\newcommand{\vsshat}{\ensuremath{\hat{V}_{ss}}}
\newcommand{\vsbhat}{\ensuremath{\hat{V}_{sb}}}

\newcommand{\vbbhat}{\ensuremath{\hat{V}_{bb}}}
\newcommand{\vssihat}{\ensuremath{\hat{V}_{ss}^{-1}}}
\newcommand{\vsbihat}{\ensuremath{\hat{V}_{sb}^{-1}}}

\newcommand{\vbbihat}{\ensuremath{\hat{V}_{bb}^{-1}}}
\newcommand{\vssi}{\ensuremath{V_{ss}^{-1}}}
\newcommand{\vsbi}{\ensuremath{V_{sb}^{-1}}}

\newcommand{\vbbi}{\ensuremath{V_{bb}^{-1}}}

\newcommand{\ws}{\ensuremath{w_{\mathrm{s}}}}

\newcommand{\ns}{\ensuremath{N_{\mathrm{s}}}}
\newcommand{\nshat}{\ensuremath{\hat{N}_{\mathrm{s}}}}
\newcommand{\nb}{\ensuremath{N_{\mathrm{b}}}}
\newcommand{\nbhat}{\ensuremath{\hat{N}_{\mathrm{b}}}}

\newcommand{\deriv}{\ensuremath{\mathrm{d}}}

\newboolean{articletitles}
\setboolean{articletitles}{true}

\usepackage{cite}
\usepackage{mciteplus}

\usepackage{enumitem}

\usepackage{etoolbox}

\newboolean{inbibliography}

\newcommand*\linenomathpatchAMS[1]{%
  \expandafter\pretocmd\csname #1\endcsname {\linenomathAMS}{}{}%
  \expandafter\pretocmd\csname #1*\endcsname{\linenomathAMS}{}{}%
  \expandafter\apptocmd\csname end#1\endcsname {\endlinenomath}{}{}%
  \expandafter\apptocmd\csname end#1*\endcsname{\endlinenomath}{}{}%
}

\expandafter\ifx\linenomath\linenomathWithnumbers
  \let\linenomathAMS\linenomathWithnumbers
  \patchcmd\linenomathAMS{\advance\postdisplaypenalty\linenopenalty}{}{}{}
\else
  \let\linenomathAMS\linenomathNonumbers
\fi

\linenomathpatchAMS{gather}
\linenomathpatchAMS{multline}
\linenomathpatchAMS{align}
\linenomathpatchAMS{alignat}
\linenomathpatchAMS{flalign}

\begin{document}
\pagenumbering{roman}
\thispagestyle{empty}

\vspace*{1.0cm}
\begin{center}
{\huge\bfseries \boldmath
Parameter uncertainties in weighted unbinned maximum likelihood fits} \\[1.0 cm]
{\Large
Christoph~Langenbruch$^{a}$, 
}\\[0.4 cm] 
{\small
$^a$ I. Physikalisches Institut B, RWTH Aachen, Sommerfeldstr.\ 14, 52074 Aachen, Germany
} \\[0.5 cm]
\small
E-Mail:
\texttt{\href{mailto:christoph.langenbruch@cern.ch}{christoph.langenbruch@cern.ch}}.
\vspace{\fill}

\vspace*{2.0cm}

\begin{abstract} 
  \noindent 
  Parameter estimation via unbinned maximum likelihood fits is central for many analyses performed in high energy physics. 
  Unbinned maximum likelihood fits using event weights, for example to statistically subtract background contributions via the {\it sPlot} formalism, or to correct for acceptance effects,
  have recently seen increasing use in the community.
  However, it is well known that the naive approach to the estimation of parameter uncertainties via the second derivative of the logarithmic likelihood
  does not yield confidence intervals with the correct coverage in the presence of event weights. 
  This paper derives the asymptotically correct expressions  
  and compares them with several commonly used approaches for the determination of parameter uncertainties,
  some of which are shown to not generally be asymptotically correct.
  In addition, the effect of uncertainties on event weights is discussed,
  including uncertainties that can arise from the presence of nuisance parameters in the determination of {\it sWeights}. 
\end{abstract} 

\vspace{\fill}

{\footnotesize 
\centerline{\copyright~C.\ Langenbruch, licence \href{http://creativecommons.org/licenses/by/4.0/}{CC-BY-4.0}.}}
\end{center}

\clearpage

\tableofcontents

\clearpage

\setcounter{page}{1}
\pagenumbering{arabic}

\section{Introduction} 
\label{sec:introduction} 
Unbinned maximum likelihood fits are an essential tool for parameter estimation in high energy physics,
due to the desirable features of the maximum likelihood estimator. 
In the asymptotic limit the maximum likelihood estimator is normally distributed around the true parameter value and its variance is equal to the minimum variance bound~\cite{Eadie:100342,James:2006zz}. 
Furthermore, in the unbinned approach no information is lost due to binning.

The inclusion of weights into the maximum likelihood formalism is desirable in many applications.
Examples are the statistical subtraction of background events in the {\it sPlot} formalism~\cite{Pivk:2004ty} through the use of per-event weights,
and the 
correction of acceptance effects via weighting by the inverse efficiency. 
However, with the inclusion of per-event weights 
the confidence intervals determined by the inverse second derivative of the negative logarithmic likelihood
(in the multidimensional case the inverse of the Hessian matrix of the negative logarithmic likelihood) 
are no longer asymptotically correct.\footnote{It should further be noted, that the inclusion of event weights involves some loss of information~\cite{doi:10.1146/annurev.ns.14.120164.002111}.}
There are several approaches that are commonly 
used to determine confidence intervals in the presence of event weights. 
However, as will be shown below, not all of these techniques are guaranteed to give asymptotically correct coverage. 
In this paper, the asymptotically correct expressions for the determination of parameter uncertainties will be derived and then compared with these approaches. 

This paper is structured as follows:
In Sec.~\ref{sec:maximumlikelihood} the unbinned maximum likelihood formalism is briefly summarised and the inclusion of event weights is discussed. 
The asymptotically correct expression for static event weights is derived in Sec.~\ref{sec:alternativemethod} and
compared with several commonly used approaches to determine uncertainties in weighted maximum likelihood fits in Sec.~\ref{sec:commonapproaches}. 
Section~\ref{sec:acceptance} details the correction of acceptance effects through event weights and discusses the impact of weight uncertainties in this context.
Section~\ref{sec:splots} derives the asymptotically correct expressions for parameter uncertainties from fits of \textit{sWeighted} data,
and also details the impact of potential nuisance parameters present in the determination of \textit{sWeights}. 
Different approaches to the determination of parameter uncertainties 
are compared and contrasted using two specific examples in Sec.~\ref{sec:examples}, 
an angular fit correcting for an acceptance effect (Sec.~\ref{sec:angularfit}) and 
the determination of a lifetime when statistically subtracting background events using {\it sWeights} (Sec.~\ref{sec:sweights}). 
Finally, conclusions are given in Sec.~\ref{sec:conclusions}. 

\section{Unbinned maximum likelihood fits and event weights}
\label{sec:maximumlikelihood} 
The maximum likelihood estimator for a set of $N_P$ parameters $\bm{\lambda}=\left\{\lambda_{1},\ldots,\lambda_{N_P}\right\}$,
given $N$
independent and identically distributed 
measurements $\bm{x}=\left\{x_1,\ldots,x_N\right\}$, is 
determined by solving (typically numerically using a software package like {\scshape Minuit}~\cite{James:1975dr}) the maximum likelihood condition
\begin{align}
  \left.\frac{\partial}{\partial\lambda_j}\ln {\cal L}(x_1,\ldots,x_N;\bm{\lambda})\right|_{\hat{\bm{\lambda}}} &= 0\nonumber\\
  \sum_{e=1}^N\left.\frac{\partial}{\partial\lambda_j} \ln {\cal P}(x_e;\bm{\lambda})\right|_{\hat{\bm{\lambda}}} &= 0,\label{eq:ml}
\end{align}
where ${\cal P}(x_e;\bm{\lambda})$ denotes the probability density function evaluated for the event $x_e$ and parameters $\bm{\lambda}$. 
Maximising the logarithmic likelihood $\ln{\cal L}$ finds the parameters $\hat{\bm{\lambda}}$ for which the measured data $\bm{x}$ becomes the most likely.
The covariance matrix $\bm{C}$ for the parameters in the absence of event weights can be calculated from the inverse matrix of second derivatives (the Hessian matrix) of the negative logarithmic likelihood 
\begin{align}
  C_{ij} &= -\left.\left(\frac{\partial^2}{\partial\lambda_i\partial\lambda_j} \ln{\cal L}(x_1,\ldots,x_N;\bm{\lambda})\right|_{\hat{\bm{\lambda}}}\right)^{-1}\nonumber\\
  &= -\left.\left(\sum_{e=1}^N\frac{\partial^2}{\partial\lambda_i\partial\lambda_j} \ln{\cal P}(x_e;\bm{\lambda})\right|_{\hat{\bm{\lambda}}}\right)^{-1}.\label{eq:variance}
\end{align}
evaluated at $\bm{\lambda}=\hat{\bm{\lambda}}$.
When including event weights $w_{e=1,\ldots,N}$ to give each measurement a specific weight 
the maximum likelihood condition becomes\footnote{It should be noted that the left-hand side of Eq.~\ref{eq:mlweighted} strictly speaking is no longer a standard logarithmic likelihood, however that does not preclude its use in parameter estimation as an estimating function.}
\begin{align}
  \sum_{e=1}^N\left.w_e\frac{\partial}{\partial\lambda_j} \ln {\cal P}(x_e;\bm{\lambda})\right|_{\hat{\bm{\lambda}}} &= 0.\label{eq:mlweighted}
\end{align}
Depending on the application, the weight $w_e$ can be a function of $x_e$ and 
other event quantities $y_e$ that ${\cal P}(x_e;\bm{\lambda})$ does not depend on directly.
For efficiency corrections the weights are given by $w_e(x_e,y_e)=1/\epsilon(x_e,y_e)$ as detailed in Sec.~\ref{sec:acceptance},
for \textit{sWeights} the weights $w_e(y_e)$ depend on the \textit{discriminating variable} $y$ as described in Sec.~\ref{sec:splots}. 
It should be noted, that the weighted inverse Hessian matrix
\begin{align}
  C_{ij} &= -\left.\left(\sum_{e=1}^N w_e\frac{\partial^2}{\partial\lambda_i\partial\lambda_j} \ln{\cal P}(x_e;\bm{\lambda})\right|_{\hat{\bm{\lambda}}}\right)^{-1}.\label{eq:covprime}
\end{align}
will generally not give asymptotically correct confidence intervals. 
This can be most easily seen when assuming constant weights $w_e=w$ 
which will result in an over-estimation ($w>1$) or under-estimation ($w<1$) of the statistical power of the sample
and confidence intervals that thus under- or overcover. 

\subsection{Asymptotically correct uncertainties in the presence of per-event weights}
\label{sec:alternativemethod} 
To derive the parameter variance in the presence of event weights, 
which for now are considered to be static, 
the simple case of a single parameter $\lambda$ is discussed first. 
In this case, 
the estimator $\hat{\lambda}$ is defined implicitly by the condition 
\begin{align}
  \sum_{e=1}^N w_e \left.\frac{\partial \ln{\cal P}(x_e;\lambda)}{\partial\lambda}\right|_{\hat{\lambda}} &= 0 \label{eq:lheqedonedim}
\end{align}
which is referred to as an \textit{estimating equation} in the statistical literature. 
A central prerequisite for the following derivation is the property\footnote{
Estimating functions fulfilling Eq.~\ref{eq:lhonedimexpectation} are referred to as \textit{unbiased estimating equations} in the statistical literature (see \eg\ Ref.~\cite{davison_2003}). 
It should be noted that the existence of an \textit{unbiased estimating equation} does not imply that the corresponding estimator itself is unbiased. 
In particular, it is well known that maximum likelihood estimators are biased for finite samples.}
\begin{align}
  E\biggl( \sum_{e=1}^N w_e \left. \frac{\partial \ln {\cal P}(x_e;\lambda)}{\partial\lambda}\right|_{\lambda_0}\biggr) &= 0,\label{eq:lhonedimexpectation}
\end{align}
which is shown for event weights to correct an acceptance effect in Sec.~\ref{sec:acceptance} (Eq.~\ref{eq:accunbiased}). 
The more complex case of \textit{sWeights} will be discussed in Sec.~\ref{sec:splots} (see Eqs.~\ref{eq:sweightsunbiasedone}--\ref{eq:sweightsunbiasedthree}). 
Using the fact that $E( w\partial^2 \ln {\cal P}/\partial\lambda^2|_{\lambda})<0$ (see Eq.~\ref{eq:accnegativehesse}) 
it can be shown that the estimator $\hat{\lambda}$ defined by Eq.~\ref{eq:lheqedonedim} is consistent~\cite{davison_2003}. 
We can then Taylor-expand Eq.~\ref{eq:lheqedonedim}
to first order around the (unknown) true value $\lambda_0$, to which $\hat{\lambda}$ converges in the asymptotic limit of large $N$:
\begin{align}
  \sum_{e=1}^N w_e \left.\frac{\partial \ln {\cal P}(x_e;\lambda)}{\partial\lambda}\right|_{\lambda_0} + \left(\hat{\lambda}-\lambda_0\right)\sum_{e=1}^N w_e \left.\frac{\partial^2 \ln {\cal P}(x_e;\lambda)}{\partial\lambda^2}\right|_{\lambda_0}  &= 0.\label{eq:tayloronedim}
\end{align}
This equation can be rewritten as
\begin{align}
  \hat{\lambda}-\lambda_0 & = -\frac{\sum_{e=1}^N w_e \left.\frac{\partial \ln {\cal P}(x_e;\lambda)}{\partial\lambda}\right|_{\lambda_0}}{\sum_{e=1}^N w_e \left.\frac{\partial^2 \ln {\cal P}(x_e;\lambda)}{\partial\lambda^2}\right|_{\lambda_0}}\label{eq:oneddifference} 
  = -\frac{\sum_{e=1}^N w_e \left.\frac{\partial \ln {\cal P}(x_e;\lambda)}{\partial\lambda}\right|_{\lambda_0}}{N E\Bigl( w \left.\frac{\partial^2 \ln {\cal P}(x_e;\lambda)}{\partial\lambda^2}\right|_{\lambda_0}\Bigr)} + {\cal O}(1/N),
\end{align}
giving the deviation of the estimator $\hat{\lambda}$ from the true value $\lambda_0$. 
Here, we used that the sum in the denominator goes to 
\begin{align}
\sum_{e=1}^N w_e \left.\frac{\partial^2 \ln {\cal P}(x_e;\lambda)}{\partial\lambda^2}\right|_{\lambda_0} &\to N E\biggl( w \left.\frac{\partial^2\ln {\cal P}(x;\lambda)}{\partial\lambda^2}\right|_{\lambda_0}\biggr)\label{eq:secondlimit}
\end{align}
in the asymptotic limit due to the law of large numbers. 
Due to the central limit theorem, the numerator converges to a Gaussian distribution with mean zero (according to Eq.~\ref{eq:lhonedimexpectation}) and variance 
\begin{align}
N \mathrm{var}\left(w\left.\frac{\partial \ln{\cal P}(x;\lambda)}{\partial\lambda}\right|_{\lambda_0}\right) &= N E\biggl( w^2 \left.\frac{\partial\ln{\cal P}(x;\lambda)}{\partial\lambda}\right|_{\lambda_0}\left.\frac{\partial\ln{\cal P}(x;\lambda)}{\partial\lambda}\right|_{\lambda_0}\biggr) .\label{eq:firstlimit}
\end{align}
Using Eqs.~\ref{eq:secondlimit} and~\ref{eq:firstlimit}, the variance in the asymptotic limit at leading order is thus given by
\begin{align}
  \mathrm{var}(\hat{\lambda}-\lambda_0) &= E\left( (\hat{\lambda}-\lambda_0)^2 \right)\nonumber\\
    &= \frac{E\left( w^2\left.\frac{\partial\ln{\cal P}(x;\lambda)}{\partial\lambda}\right|_{\lambda_0}\left.\frac{\partial\ln{\cal P}(x;\lambda)}{\partial\lambda}\right|_{\lambda_0}\right) }{N E\left( w\left.\frac{\partial^2\ln {\cal P}(x;\lambda)}{\partial\lambda^2}\right|_{\lambda_0}\right)^{2}}\label{eq:godambe}
\end{align}
The right-hand side of Eq.~\ref{eq:godambe} is the inverse Godambe information~\cite{godambe1960,godambe1978}, which is central to the theory of estimating equations.
As the estimator is consistent, we replace $\lambda_0$ with $\hat{\lambda}$ in the asymptotic limit,
and further estimate the expectation values through the sample, 
resulting in 
\begin{align}
\mathrm{var}(\hat{\lambda}-\lambda_0) &= \frac{\sum_{e=1}^N w_e^2 \left.\frac{\partial \ln{\cal P}(x_e;\lambda)}{\partial\lambda}\right|_{\hat{\lambda}}\left.\frac{\partial \ln{\cal P}(x_e;\lambda)}{\partial\lambda}\right|_{\hat{\lambda}}}{\left(\sum_{e=1}^N w_e\left.\frac{\partial^2 \ln{\cal P}(x_e;\lambda)}{\partial\lambda^2}\right|_{\hat{\lambda}} \right)\left(\sum_{e=1}^N w_e\left.\frac{\partial^2 \ln{\cal P}(x_e;\lambda)}{\partial\lambda^2}\right|_{\hat{\lambda}}\right)}.  
\end{align}
This expression is also known as the {\it sandwich estimator}. 
In the case where event weights are absent ($w_e=1$), 
the numerator in Eq.~\ref{eq:godambe} cancels with one of the inverse Hessian matrices as in this case 
\begin{align}
E\biggl(
    \left.\frac{\partial \ln {\cal P}(x;\lambda)}{\partial \lambda}\right|_{\lambda_0}
  \left.\frac{\partial \ln {\cal P}(x;\lambda)}{\partial \lambda}\right|_{\lambda_0}
\biggr)
  &= 
  -E\biggl(
  \left.\frac{\partial^2 \ln {\cal P}(x;\lambda)}{\partial\lambda^2}\right|_{\lambda_0}
  \biggr).
\end{align}
For $w_e=1$ the Godambe information thus simplifies to the well known Fisher information. 

For the multidimensional case we analogously Taylor-expand Eq.~\ref{eq:mlweighted} to first order, resulting in
\begin{align}
\sum_{e=1}^N w_e \left.\frac{\partial}{\partial\lambda_i}\ln {\cal P}(x_e;\bm{\lambda})\right|_{\bm{\lambda}_0} + \sum_{e=1}^N w_e \sum_{j=1}^{N_P} (\hat{\lambda}_j-\lambda_{0j})\left.\frac{\partial^2}{\partial\lambda_j\partial\lambda_i}\ln {\cal P}(x_e;\bm{\lambda})\right|_{\bm{\lambda}_0}
= 0,
\end{align}
which can be written as a matrix equation
\begin{align}
\left.d_i\right|_{\bm{\lambda}_0} &= - \sum_{j=1}^{N_P} \left.{H}_{ij}\right|_{\bm{\lambda}_0}\left(\hat{\lambda}_j-\lambda_{0j}\right)~~~\text{with~the definitions}\label{eq:definitions}\\
d_i &= \sum_{e=1}^N w_e\frac{\partial \ln {\cal P}(x_e;\bm{\lambda})}{\partial\lambda_i} ~~~\text{and}\nonumber\\
{H}_{ij} &= \sum_{e=1}^N w_e\frac{\partial^2}{\partial\lambda_i\partial\lambda_j} \ln{\cal P}(x_e;\bm{\lambda})\nonumber 
\end{align}
Matrix inversion yields an expression for the deviation of the estimator $\hat{\lambda}_i$ from the true value $\lambda_{0i}$ 
\begin{align}
  \hat{\lambda}_i-\lambda_{0i} &= -\sum_{j=1}^{N_P}\left.{H}_{ij}^{-1} d_j\right|_{\bm{\lambda}_0}.\label{eq:differencemultidim}
\end{align}
The covariance matrix $\bm{C}$ is then given by
  \begin{align}
    {C}_{ij} =& E\left(\hat{\lambda}_i-\lambda_{0i})(\hat{\lambda}_j-\lambda_{0j})\right)\nonumber\\
    =& \sum_{k,l=1}^{N_P} \left.{H}^{-1}_{ik} E(d_k d_l) {H}^{-1}_{lj}\right|_{\hat{\bm{\lambda}}}\nonumber\\
    =& \sum_{k,l=1}^{N_P} \left.{H}^{-1}_{ik} \left(\sum_{e=1}^N w_e^2\frac{\partial\ln {\cal P}(x_e;\bm{\lambda})}{\partial\lambda_k}\frac{\partial\ln {\cal P}(x_e;\bm{\lambda})}{\partial\lambda_l}\right) {H}^{-1}_{lj}\right|_{\hat{\bm{\lambda}}},
\end{align}
which can be compactly written as
\begin{align}
{C}_{ij} =& \sum_{k,l=1}^{N_P} \left.{H}^{-1}_{ik} {D}_{kl} {H}^{-1}_{lj}\right|_{\hat{\bm{\lambda}}} ~~~\text{with}\label{eq:correctcovariancemultidim}\\
{D}_{kl} =& \sum_{e=1}^N w_e^2\frac{\partial\ln {\cal P}(x_e;\bm{\lambda})}{\partial\lambda_k}\frac{\partial\ln {\cal P}(x_e;\bm{\lambda})}{\partial\lambda_l}\nonumber.
\end{align}
The above expressions are 
familiar from the derivation of Eq.~\ref{eq:variance} (in the absence of event weights) in standard textbooks (\eg\ Ref.~\cite{Barlow:213033}). 
Equation~\ref{eq:correctcovariancemultidim} has been previously discussed in Ref.~\cite{doi:10.1146/annurev.ns.14.120164.002111} in the context of event weights for efficiency correction. 
However, it does not seem to be commonly used and often one of the approaches detailed below in Sec.~\ref{sec:commonapproaches} is employed instead. 

\subsection{Commonly used approaches to uncertainties in weighted fits}
\label{sec:commonapproaches}
Instead of using the asymptotically correct approach for static event weights given by Eq.~\ref{eq:correctcovariancemultidim}, 
often other techniques are used to determine parameter uncertainties in weighted unbinned maximum likelihood fits which are presented below. 
We stress that of the techniques (a)--(c) listed below only the bootstrapping approach (c) will result in generally asymptotically correct uncertainties.
\begin{enumerate}[label=(\alph*)]
\item A simple approach used sometimes (\eg\ in Ref.~\cite{Aaij:2013oba}) is to rescale the weights $w_e$ according to
  \begin{align}
    w_e^\prime &= w_e\frac{\sum_{e=1}^N w_e}{\sum_{e=1}^N w_e^2}\label{eq:scaling}
  \end{align}
  and to use Eq.~\ref{eq:covprime} with the weights $w_e^\prime$. 
  This will rescale the weights such that their sum 
  corresponds to Kish's effective sample size~\cite{kish1965survey}, 
  however, this approach will not generally reproduce the result in Eq.~\ref{eq:correctcovariancemultidim}.
\item A method proposed in Refs.~\cite{Eadie:100342,James:2006zz} is
  to correct the covariance matrix according to
  \begin{align}
    {C}_{ij}^\prime &= \sum_{k,l=1}^{N_P} \left.{H}_{ik}^{-1} {W}_{kl} {H}_{lj}^{-1}\right|_{\hat{\bm{\lambda}}},\label{eq:approximate}
  \end{align}
where $\bm{H}$ is the weighted Hessian matrix defined in Eq.~\ref{eq:definitions} and $\bm{W}$ is the Hessian matrix determined using squared weights $w_e^2$ according to
  \begin{align}
  {W}_{kl} &= -\sum_{e=1}^N w_e^2\frac{\partial^2\ln{\cal P}(x_e;\bm{\lambda})}{\partial\lambda_k\partial\lambda_l}.\label{eq:squaredcov}
\end{align}
This method is the nominal method used in the {\scshape Roofit} software package when using weighted events~\cite{Verkerke:2003ir} and is thus widely used in particle physics.
It corresponds to the result in Eq.~\ref{eq:correctcovariancemultidim} only if
\begin{align}
E\biggl(  \sum_{e=1}^N w_e^2 \left.\frac{\partial\ln {\cal P}(x_e;\bm{\lambda})}{\partial\lambda_k}\right|_{\hat{\bm{\lambda}}}
  \left.\frac{\partial\ln {\cal P}(x_e;\bm{\lambda})}{\partial\lambda_l}\right|_{\hat{\bm{\lambda}}} \biggr) &=
-E\biggl( \sum_{e=1}^N w_e^2 \left.\frac{\partial^2\ln {\cal P}(x_e;\bm{\lambda})}{\partial\lambda_k\partial\lambda_l}\right|_{\hat{\bm{\lambda}}}\biggr)\label{eq:squaredinequality}
\end{align}
This is however not generally the case.
This becomes more clear when rewriting the left- and right-hand side of Eq.~\ref{eq:squaredinequality} according to
\begin{align}
\sum_{e=1}^N w_e^2 \left.\frac{\partial\ln {\cal P}(x_e;\bm{\lambda})}{\partial\lambda_k}\right|_{\hat{\bm{\lambda}}}
\left.\frac{\partial\ln {\cal P}(x_e;\bm{\lambda})}{\partial\lambda_l}\right|_{\hat{\bm{\lambda}}} =&
\sum_{e=1}^N \frac{w_e^2}{{\cal P}^2(x_e;\bm{\lambda})}
\left.\frac{\partial {\cal P}(x_e;\bm{\lambda})}{\partial\lambda_k}\right|_{\hat{\bm{\lambda}}}
\left.\frac{\partial {\cal P}(x_e;\bm{\lambda})}{\partial\lambda_l}\right|_{\hat{\bm{\lambda}}}~~~{\rm and}\\
-\sum_{e=1}^N w_e^2 \left.\frac{\partial^2\ln {\cal P}(x_e;\bm{\lambda})}{\partial\lambda_k\partial\lambda_l}\right|_{\hat{\bm{\lambda}}}
=& \sum_{e=1}^N \frac{w_e^2}{{\cal P}^2(x_e;\bm{\lambda})}
\left.\frac{\partial {\cal P}(x_e;\bm{\lambda})}{\partial\lambda_k}\right|_{\hat{\bm{\lambda}}}
\left.\frac{\partial {\cal P}(x_e;\bm{\lambda})}{\partial\lambda_l}\right|_{\hat{\bm{\lambda}}}\nonumber\\
&-\sum_{e=1}^N\frac{w_e^2}{{\cal P}(x_e;\bm{\lambda})}
  \left.\frac{\partial^2 {\cal P}(x_e;\bm{\lambda})}{\partial\lambda_k\partial\lambda_l}\right|_{\hat{\bm{\lambda}}}
    \label{eq:expectationnonzero}
\end{align}
The expectation value of the second part on the right-hand side of Eq.~\ref{eq:expectationnonzero} is not generally zero. 
While Refs.~\cite{Eadie:100342,James:2006zz} correctly derive that the expectation value
\begin{align}
  E\biggl(\frac{w}{{\cal P}(x;\bm{\lambda})}\frac{\partial^2{\cal P}(x;\bm{\lambda})}{\partial\lambda_k\partial\lambda_l}\biggr) &=
  0\label{eq:iszero}
\end{align}
for an efficiency correction $\epsilon_e=1/w_e$, this is not generally the case for the expression with squared weights
\begin{align}
  E\biggl(\frac{w^2}{{\cal P}(x;\bm{\lambda})}\frac{\partial^2{\cal P}(x;\bm{\lambda})}{\partial\lambda_k\partial\lambda_l}\biggr) \label{eq:expectation}
\end{align}
resulting in confidence intervals that are not generally asymptotically correct when using this approach. 
This will be detailed for efficiency corrections in Sec.~\ref{sec:acceptance}.
For the specific example discussed in Sec.~\ref{sec:angularfit}, the corresponding expectation values are calculated explicitly in App.~\ref{app:nonzeroacceptance}. 
\item A general approach for the determination of parameter uncertainties is to bootstrap the data~\cite{10.2307/2958830}. 
  Repeatedly resampling the data set with replacement allows new samples to be generated that can in turn be used to estimate the parameters $\bm{\lambda}$ using the maximum likelihood method.
  The width of the distribution of estimated parameter values 
  can then be used as estimator for the parameter uncertainty.
  This approach is generally valid, however 
  repeatedly (typically ${\cal O}(10^3)$ times) solving Eq.~\ref{eq:mlweighted} numerically can be computationally expensive and thus this approach is often unfeasible. 
\end{enumerate}

\section{Event weights and inclusion of weight uncertainties}
\label{sec:eventweights}
\subsection{Acceptance corrections}
\label{sec:acceptance}
Following the notation in Refs.~\cite{Eadie:100342,James:2006zz}, this section details the correction of acceptance effects using event weights. 
Acceptance of events with a certain probability $\epsilon$, depending on the measurements $x_e$ and $y_e$,
can be accounted for in unbinned maximum likelihood fits by using event weights $w_e=1/\epsilon(x_e,y_e)$ in Eq.~\ref{eq:mlweighted}. 
The efficiency $\epsilon(x,y)$ should be positive over the full phasespace considered, regions of phasespace where the efficiency is zero should be excluded from the analysis\footnote{To exclude pathological cases we furthermore require $\sum_{e=1}^N w_e=\sum_{e=1}^N 1/\epsilon_e$ to be of ${\cal O}(N)$.}. 
Here, we differentiate  between the quantities $x$ that the probability density function ${\cal P}(x;\bm{\lambda})$ in Eq.~\ref{eq:mlweighted} depends on directly,
and potential additional quantities $y$ that can depend on $x$. 
Using event weights 
can be advantageous when it is difficult or computationally expensive to determine the norm of the probability density function when including the efficiency as an explicit additional multiplicative factor $\epsilon(x,y)$. 
The covariance in this case can be estimated using Eq.~\ref{eq:correctcovariancemultidim} as previously suggested in Ref.~\cite{doi:10.1146/annurev.ns.14.120164.002111}. 

To determine expectation values it is necessary to 
include the acceptance effect in the probability density function.  
The probability density function ${\cal P}(x,y;\bm{\lambda})$ gives the probability to find the measurements $x$ and $y$ depending on the parameters $\bm{\lambda}$ with
\begin{align}
  {\cal P}(x,y;\bm{\lambda}) &= {\cal P}(x;\bm{\lambda}){\cal Q}(y;x)
\end{align}
and the proper normalisation $\int {\cal P}(x;\bm{\lambda}){\textrm d}x=1$ and $\int {\cal Q}(y;x){\textrm d}y=1$. 
The resulting total probability density function including the acceptance effect is then given by
\begin{align}
  {\cal G}(x,y;\bm{\lambda}) &= \frac{{\cal P}(x;\bm{\lambda}){\cal Q}(y;x)\epsilon(x,y)}{\int {\cal P}(x;\bm{\lambda}){\cal Q}(y;x)\epsilon(x,y){\textrm d}x{\textrm d}y} = {\cal P}(x;\bm{\lambda}){\cal Q}(y;x)\epsilon(x,y)/{\cal N}
\end{align}
with normalisation ${\cal N}$. 
This is the probability density function that needs to be used when determining expectation values. 
For the likelihood condition we find
\begin{align}
  E\biggl( w(x,y)\frac{\partial\ln {\cal P}(x;\bm{\lambda})}{\partial\lambda_j} \biggr) &= E\biggl( w(x,y)\frac{1}{{\cal P}(x;\bm{\lambda})}\frac{\partial {\cal P}(x;\bm{\lambda})}{\partial\lambda_j}\biggr)\nonumber\\
  &= \int \frac{1}{\epsilon(x,y)} \frac{1}{{\cal P}(x;\bm{\lambda})} \frac{\partial {\cal P}(x;\bm{\lambda})}{\partial\lambda_j} {\cal P}(x;\bm{\lambda}){\cal Q}(y;x)\epsilon(x,y){\textrm d}x{\textrm d}y /{\cal N}\nonumber\\
  &= \frac{\partial}{\partial\lambda_j}\int {\cal P}(x;\bm{\lambda}){\cal Q}(y;x){\textrm d}x{\textrm d}y /{\cal N} 
  = \frac{\partial}{\partial\lambda_j} 1 /{\cal N}  =0,\label{eq:accunbiased}
\end{align}
confirming the central property of Eq.~\ref{eq:lhonedimexpectation}.  
Further, we obtain
\begin{align}
  E\biggl( w(x,y)\frac{\partial^2\ln {\cal P}(x;\bm{\lambda})}{\partial\lambda_i\partial\lambda_j} \biggr) 
  =& E\biggl( w(x,y) \frac{1}{{\cal P}(x;\bm{\lambda})}\frac{\partial^2 {\cal P}(x;\bm{\lambda})}{\partial\lambda_i\partial\lambda_j}\biggr)
  - E\biggl( w(x,y) \frac{1}{{\cal P}^2(x;\bm{\lambda})}\frac{\partial {\cal P}(x;\bm{\lambda})}{\partial\lambda_i} \frac{\partial {\cal P}(x;\bm{\lambda})}{\partial \lambda_j}\biggr)\nonumber\\
  =& \int \frac{1}{\epsilon(x,y)} \frac{1}{{\cal P}(x;\bm{\lambda})} \frac{\partial^2 {\cal P}(x;\bm{\lambda})}{\partial\lambda_i\partial\lambda_j} {\cal P}(x;\bm{\lambda}){\cal Q}(y;x)\epsilon(x,y){\textrm d}x{\textrm d}y /{\cal N}\nonumber\\
  &-\int \frac{1}{\epsilon(x,y)} \frac{1}{{\cal P}^2(x;\bm{\lambda})} \frac{\partial {\cal P}(x;\bm{\lambda})}{\partial\lambda_i}\frac{\partial {\cal P}(x;\bm{\lambda})}{\partial\lambda_j} {\cal P}(x;\bm{\lambda}){\cal Q}(y;x)\epsilon(x,y){\textrm d}x{\textrm d}y /{\cal N}\nonumber\\
  =& -\int \frac{1}{{\cal P}(x;\bm{\lambda})} \frac{\partial {\cal P}(x;\bm{\lambda})}{\partial\lambda_i}\frac{\partial {\cal P}(x;\bm{\lambda})}{\partial\lambda_j} {\textrm d}x /{\cal N},\label{eq:accnegativehesse}
\end{align}
where also the expectation value in Eq.~\ref{eq:iszero} is shown, and
\begin{align}
E\biggl( w(x,y)\frac{\partial\ln {\cal P}(x;\bm{\lambda})}{\partial\lambda_i}\frac{\partial\ln {\cal P}(x;\bm{\lambda})}{\partial\lambda_j} \biggr) =& E\biggl( w(x,y) \frac{1}{{\cal P}^2(x;\bm{\lambda})}\frac{\partial {\cal P}(x;\bm{\lambda})}{\partial \lambda_i}\frac{\partial {\cal P}(x;\bm{\lambda})}{\partial \lambda_j}\biggr)\nonumber\\
=& \int \frac{1}{\epsilon(x,y)} \frac{1}{{\cal P}^2(x;\bm{\lambda})} \frac{\partial {\cal P}(x;\bm{\lambda})}{\partial\lambda_i}\frac{\partial {\cal P}(x;\bm{\lambda})}{\partial\lambda_j} {\cal P}(x;\bm{\lambda}){\cal Q}(y;x)\epsilon(x,y){\textrm d}x{\textrm d}y /{\cal N}\nonumber\\
=& -E\biggl( w(x,y)\frac{\partial^2\ln {\cal P}(x;\bm{\lambda})}{\partial\lambda_i\partial\lambda_j} \biggr).
\end{align}
However, the equality derived above is not generally fulfilled for squared weights. In this case, we find
\begin{align}
  E\biggl( w^2(x,y)\frac{\partial^2\ln {\cal P}(x;\bm{\lambda})}{\partial\lambda_i\partial\lambda_j} \biggr) 
  =& E\biggl( w^2(x,y) \frac{1}{{\cal P}(x;\bm{\lambda})}\frac{\partial^2 {\cal P}(x;\bm{\lambda})}{\partial\lambda_i\partial\lambda_j}\biggr)
  - E\biggl( w^2(x,y) \frac{1}{{\cal P}^2(x;\bm{\lambda})}\frac{\partial {\cal P}(x;\bm{\lambda})}{\partial\lambda_i} \frac{\partial {\cal P}(x;\bm{\lambda})}{\partial\lambda_j}\biggr)\nonumber\\
  =& \frac{\partial^2}{\partial\lambda_i\partial\lambda_j} \int \frac{1}{\epsilon(x,y)} {\cal P}(x;\bm{\lambda}) {\cal Q}(y;x){\textrm d}x{\textrm d}y /{\cal N}\nonumber\\
  &-\int \frac{1}{\epsilon(x,y)}\frac{1}{{\cal P}(x;\bm{\lambda})} \frac{\partial {\cal P}(x;\bm{\lambda})}{\partial\lambda_i}\frac{\partial {\cal P}(x;\bm{\lambda})}{\partial\lambda_j} {\cal Q}(y;x){\textrm d}x{\textrm d}y /{\cal N}
\end{align}
and
\begin{align}
E\biggl( w^2(x,y)\frac{\partial\ln {\cal P}(x;\bm{\lambda})}{\partial\lambda_i}\frac{\partial\ln {\cal P}(x;\bm{\lambda})}{\partial\lambda_j} \biggr) =& E\biggl( w^2(x,y) \frac{1}{{\cal P}^2(x;\bm{\lambda})}\frac{\partial {\cal P}(x;\bm{\lambda})}{\partial\lambda_i}\frac{\partial {\cal P}(x;\bm{\lambda})}{\partial\lambda_j}\biggr)\nonumber\\
=& \int \frac{1}{\epsilon(x,y)} \frac{1}{{\cal P}(x;\bm{\lambda})} \frac{\partial {\cal P}(x;\bm{\lambda})}{\partial\lambda_i}\frac{\partial {\cal P}(x;\bm{\lambda})}{\partial\lambda_j} {\cal Q}(y;x){\textrm d}x{\textrm d}y /{\cal N}\nonumber\\
=& -E\biggl( w^2(x,y)\frac{\partial^2\ln {\cal P}(x;\bm{\lambda})}{\partial\lambda_i\partial\lambda_j} \biggr)\nonumber\\
& +\frac{\partial^2}{\partial\lambda_i\partial\lambda_j} \int \frac{1}{\epsilon(x,y)} {\cal P}(x;\bm{\lambda}) {\cal Q}(y;x){\textrm d}x{\textrm d}y /{\cal N},\label{eq:expectationnonzeroacc}
\end{align}
where the term in the last line of Eq.~\ref{eq:expectationnonzeroacc}, which corresponds to the expectation value in Eq.~\ref{eq:expectation}, is not generally zero,
as the integral in the numerator can retain a dependence on $\bm{\lambda}$. 
For the example discussed in Sec.~\ref{sec:angularfit} this is explicitly calculated in App.~\ref{app:nonzeroacceptance}. 
This shows that parameter uncertainties determined using Eq.~\ref{eq:approximate} 
are not generally asymptotically correct when performing weighted fits to account for acceptance corrections. 

\subsubsection{Weight uncertainties}
If the weights to 
correct for an acceptance effect 
are only known to a certain precision, \ie\ they are not fixed as assumed in Sec.~\ref{sec:alternativemethod}, 
this induces an additional variance that is not included in Eq.~\ref{eq:correctcovariancemultidim} and that needs to be accounted for. 
This additional covariance can be estimated using standard error propagation, starting from Eq.~\ref{eq:differencemultidim}.
For weights depending on the $N_T$ parameters $p_m$ with covariance matrix $\bm{M}$, this results in
\begin{align}
{C}_{ij}^{\prime\prime} 
=& \sum_{m,n=1}^{N_T} \frac{\partial(\hat{\lambda}_i-\lambda_{0i})}{\partial p_m} {M}_{mn} \frac{\partial(\hat{\lambda}_j-\lambda_{0j})}{\partial p_n} ~~~\text{with}\label{eq:firstterm}\\
\frac{\partial(\hat{\lambda}_i-\lambda_{0i})}{\partial p_m} =& -\sum_{j=1}^{N_P} {H}_{ij}^{-1}\frac{\partial d_j}{\partial p_m}\biggr|_{\bm{\lambda}_0}
-\sum_{j=1}^{N_P}\frac{\partial {H}_{ij}^{-1}}{\partial p_m}d_j\biggr|_{\bm{\lambda}_0},\label{eq:secondterm}
\end{align}
where $d_j$ and ${H}_{ij}$ are defined in Eq.~\ref{eq:definitions}. 
Due to the likelihood condition $d_j\bigr|_{\hat{\bm{\lambda}}}=0$, the second term in Eq.~\ref{eq:secondterm}
behaves as ${\cal O}(1/\sqrt{N})$ and only the first term needs to be considered. 

For the case of an efficiency histogram, where the efficiency is given in $N_B$ bins with weight uncertainty $\sigma_{m=1,\ldots,N_B}$, 
weights inside a bin are fully correlated, but typically uncorrelated with other bins. 
In this case, 
the additional covariance matrix that needs to be added to account for the weight uncertainties is given by
\begin{align}
{C}_{ij}^{\prime\prime} 
=& 
\sum_{k,l=1}^{N_P} {H}_{ik}^{-1}
\sum_{m=1}^{N_B}\left(
\left[\sum_{e\in {\rm bin}\,m} 
  \left.\frac{\partial\ln {\cal P}(x_e;\bm{\lambda})}{\partial\lambda_k}\right|_{\hat{\bm{\lambda}}}\right]
\sigma_m^2
\left[\sum_{e\in {\rm bin}\,m} 
\left.\frac{\partial\ln {\cal P}(x_e;\bm{\lambda})}{\partial\lambda_l}\right|_{\hat{\bm{\lambda}}}\right]
\right)
{H}_{lj}^{-1}
\end{align}

If the efficiency is modelled analytically, for example by a parameterisation that is fit to simulated samples,
the impact of the uncertainty of the parameters $\bm{p}$ 
on the event weights $w_e(\bm{p})=1/\epsilon_e(\bm{p})$ 
needs to be accounted for. 
For $N_T$ parameters with covariance $\bm{M}$ the additional covariance matrix that needs to be added to Eq.~\ref{eq:correctcovariancemultidim} is given by
\begin{align}
{C}_{ij}^{\prime\prime} 
=& 
\sum_{k,l=1}^{N_P}{H}_{ik}^{-1}
\sum_{m,n=1}^{N_T}\biggl(
\sum_{e=1}^N \frac{\partial w_e}{\partial p_m} \left.\frac{\partial\ln {\cal P}(x_e;\bm{\lambda})}{\partial\lambda_k}\right|_{\hat{\bm{\lambda}}}
\biggr){M}_{mn}\biggl(
\sum_{e=1}^N \frac{\partial w_e}{\partial p_n} \left.\frac{\partial\ln {\cal P}(x_e;\bm{\lambda})}{\partial\lambda_l}\right|_{\hat{\bm{\lambda}}}
\biggr) 
{H}_{lj}^{-1}\label{eq:effsaddcovariance}
\end{align}
Identical results are obtained when using the systematic approach to error propagation that is employed for \textit{sWeights} in the next section, 
which is based on combining the estimating equations for the parameters $\bm{p}$ and $\bm{\lambda}$ in a single vector.  

\subsection[The \textit{sPlot} formalism]{The \textit{\textbf{sPlot}} formalism}
\label{sec:splots}
The {\it sPlot} formalism 
was introduced in Ref.~\cite{Pivk:2004ty} to  statistically separate different event species in a data sample 
using per-event weights, the so-called {\it sWeights}, that are determined using a {\it discriminating variable} (in the following denoted by $y$). 
The \textit{sWeights} allow to reconstruct the distribution of the different species in a \textit{control variable} (in the following denoted by $x$),
assuming the \textit{control} and \textit{discriminating} variables are statistically independent for each species. 
In this section, only a brief recap of the {\it sPlot} formalism is given, it is described in more detail in Ref.~\cite{Pivk:2004ty}. 

The {\it sWeights} are determined using an extended unbinned maximum likelihood fit of the {\it discriminating variable} $y$ where the $N_S$ different event species are well separated.
An example of a discriminating variable (which will be discussed in more detail in Sec.~\ref{sec:sweights}) would be the reconstructed mass of a particle 
which is flat for the background components and peaks clearly for the signal component. 
For the typical use case of a signal component of interest and a single background component, 
the \textit{sWeight} for event $e$ is given by
\begin{align}
  \ws(y_{e}) &= 
  \frac{\vsshat\ps(y_e)+\vsbhat\pb(y_e)}{\nshat\ps(y_e)+\nbhat\pb(y_e)}\nonumber\\
  &= \frac{\vbbihat\ps(y_e)-\vsbihat\pb(y_e)}{(\vbbihat-\vsbihat)\ps(y_e)+(\vssihat-\vsbihat)\pb(y_e)},\label{eq:sweight}
\end{align}
where $\nshat=\vsshat+\vsbhat$ and $\nbhat=\vbbhat+\vsbhat$ is used~\cite{Pivk:2004ty}. 
Retaining only the dependency on the inverse covariance matrix elements \viji\ simplifies the following derivations. 
The estimates for the inverse covariance matrix elements are given by 
\begin{align}
  \vijihat &= \sum_{e=1}^N \frac{\calpi(y_e)\calpj(y_e)}{\bigl(\nshat\ps(y_e)+\nbhat\pb(y_e)\bigr)^2}, ~~~\mathrm{with~expectation}\label{eq:viji}\\
  \viji &= \int \frac{\calpi(y)\calpj(y)}{\ns\ps(y)+\nb\pb(y)}\deriv y.\nonumber
\end{align}
Using the {\it sWeights} in a weighted unbinned maximum likelihood fit allows to statistically subtract events originating from species not of interest~\cite{2009arXiv0905.0724X}, by including them as event weights in Eq.~\ref{eq:mlweighted}, resulting in the estimating equations
\begin{align}
\sum_{e=1}^N \ws(y_e;\vssihat,\vsbihat,\vbbihat) \frac{\partial\ln\calp(x_e;\bm{\lambda})}{\partial\lambda_i}\biggr|_{\hat{\bm{\lambda}}} = 0.
\end{align}
The \textit{sWeights} depend on estimates for the inverse covariance matrix elements \viji\ (Eq.~\ref{eq:sweight}), which in turn depend on estimates for the signal and background yields (Eq.~\ref{eq:viji}) determined from the same sample\footnote{
Problems of this type are discussed in the statistical literature as \textit{two-step M-estimation}, see for example Refs.~\cite{wooldridge2010econometric,neweymcfadden,murphytopel}.
}. 
To account for this effect, 
\ie\ in order to systematically perform error propagation, 
it is useful to combine the estimating equations for the yields \ns\ and \nb, the inverse covariance matrix elements \viji, and the parameters of interest $\bm{\lambda}$ in a single vector

{\small
\begin{align}
  {\bm{g}}(\bm{x},\bm{y};\bm{\theta}) &= \left(
  \renewcommand{\arraystretch}{1.3}
  \begin{array}{c}
    \varphi_s(\bm{y};\ns,\nb)\\
    \varphi_b(\bm{y};\ns,\nb)\\
    \psi_{ss}(\bm{y};\vssi,\ns,\nb)\\
    \psi_{sb}(\bm{y};\vsbi,\ns,\nb)\\
    \psi_{bb}(\bm{y};\vbbi,\ns,\nb)\\
    \xi_i(\bm{x},\bm{y};\bm{\lambda},\vssi,\vsbi,\vbbi)
  \end{array}
  \right)  
  = \left(
  \begin{array}{c}
\sum_e\frac{\partial}{\partial\ns}\bigl[\ln(\ns\ps(y_e)+\nb\pb(y_e)) - \frac{\ns+\nb}{N} \bigr]\\
\sum_e\frac{\partial}{\partial\nb}\bigl[\ln(\ns\ps(y_e)+\nb\pb(y_e)) - \frac{\ns+\nb}{N} \bigr]\\
\sum_e \bigl[\frac{\ps(y_e)\ps(y_e)}{(\ns\ps(y_e)+\nb\pb(y_e))^2} - \frac{\vssi}{N}\bigr]\\    
\sum_e \bigl[\frac{\ps(y_e)\pb(y_e)}{(\ns\ps(y_e)+\nb\pb(y_e))^2} - \frac{\vsbi}{N}\bigr]\\    
\sum_e \bigl[\frac{\pb(y_e)\pb(y_e)}{(\ns\ps(y_e)+\nb\pb(y_e))^2} - \frac{\vbbi}{N}\bigr]\\    
\sum_e \ws(y_e;\vssi,\vsbi,\vbbi) \frac{\partial\ln\calp(x_e;\bm{\lambda})}{\partial\lambda_i}
    \end{array}
  \right),  \label{eq:vectorg}
\end{align}}
where $\bm{\theta}$ denotes the vector of parameters $\bm{\theta}=\{\ns,\nb,\vssi,\vsbi,\vbbi,\bm{\lambda}\}$. 
It should be noted that solving $\bm{g}(\bm{x},\bm{y};\bm{\theta})|_{\hat{\bm{\theta}}}=0$ is equivalent to solving the estimating equations for the yields, the inverse covariance matrix elements, and the parameters of interest $\bm{\lambda}$ sequentially. 
It can be shown that $E\bigl(\bm{g}(\bm{x},\bm{y};\bm{\theta})|_{\bm{\theta}_0}\bigr)=0$, \ie\ the estimating equations are unbiased\footnote{Note that the expectation values are evaluated at the true parameter values $\bm{\theta}_0$ which simplifies their calculation significantly.}, as
\begin{align}
  E\biggl(\varphi_i(\bm{y};\ns,\nb)\bigr|_{\bm{\theta}_0}\biggr) &= \int \frac{\calpi(\ns\ps+\nb\pb)}{\ns\ps+\nb\pb}\deriv y - 1 = 0\label{eq:sweightsunbiasedone}\\
  E\biggl(\psi_{ij}(\bm{y};\viji,\ns,\nb)\bigr|_{\bm{\theta}_0}\biggr) &= \int \frac{\calpi\calpj}{\ns\ps+\nb\pb}\deriv y - E(\viji) = 0, \label{eq:sweightsunbiasedtwo}
\end{align}
and
\begin{align}
E\biggl(\xi_i(\bm{x},\bm{y};\bm{\lambda},\vssi,\vsbi,\vbbi)\bigr|_{\bm{\theta}_0}\biggr)  &= E\biggl(\sum_e\ws(y_e)\frac{\partial\ln\ps(x_e;\bm{\lambda})}{\partial\lambda_i}\biggr)\nonumber\\
&= \int \frac{\vbbi\ps(y)-\vsbi\pb(y)}{(\vbbi-\vsbi)\ps(y)+(\vssi-\vsbi)\pb(y)}
\frac{\partial\ln\ps(x;\bm{\lambda})}{\partial\lambda_i}\nonumber\\
&\hphantom{=} 
     \times\bigl(\ns\ps(x)\ps(y)+\nb\pb(x)\pb(y)\bigr)\deriv x\deriv y\nonumber\\
     &= \int \frac{\vbbi\ps(y)-\vsbi\pb(y)}{(\vbbi-\vsbi)\ps(y)+(\vssi-\vsbi)\pb(y)}\pb(y)\deriv y\nonumber\\
     &\hphantom{=}\times \underbrace{\int \nb\frac{\pb(x)}{\ps(x;\bm{\lambda})}\frac{\partial\ps(x;\bm{\lambda})}{\partial\lambda_i}\deriv x}_{\kappa_i}\nonumber\\
    &= \frac{\vbbi\vsbi-\vsbi\vbbi}{\vssi\vbbi-\vsbi\vsbi} \times \kappa_i=0~.\label{eq:sweightsunbiasedthree}
\end{align}
The covariance matrix for the full system in the asymptotic limit is given by\footnote{Here and in the following the superscript ${}^{-T}$ refers to the transposed inverted matrix.}~\cite{davison_2003,van2000asymptotic}
\begin{align}
  {\bm{C}}_{\bm{\theta}} &= E\biggl(\frac{\partial\bm{g}(\bm{x},\bm{y};\bm{\theta})}{\partial\bm{\theta}^T}\biggr)^{-1}
  \times
    E\bigl(\bm{g}(\bm{x},\bm{y};\bm{\theta})\bm{g}(\bm{x},\bm{y};\bm{\theta})^T\bigr)
  \times
  E\biggl(\frac{\partial\bm{g}(\bm{x},\bm{y};\bm{\theta})}{\partial\bm{\theta}^T}\biggr)^{-T}.\label{eq:correctvar}
\end{align}
The covariance can be estimated from the sample by replacing the expectation values $E\bigl(\partial g_i(\bm{x},\bm{y};\bm{\theta})/\partial\theta_j\bigr)$ and $E\bigl(g_i(\bm{x},\bm{y};\bm{\theta})g_j(\bm{x},\bm{y};\bm{\theta})\bigr)$ by their sample estimates, which are given in Apps.~\ref{app:unbinneddenom} and~\ref{app:unbinnednum}, respectively. 

For the case of classic \textit{sWeights}, where the shapes of all probability density functions are known, the above expression can be simplified further. 
The detailed calculation is given in App.~\ref{app:mestimationunbinned}. For the covariance of the parameters of interest $\bm{\lambda}$ it results in 
\begin{align}
  \bm{C}_{\bm{\lambda}} &= \bm{H}^{-1}\bm{H}^\prime\bm{H}^{-T} - \bm{H}\bm{E}\bm{C}^\prime \bm{E}^T\bm{H}^{-T} ~~~\mathrm{with}\label{eq:sweightsvariance}\\
  {H}_{ij} &= \sum_e \ws(y_e)\frac{\partial^2\ln \ps(x_e;\bm{\lambda})}{\partial\lambda_i\partial\lambda_j} \nonumber\\
  {H}_{ij}^\prime &= \sum_e \ws^2(y_e)\frac{\partial\ln \ps(x_e;\bm{\lambda})}{\partial\lambda_i}\frac{\partial\ln \ps(x_e;\bm{\lambda})}{\partial\lambda_j} \nonumber\\
  {E}_{k(ij)} &=  \sum_e \frac{\partial\ws(y_e)}{\partial\viji}\frac{\partial\ln \ps(x_e;\bm{\lambda})}{\partial\lambda_k} \nonumber\\
  {C}_{(ij)(kl)}^\prime &=  \sum_e \frac{\calpi(y_e)\calpj(y_e)\calpk(y_e)\calpl(y_e)}{\bigl(\ns\ps(y_e)+\nb\pb(y_e)\bigr)^4}\nonumber
  \nonumber
\end{align}
Note that using only the first term in Eq.~\ref{eq:sweightsvariance} (which corresponds to Eq.~\ref{eq:correctcovariancemultidim}) 
would not generally be asymptotically correct but instead conservative for the parameter variances, as the matrix $\bm{C}^\prime$ is positive definite. 

The same technique used for the calculation of Eq.~\ref{eq:sweightsvariance} can also be used to determine the (co)variance of the sum of \textit{sWeights} in non-overlapping bins in the control variable $x$, 
which is needed to perform $\chi^2$ fits of binned \textit{sWeighted} data. 
The detailed calculation for the covariance of $\bm{S}$ (with $S_i=\sum_{e\,\in\,\mathrm{bin}\,i}\ws(y_e)$ for bin $i$) 
is given in App.~\ref{app:mestimationbinned} and results in
\begin{align}
  \bm{C}_{\bm{S}} &= \bm{H}^\prime -\bm{E}\bm{C}^\prime\bm{E}^T\label{eq:binnedvariance} ~~~\mathrm{where}\\
     {H}^\prime_{ij} &= \delta_{ij}\sum_{e\,\in\,\mathrm{bin}\,i} \ws^2(y_e)\nonumber\\
     E_{k(ij)} &= \sum_{e\,\in\,\mathrm{bin}\,k} \frac{\partial\ws(y_e)}{\partial\viji}\nonumber\\
     C_{(ij)(kl)}^\prime &= \sum_e \frac{\calpi(y_e)\calpj(y_e)\calpk(y_e)\calpl(y_e)}{\bigl(\ns\ps(y_e)+\nb\pb(y_e)\bigr)^4}\nonumber
\end{align}
As is apparent, using only the first term in Eq.~\ref{eq:binnedvariance}, \ie\ using $\sum_{e\,\in\,\mathrm{bin}\,i}\ws^2(y_e)$ as estimate for the variance of the content of bin $i$, is not generally asymptotically correct but conservative, as $\bm{C}^\prime$ is positive definite. 
The second term in Eq.~\ref{eq:binnedvariance} also induces correlations between bins which should be accounted for in a binned $\chi^2$ fit. 

\subsubsection[\textit{sWeights} and nuisance parameters]{\textit{\textbf{sWeights}} and nuisance parameters}
\label{sec:sweightnuisances}
Additional nuisance parameters $\bm{\alpha}$ present in the extended maximum likelihood fit of the event yields (\eg\ shape parameters of \ps\ or \pb) can be easily included in the formalism used in the previous section.
The estimating equations $\varphi_i(\bm{y};\bm{\alpha},\ns,\nb)$ for the parameters $\bm{\alpha}$ need to be added to the vector $\bm{g}$ defined in Eq.~\ref{eq:vectorg}, resulting in a modified

{\footnotesize
\begin{align}
  {\bm{g}^\prime}(\bm{x},\bm{y};\bm{\theta}^\prime) &= \left(
  \renewcommand{\arraystretch}{1.3}
  \begin{array}{c}
    \varphi_s(\bm{y};\bm{\alpha},\ns,\nb)\\
    \varphi_b(\bm{y};\bm{\alpha},\ns,\nb)\\
    \varphi_i(\bm{y};\bm{\alpha},\ns,\nb)\\
    \psi_{ss}(\bm{y};\vssi,\bm{\alpha},\ns,\nb)\\
    \psi_{sb}(\bm{y};\vsbi,\bm{\alpha},\ns,\nb)\\
    \psi_{bb}(\bm{y};\vbbi,\bm{\alpha},\ns,\nb)\\
    \xi_i(\bm{x},\bm{y};\bm{\lambda},\vssi,\vsbi,\vbbi,\bm{\alpha})
  \end{array}
  \right)  
  = \left(
  \begin{array}{c}
\sum_e\frac{\partial}{\partial\ns}\bigl[\ln(\ns\ps(y_e;\bm{\alpha})+\nb\pb(y_e;\bm{\alpha})) - \frac{\ns+\nb}{N} \bigr]\\
\sum_e\frac{\partial}{\partial\nb}\bigl[\ln(\ns\ps(y_e;\bm{\alpha})+\nb\pb(y_e;\bm{\alpha})) - \frac{\ns+\nb}{N} \bigr]\\
\sum_e\frac{\partial}{\partial\alpha_i}\bigl[\ln(\ns\ps(y_e;\bm{\alpha})+\nb\pb(y_e;\bm{\alpha})) - \frac{\ns+\nb}{N} \bigr]\\
\sum_e \bigl[\frac{\ps(y_e;\bm{\alpha})\ps(y_e;\bm{\alpha})}{(\ns\ps(y_e;\bm{\alpha})+\nb\pb(y_e;\bm{\alpha}))^2} - \frac{\vssi}{N}\bigr]\\    
\sum_e \bigl[\frac{\ps(y_e;\bm{\alpha})\pb(y_e;\bm{\alpha})}{(\ns\ps(y_e;\bm{\alpha})+\nb\pb(y_e;\bm{\alpha}))^2} - \frac{\vsbi}{N}\bigr]\\    
\sum_e \bigl[\frac{\pb(y_e;\bm{\alpha})\pb(y_e;\bm{\alpha})}{(\ns\ps(y_e;\bm{\alpha})+\nb\pb(y_e;\bm{\alpha}))^2} - \frac{\vbbi}{N}\bigr]\\    
\sum_e \ws(y_e;\vssi,\vsbi,\vbbi,\bm{\alpha}) \frac{\partial\ln\calp(x_e;\bm{\lambda})}{\partial\lambda_i}
    \end{array}
  \right),  \label{eq:vecorgprime}
\end{align}}
where the vector of parameters is $\bm{\theta}^\prime=\{\ns,\nb,\bm{\alpha},\vssi,\vsbi,\vbbi,\bm{\lambda}\}$. 
The covariance matrix in the asymptotic limit is then given by
\begin{align}
  {\bm{C}}_{\bm{\theta}^\prime} &= E\biggl(\frac{\partial\bm{g}^\prime(\bm{x},\bm{y};\bm{\theta}^\prime)}{\partial\bm{\theta}^{\prime T}}\biggr)^{-1}
  \times
    E\bigl(\bm{g}^\prime(\bm{x},\bm{y};\bm{\theta}^\prime)\bm{g}^\prime(\bm{x},\bm{y};\bm{\theta}^\prime)^T\bigr)
  \times
  E\biggl(\frac{\partial\bm{g}^\prime(\bm{x},\bm{y};\bm{\theta}^\prime)}{\partial\bm{\theta}^{\prime T}}\biggr)^{-T}.\label{eq:correctvarnuisance}
\end{align}
The covariance can again be estimated from the sample by replacing the expectation values $E\bigl(\partial g_i^\prime(\bm{x},\bm{y};\bm{\theta}^\prime)/\partial\theta_j^\prime\bigr)$ and $E\bigl(g_i^\prime(\bm{x},\bm{y};\bm{\theta}^\prime)g_j^\prime(\bm{x},\bm{y};\bm{\theta}^\prime)\bigr)$ by their sample estimates.
It should be noted that the nuisance parameters $\bm{\alpha}$ in this case will induce additional covariance terms beyond Eq.~\ref{eq:sweightsvariance}\footnote{
For the binned case nuisance parameters $\bm{\alpha}$ induce additional covariance terms beyond Eq.~\ref{eq:binnedvariance} as well.
}. 

\section{Examples}
\label{sec:examples}
\subsection{Correcting for an acceptance effect with event weights}
\label{sec:angularfit}
\begin{figure}
  \centering
  \subfloat[$\epsilon(\cos\theta)=1.0-0.7\cos^2\theta$\label{fig:eps1}]{
    \includegraphics[width=0.49\textwidth]{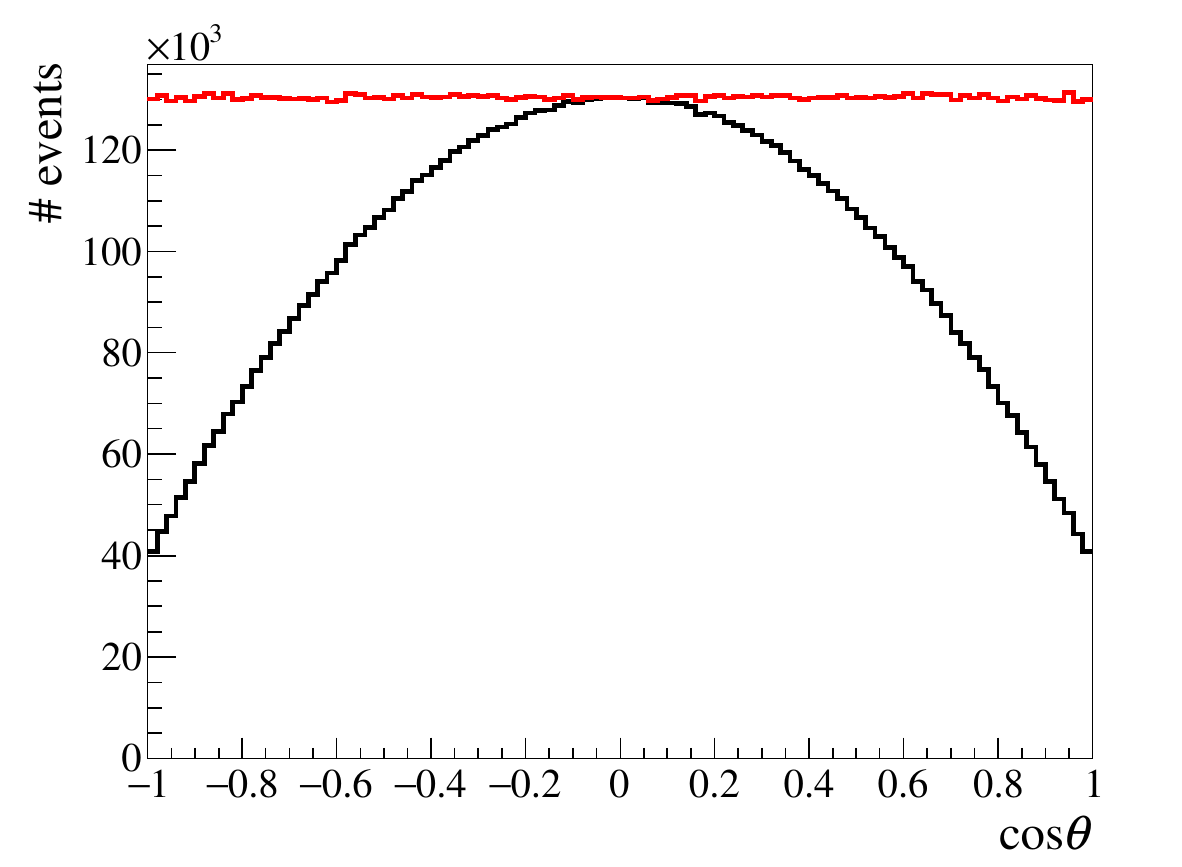}
  }
  \subfloat[$\epsilon(\cos\theta)=0.3+0.7\cos^2\theta$\label{fig:eps2}]{
    \includegraphics[width=0.49\textwidth]{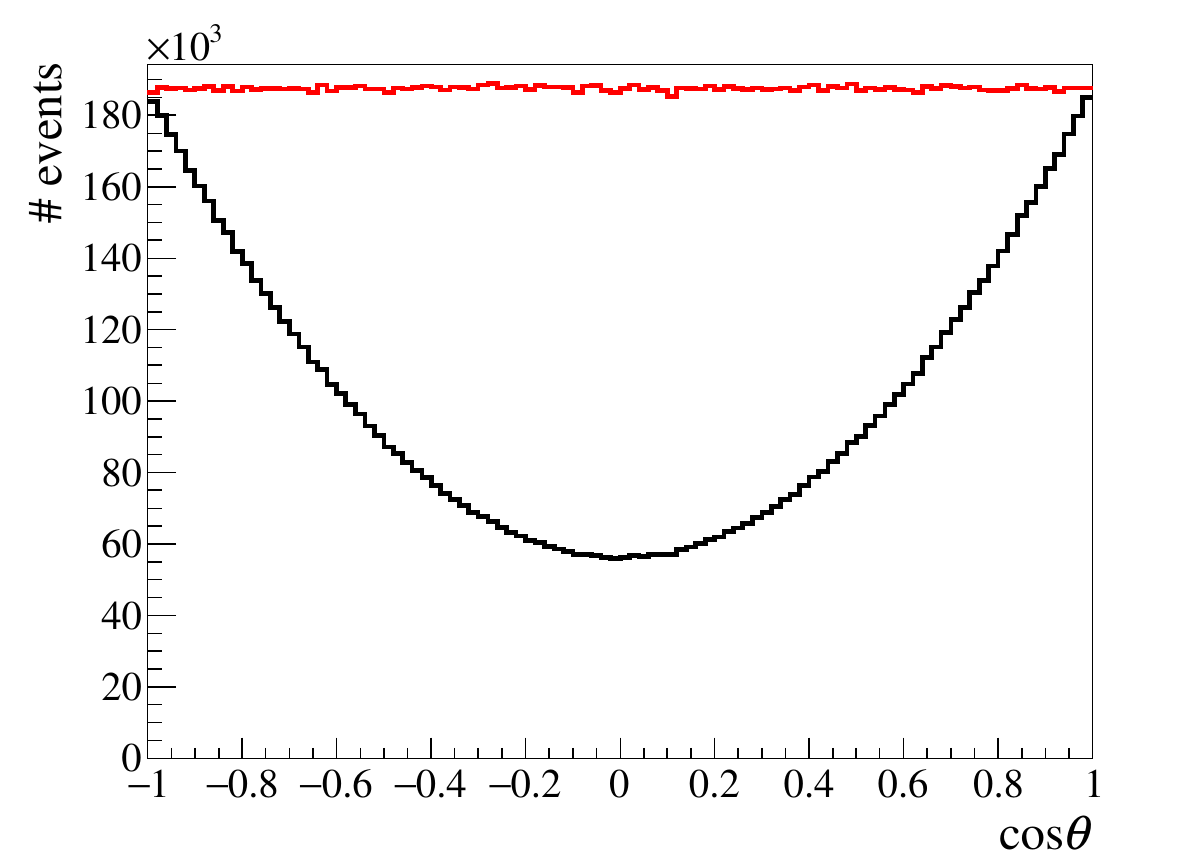}
  }  
  \caption{Angular $\cos\theta$ distribution of (black) data and (red) efficiency corrected events for $10\,000$ pseudoexperiments consisting of $1\,000$ events each.\label{fig:eff}}
\end{figure}

The first example discussed in this paper is the fit of an angular distribution to determine angular coefficients, using event weights to correct for an acceptance effect. 
The probability density function used to generate and fit the pseudoexperiments is a simple second order polynomial in the angle $\cos\theta$:
\begin{align}
{\cal P}(\cos\theta;c_0,c_1) &= \left(1+c_0\cos\theta+c_1\cos^2\theta\right)/{\cal N}~~~{\rm with}\nonumber\\
{\cal N} &= \int_{-1}^{+1} \left(1+c_0\cos\theta+c_1\cos^2\theta\right) {\rm dcos}\theta=2+\tfrac{2}{3}c_1
\end{align}
In the generation, the values $c_0^{\rm gen}=0$ and $c_1^{\rm gen}=0$ are used. 
Events are generated using a $\cos\theta$-dependent efficiency $\epsilon(\cos\theta)$. 
Two efficiencies shapes are studied, given by
\begin{enumerate}[label=(\alph*)]
\item $\epsilon(\cos\theta) = 1.0-0.7\cos^2\theta$~~~and
\item $\epsilon(\cos\theta) = 0.3+0.7\cos^2\theta$.
\end{enumerate}
For simplicity, no uncertainty is assumed on the description of the acceptance effect by $\epsilon(\cos\theta)$,
otherwise the effect of uncertainties on event weights would need to be included as described in Sec.~\ref{sec:acceptance}. 
Figure~\ref{fig:eff} shows the generated data (including the acceptance effect) in black and the efficiency corrected distributions,
weighted by $w_e=1/\epsilon(\cos\theta_e)$ in red. 

The parameters $c_0$ and $c_1$ are determined using a weighted unbinned maximum likelihood fit, solving Eq.~\ref{eq:mlweighted}. 
The uncertainties on the parameters $c_0$ and $c_1$ are determined using the approaches to determine parameter uncertainties that are discussed in Sec.~\ref{sec:maximumlikelihood}. 
The following methods are studied:
\begin{enumerate}[label=(\alph*)]
\item The method of using the uncertainties determined according to Eq.~\ref{eq:covprime} without any correction, denoted as {\it wFit} in this section.
\item Scaling the weights according to Eq.~\ref{eq:scaling}. This approach is referred to as {\it scaled weights}. 
\item Determining the covariance matrix using Eq.~\ref{eq:approximate}. This method is referred to as {\it squared correction} in the following. 
\item Bootstrapping the data (using 1000 bootstraps) with replacement, denoted as {\it bootstrapping}.
\item The method to determine the covariance according to Eq.~\ref{eq:correctcovariancemultidim} as discussed in Sec.~\ref{sec:alternativemethod}, 
  referred to as {\it asymptotic} method. 
\item A conventional fit ({\it cFit}) modelling the efficiency correction effect in the probability density function (and its normalisation) instead of 
  using event weights.   
\end{enumerate}
The performance of the methods is compared using pseudoexperiments,
with each study consisting of 10\,000 toy data samples.
The same data samples are used for every method. 
The distribution of the pull, defined as $p_i(c_0)=(c_{0,i}-c_{0}^{\rm gen})/\sigma_i(c_0)$ (and analogously for parameter $c_1$),
is used to test the different methods for uncertainty estimation.
Here, the fitted value for parameter $c_0$ in pseudoexperiment $i$ is denoted as $c_{0,i}$, the corresponding uncertainty is denoted as $\sigma_{i}(c_0)$ and the generated value as $c_0^{\rm gen}$. 
If the fit is unbiased and the uncertainties are determined correctly, the pull distribution is expected to be
a Gaussian distribution with a mean compatible with zero and a width compatible with one. 
Different event yields per data sample ($N=500$, $1\,000$, $2\,000$, $5\,000$, $10\,000$, $20\,000$, $50\,000$) are studied to investigate the influence of statistics. 

\begin{figure}
\centering
  \subfloat[$c_0$ pull distributions\label{fig:c0pullsa}]{
    \includegraphics[width=0.7\textwidth]{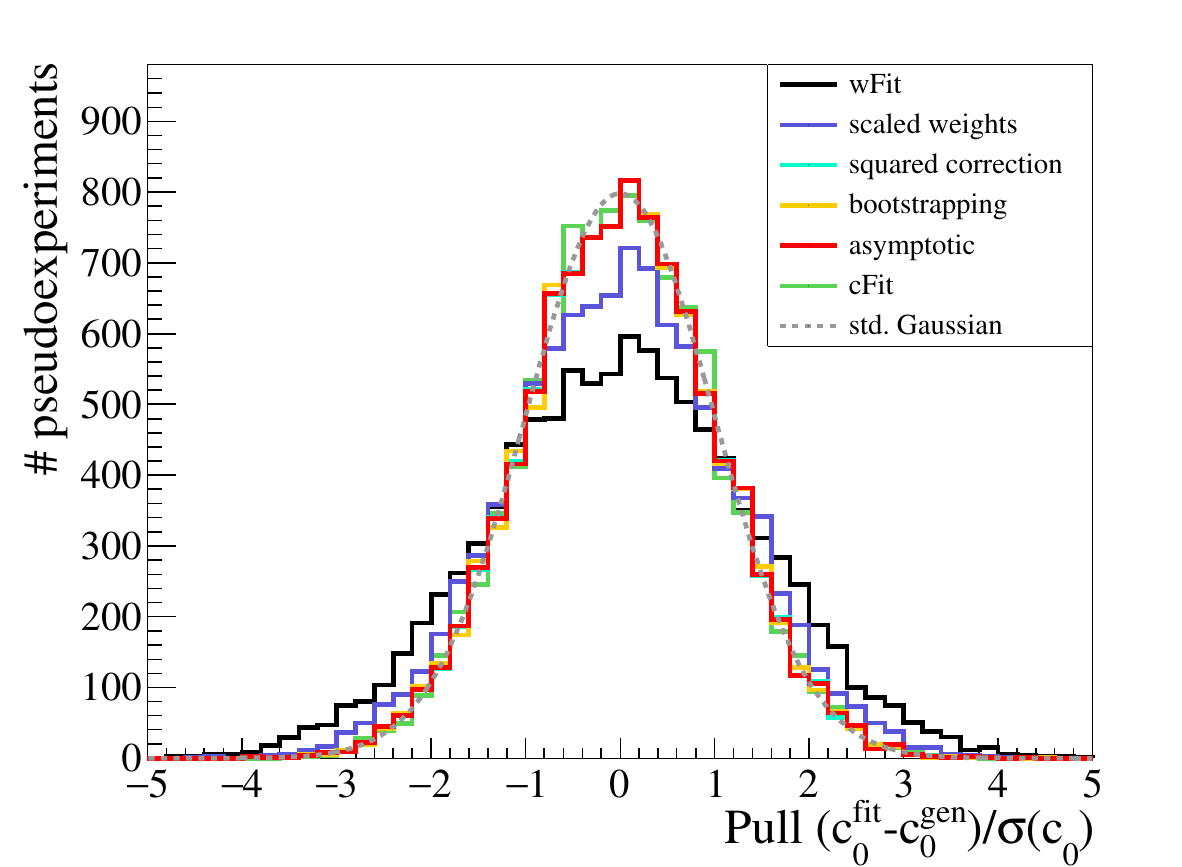}
  }\\
  \subfloat[$c_1$ pull distributions\label{fig:c1pullsa}]{
    \includegraphics[width=0.7\textwidth]{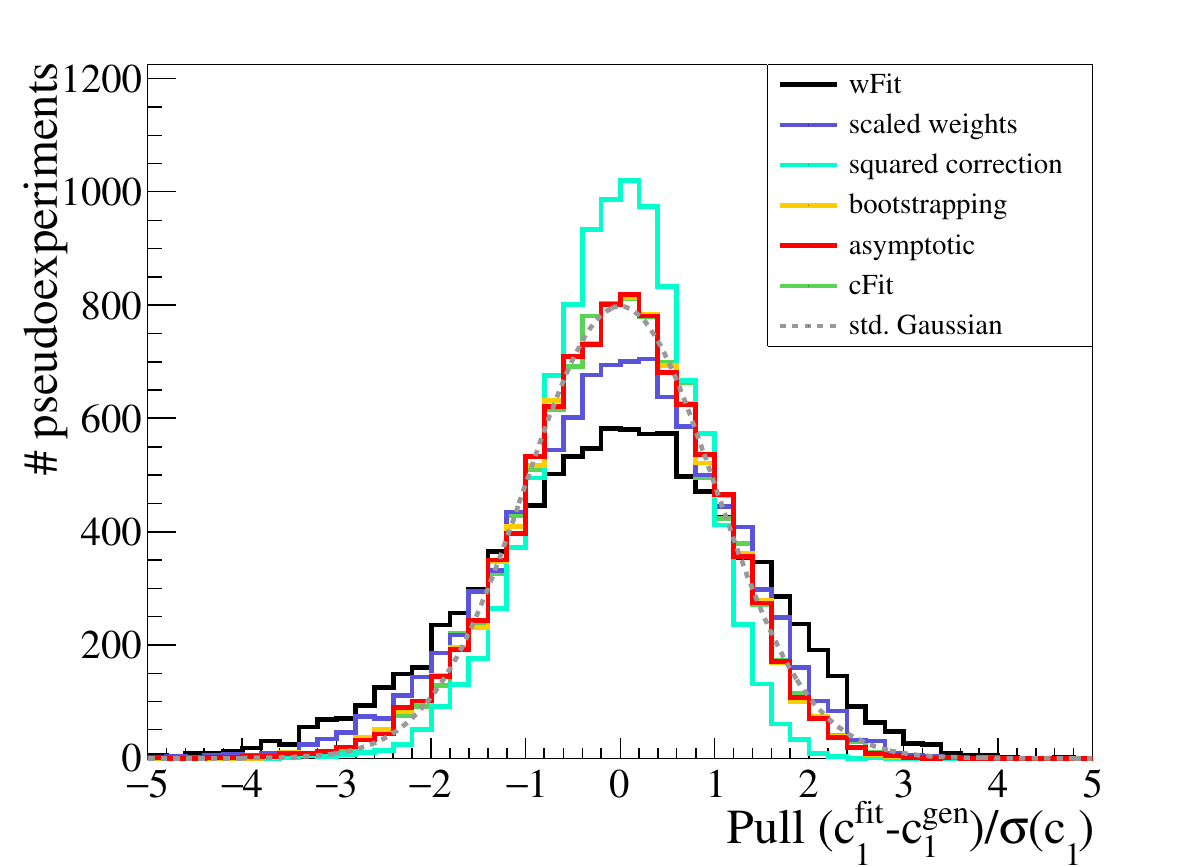}
  }
  \caption{Pull distributions from 10\,000 pseudoexperiments for the different approaches to the uncertainty estimation for the efficiency correction $\epsilon(\cos\theta)=1.0-0.7\cos^2\theta$ at a total yield of $2\,000$ events for each pseudoexperiment.\label{fig:pullsea}}  
\end{figure}

\begin{figure}
\centering
  \subfloat[$c_0$ pull distributions\label{fig:c0pullsb}]{
    \includegraphics[width=0.7\textwidth]{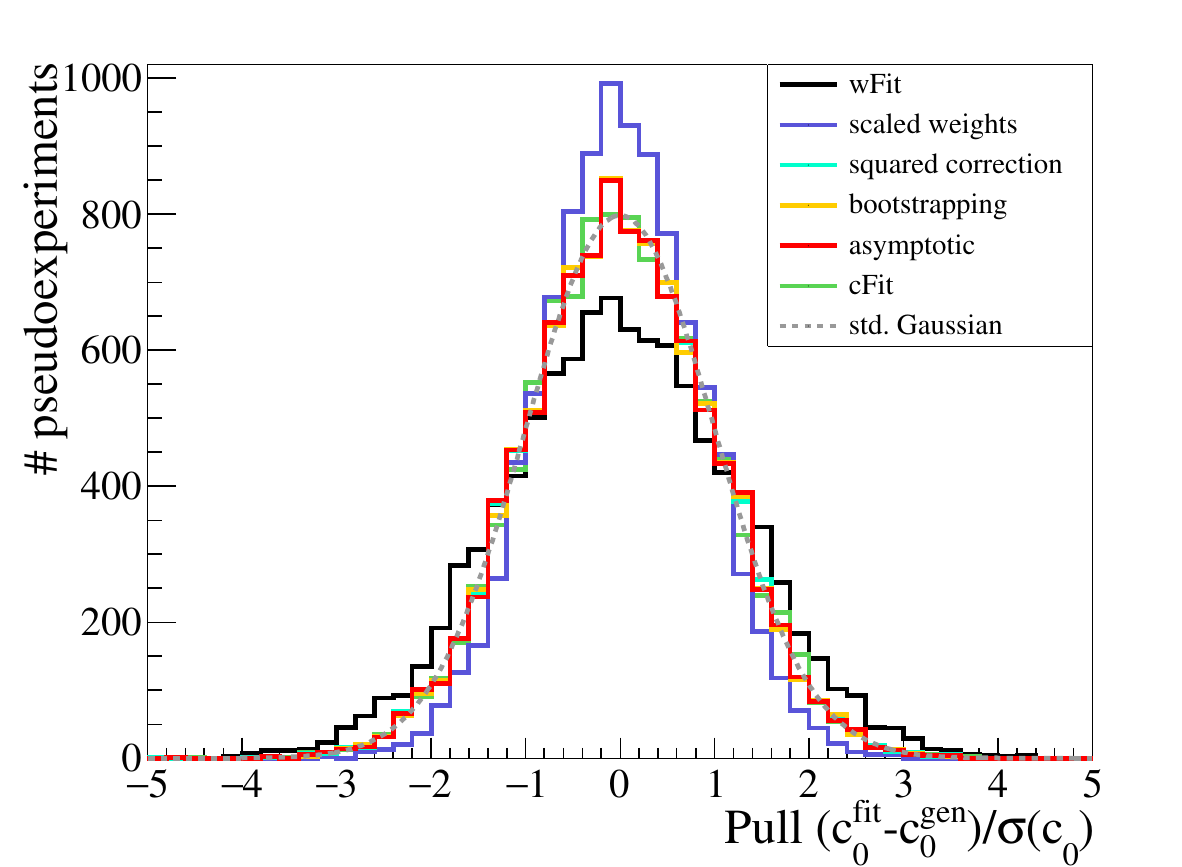}
  }\\
  \subfloat[$c_1$ pull distributions\label{fig:c1pullsb}]{
    \includegraphics[width=0.7\textwidth]{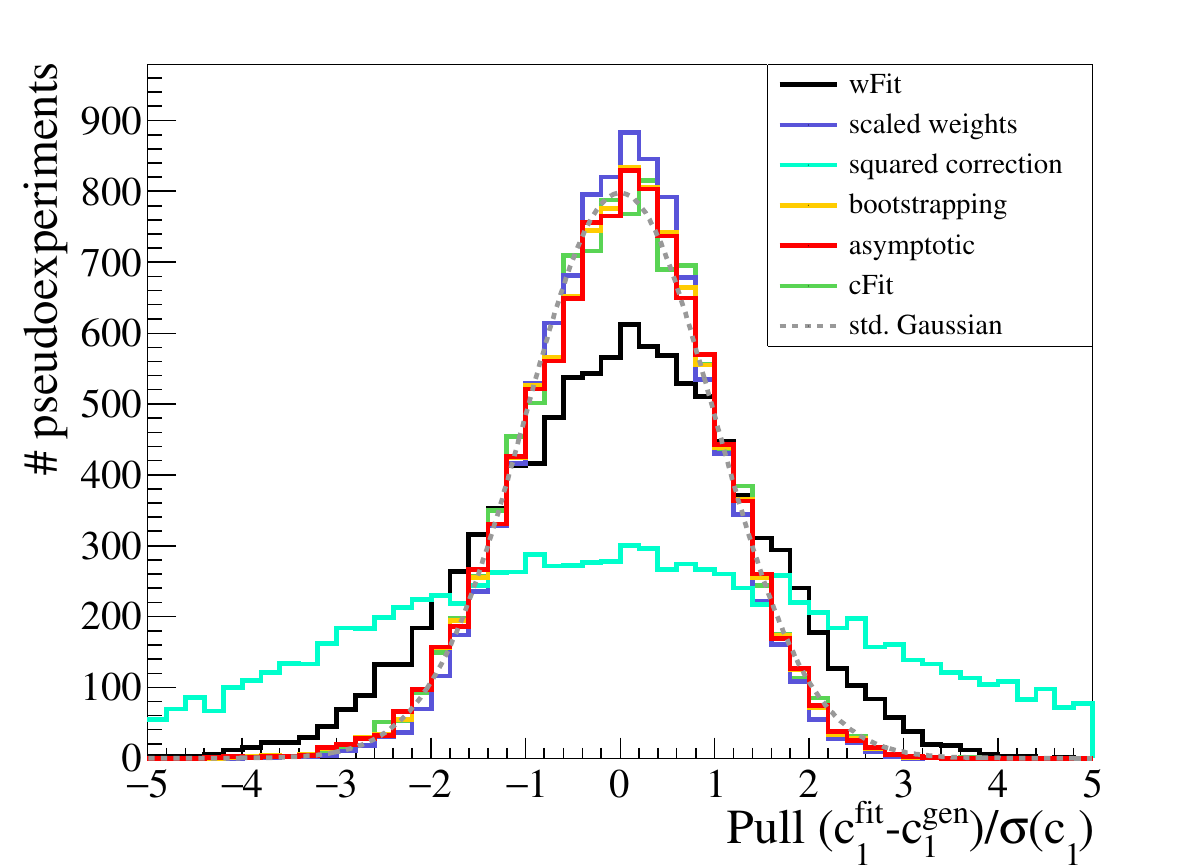}
  }
  \caption{Pull distributions from 10\,000 pseudoexperiments for the different approaches to the uncertainty estimation for the efficiency correction $\epsilon(\cos\theta)=0.3+0.7\cos^2\theta$ at a total yield of $2\,000$ events for each pseudoexperiment.\label{fig:pullseb}}  
\end{figure}

\begin{figure}
\centering
  \subfloat[$c_0$ pull mean and width\label{fig:c0meanwidtha}]{
    \includegraphics[width=0.49\textwidth]{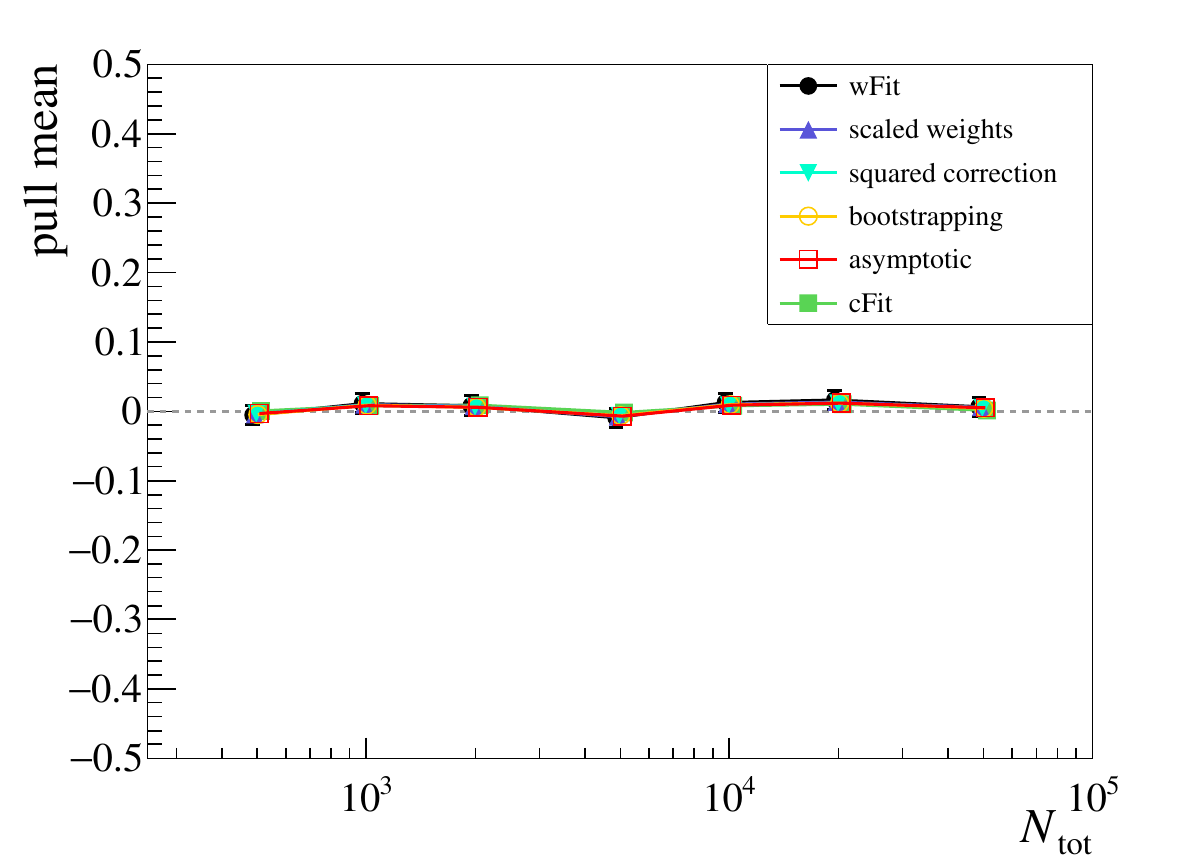}
    \includegraphics[width=0.49\textwidth]{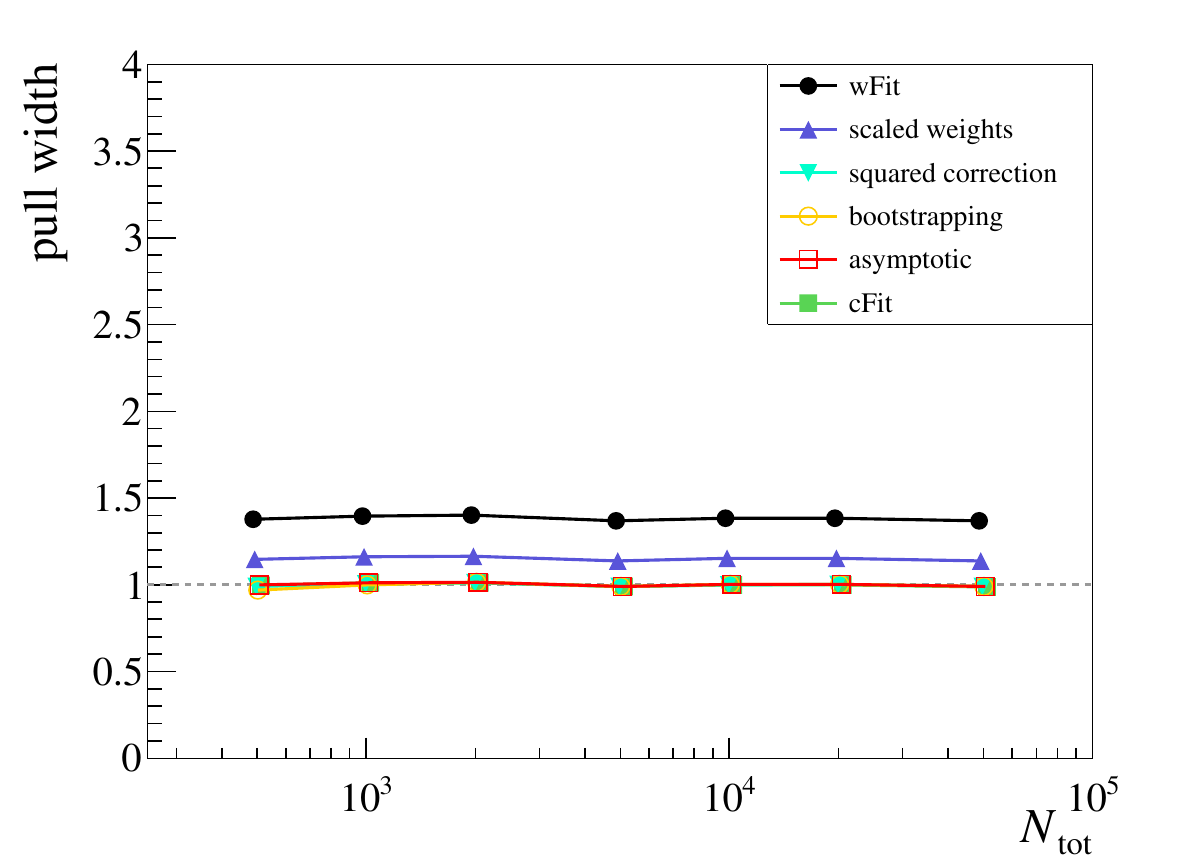}
  }\\
  \subfloat[$c_1$ pull mean and width\label{fig:10meanwidtha}]{
    \includegraphics[width=0.49\textwidth]{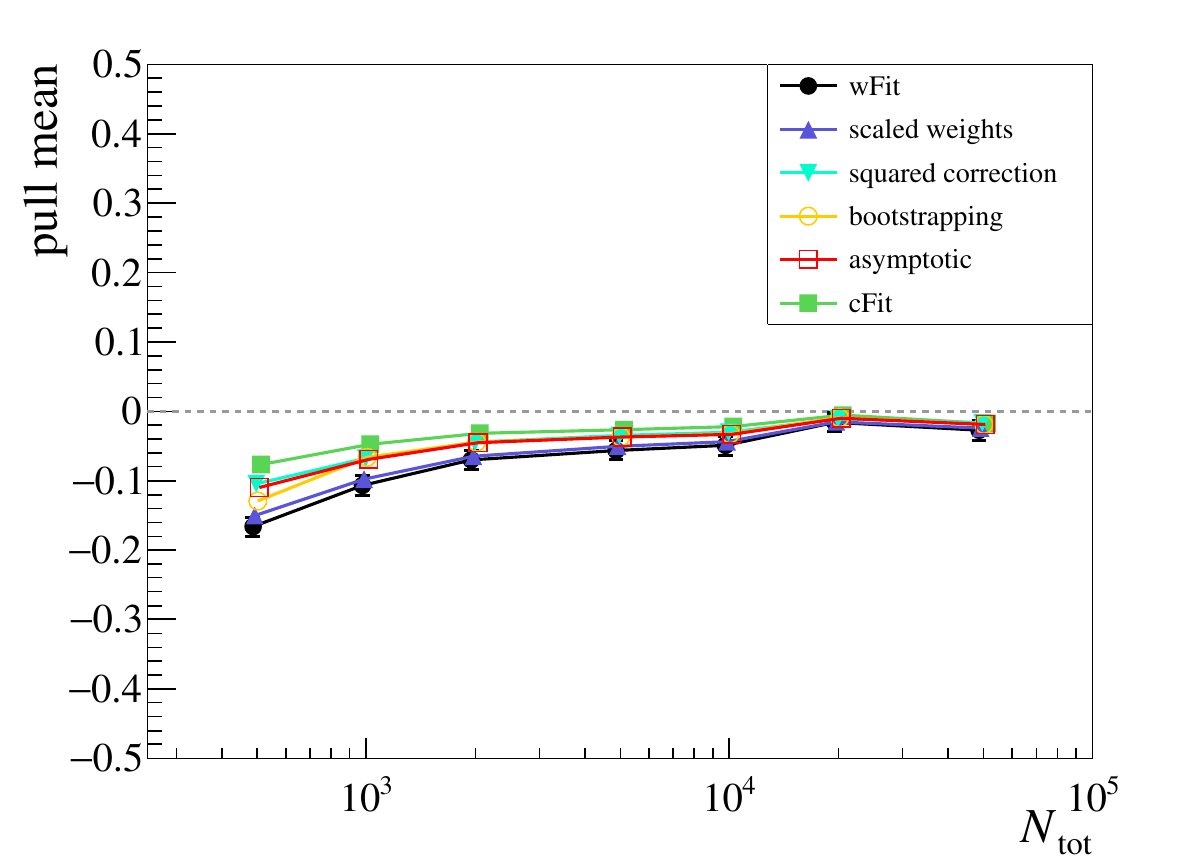}
    \includegraphics[width=0.49\textwidth]{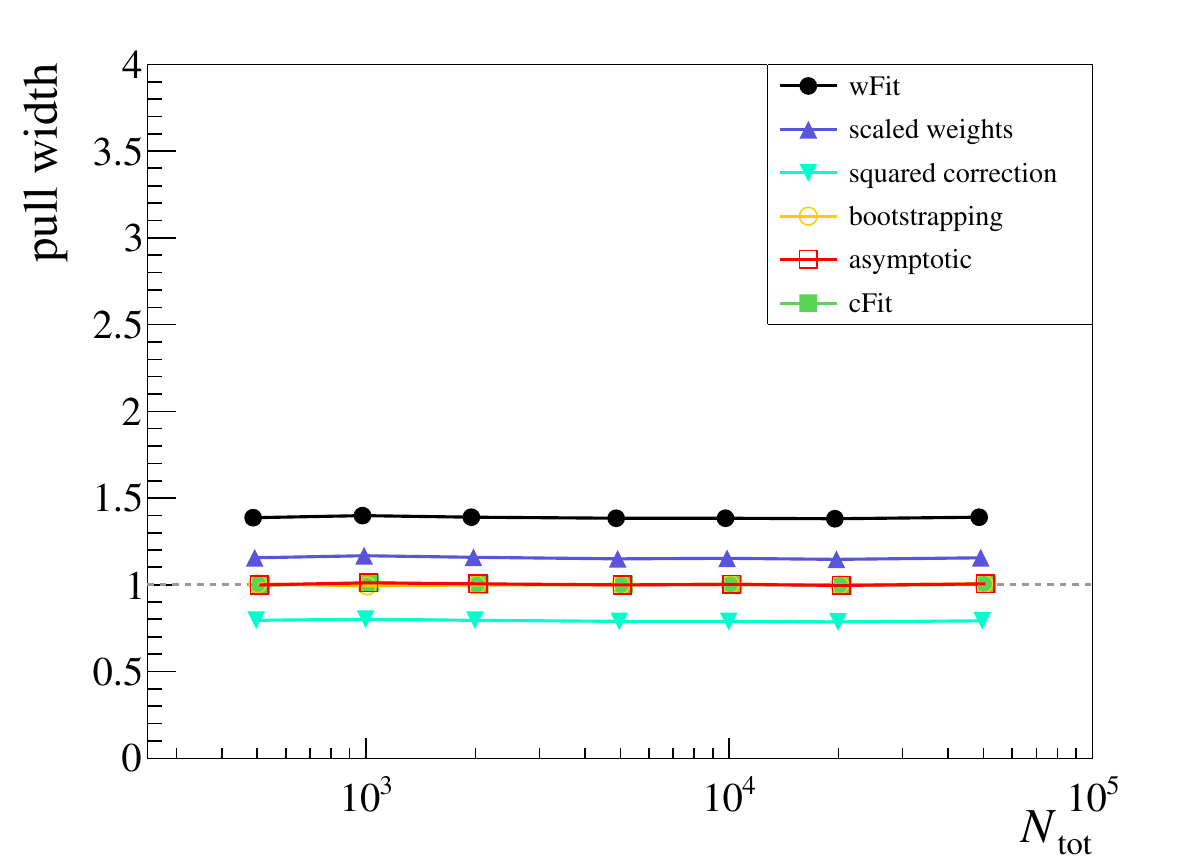}
  }
  \caption{(Left) pull means and (right) pull widths for the efficiency correction $\epsilon(\cos\theta)=1.0-0.7\cos^2\theta$, depending on total event yield $N_{\rm tot}$.
    The markers are slightly horizontally staggered to improve readability.\label{fig:pullsfa}}  
\end{figure}

\begin{figure}
\centering
  \subfloat[$c_0$ pull mean and width\label{fig:c0meanwidthb}]{
    \includegraphics[width=0.49\textwidth]{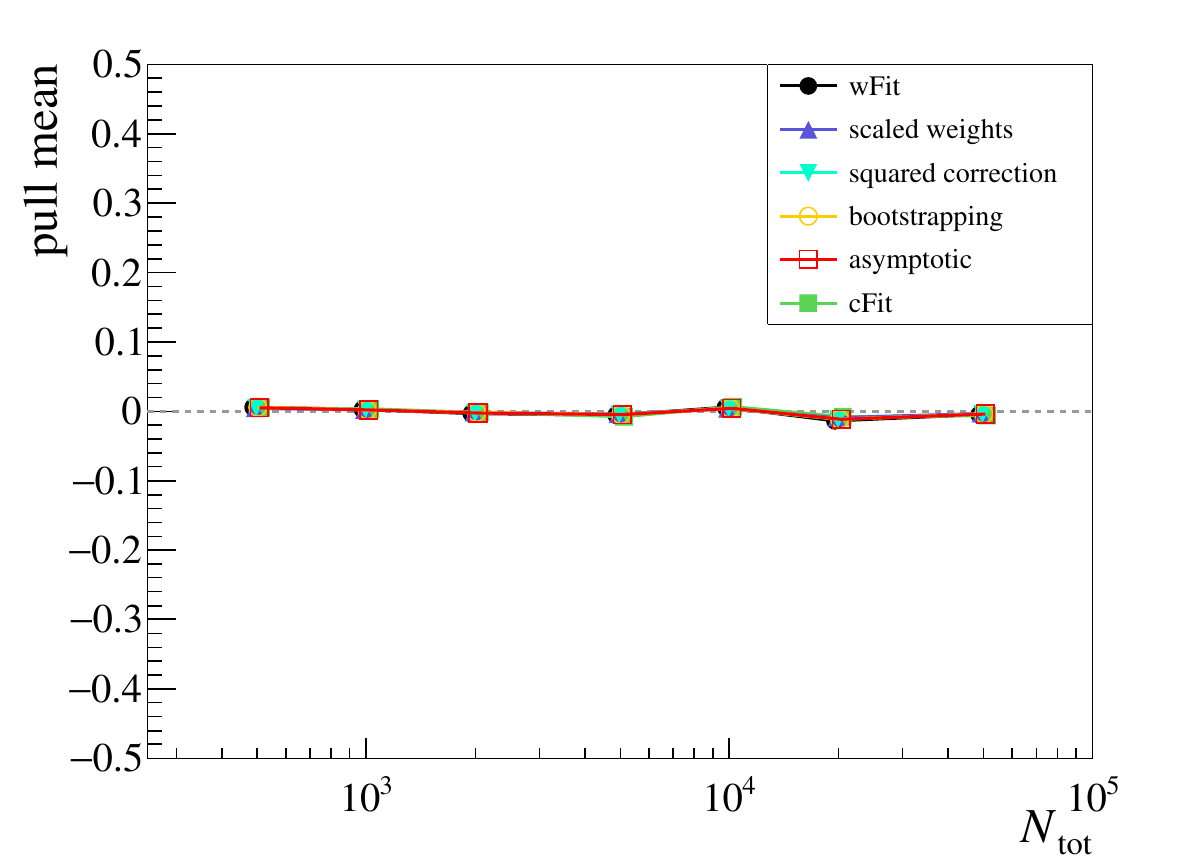}
    \includegraphics[width=0.49\textwidth]{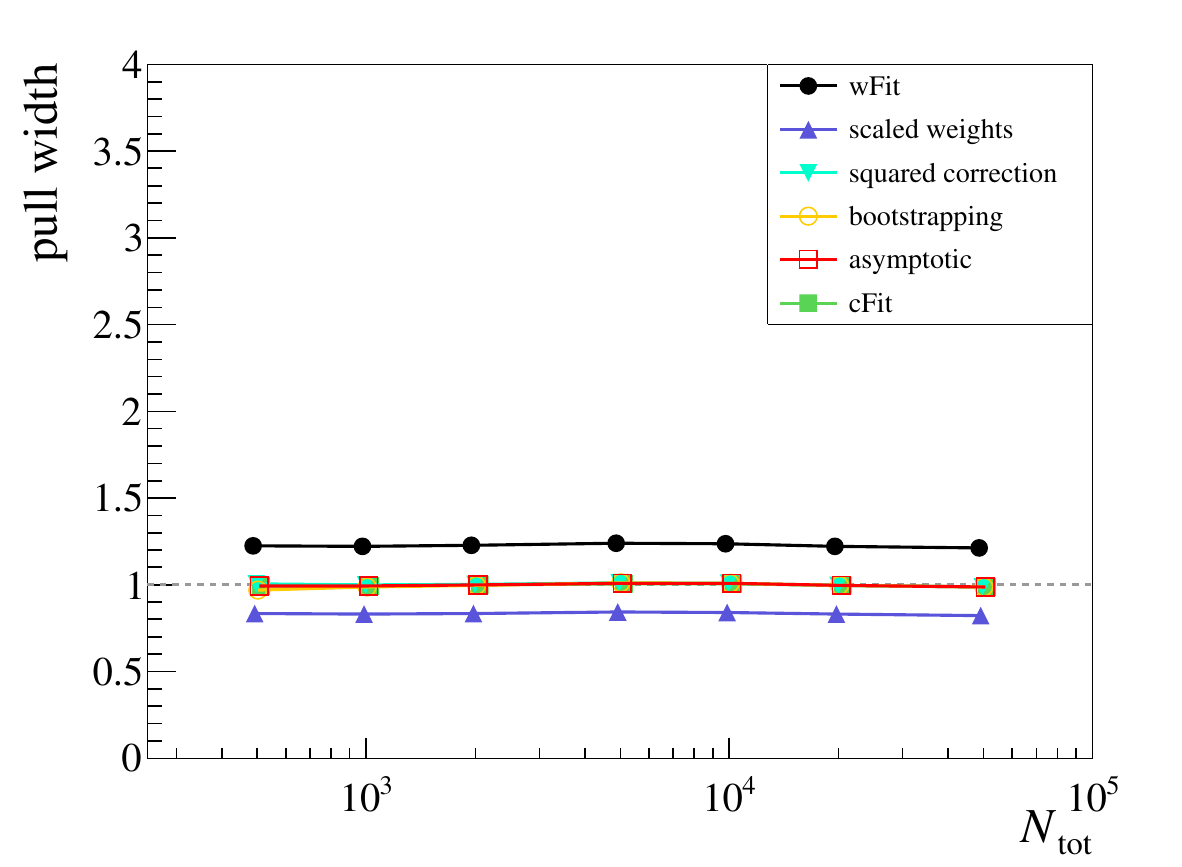}
  }\\
  \subfloat[$c_1$ pull mean and width\label{fig:10meanwidthb}]{
    \includegraphics[width=0.49\textwidth]{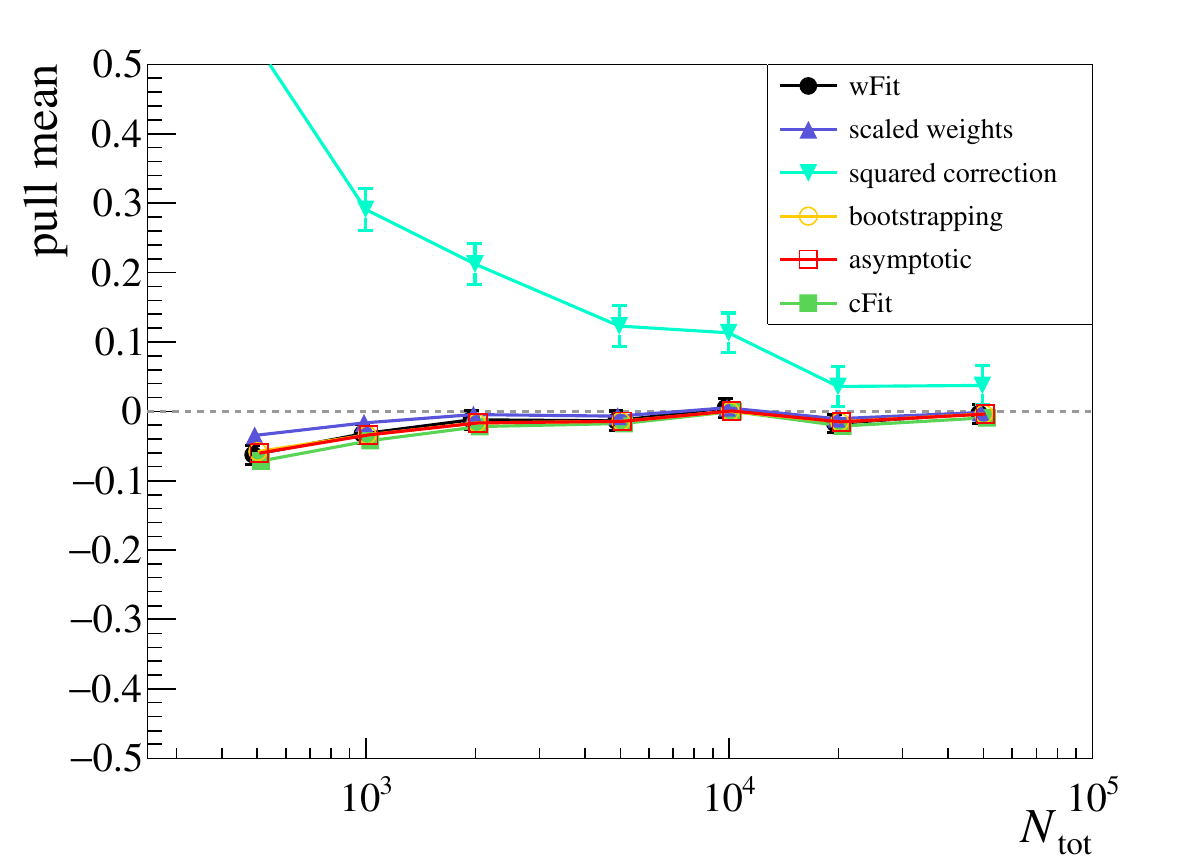}
    \includegraphics[width=0.49\textwidth]{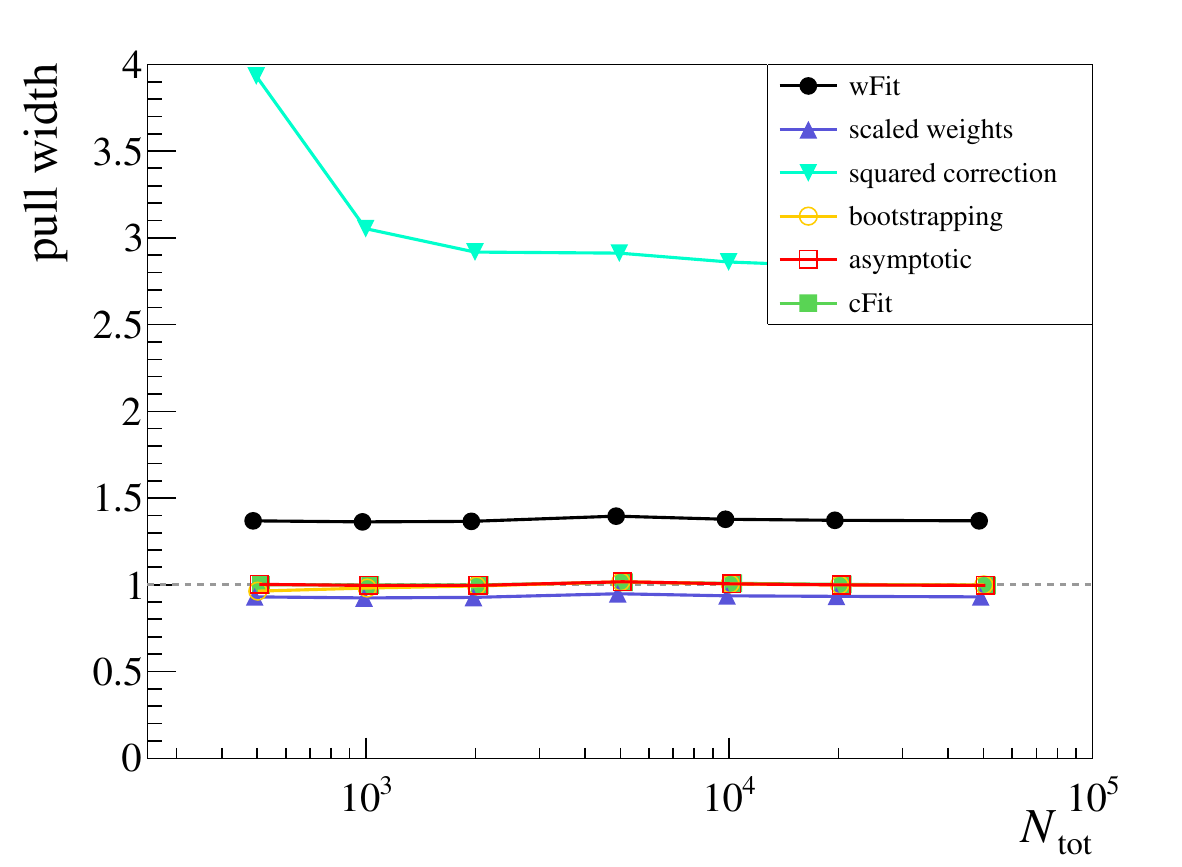}
  }
  \caption{(Left) pull means and (right) pull widths for the efficiency correction $\epsilon(\cos\theta)=0.3+0.7\cos^2\theta$, depending on total event yield $N_{\rm tot}$.
    The markers are slightly horizontally staggered to improve readability.\label{fig:pullsfb}}  
\end{figure}

The pull distributions for the parameters $c_0$ and $c_1$ for 2\,000 events are shown in Figs.~\ref{fig:pullsea} and~\ref{fig:pullseb}. 
The pull means and widths depending on statistics are given in Figs.~\ref{fig:pullsfa} and~\ref{fig:pullsfb}. 
Numerical values are given in Tabs.~\ref{tab:effpullsa} and~\ref{tab:effpullsb} in App.~\ref{sec:appeffs}. 
A few remarks are in order. 

The {\it wFit} method is unbiased but shows significant undercoverage for $c_0$ and $c_1$ for both acceptance corrections tested.
The {\it scaled weights} approach shows significant undercoverage for both $c_0$ and $c_1$ for the acceptance~(a) and overcoverage for acceptance~(b).
In both cases, the coverage remains incorrect even for high statistics. 
Both the use of the {\it wFit} as well as the {\it scaled weights} methods are therefore strongly disfavoured to determine the parameter uncertainties in this example for a simple efficiency correction.
The {\it squared method} shows good behaviour for parameter $c_0$ but incorrect coverage for parameter $c_1$. 
For parameter $c_1$ the method shows overcoverage for acceptance~(a) and very significant undercoverage (more severe than even the {\it wFit}) for acceptance~(b).
The reason for this behaviour is the different expectation value of Eq.~\ref{eq:expectation} with respect to the second derivatives to $c_0$ and $c_1$, as detailed in App.~\ref{app:nonzeroacceptance}.
This illustrates that the {\it squared correction} method, which is widely used in particle physics,
in general does not provide asymptotically correct confidence intervals when using event weights to correct for acceptance effects. 
{\it Bootstrapping} the data sample or using the {\it asymptotic} approach results in 
pull distributions with correct coverage 
for both $c_0$ and $c_1$ and both acceptance effects. 
No bias is observed for parameter $c_0$ and only a small bias is found for $c_1$ at low statistics.  
This paper therefore advocates for the use of the {\it asymptotic} method (or alternatively {\it bootstrapping}) when using event weights to account for acceptance corrections. 
The pull distributions for the {\it cFit} also show, as expected, good behaviour.
As there is no loss of information for the {\it cFit}, 
it can result in better sensitivity, as shown by the relative efficiencies given in Tabs.~\ref{tab:effpullsa} and~\ref{tab:effpullsb}, and its use should be strongly considered, where feasible. 

\subsection[Background subtraction using \textit{sWeights}]{Background subtraction using \textit{\textbf{sWeights}}}
\label{sec:sweights}
\begin{figure}
  \centering
  \includegraphics[width=0.49\textwidth]{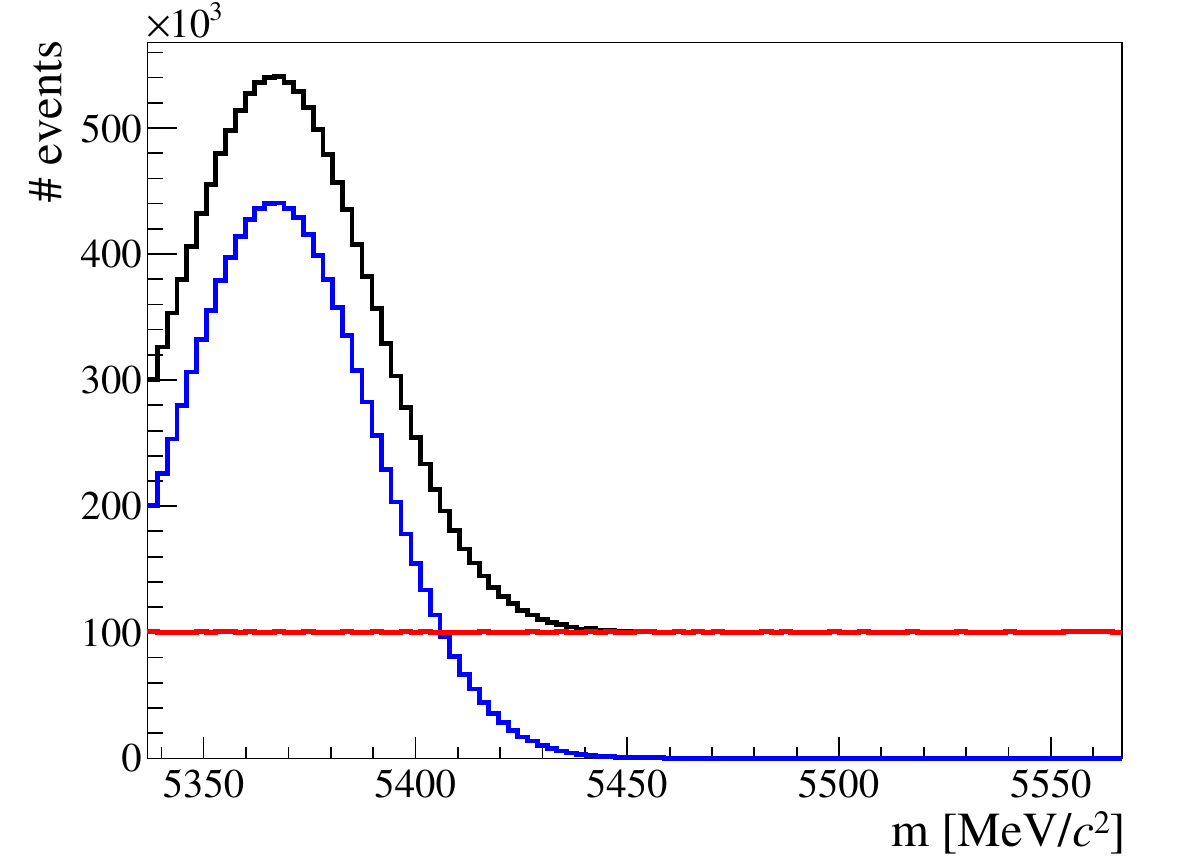}
  \caption{Discriminating mass distribution for (black) the full data, (blue) signal and (red) background.\label{fig:mass}}
\end{figure}

As second specific example for the determination of confidence intervals in the presence of event weights,
the determination of the lifetime $\tau$ of an exponential decay in the presence of background is discussed. 
The {\it sPlot} method~\cite{Pivk:2004ty} is used to statistically subtract the background component. 
As {\it discriminating variable}, the reconstructed mass is used.
In this example, the signal is distributed according to a Gaussian in the reconstructed mass,
and the background is described by a single Exponential function with slope $\alpha_{\rm bkg}$. 
Figure~\ref{fig:mass} shows the mass distribution for signal and background components.
The parameters used in the generation of the pseudoexperiments are listed in Tab.~\ref{tab:parameters}. 
The configuration is purposefully chosen such that there is a significant correlation between the yields and the slope of the background exponential,
to illustrate the effect of fixing nuisance parameters in the {\it sPlot} formalism, as discussed in Sec.~\ref{sec:splots}. 
The resulting mean correlation matrix for the mass fit is shown in Tab.~\ref{tab:correlation}. 
The simpler case, where no significant correlation between $\alpha_{\rm bkg}$ and the event yields is present, due to a different choice of mass range,
is discussed in App.~\ref{sec:appsweightsnocorr}.
\begin{table}
  \centering
  \begin{minipage}[t]{0.59\textwidth}
  \subfloat[Parameters used in the generation\label{tab:parameters}]{
    \begin{tabular}{l|r}\hline
      parameter & value\\\hline\hline
      $f_{\rm sig}=N_{\rm sig}/(N_{\rm sig}+N_{\rm bkg})$ & $0.5$\\
      $\alpha_{\rm bkg}$ & $0.0$\\ \hline
      $m(B)$ & $ 5.367\gevcc$\\
      $\sigma(m)$ & $23\mevcc$\\
      mass range & $[5\,337, 5\,567]\mevcc$\\\hline
      $\tau_{\rm sig}^{\rm gen}$ & $1.5\,{\mathrm ps}$\\
      $t$ range & $[0,10]\,{\mathrm ps}$\\
    \hline\end{tabular}
  }
  \end{minipage}
  \begin{minipage}[t]{0.39\textwidth}
  \subfloat[Mean correlation matrix\label{tab:correlation}]{
    \begin{tabular}{l|rrr}\hline
      & $N_{\rm sig}$ & $N_{\rm bkg}$ & $\alpha_{\rm bkg}$\\\hline\hline
      $N_{\rm sig}$ & $1.00$ & $-0.63$ & $-0.67$\\
      $N_{\rm bkg}$ & $-0.63$ & $1.00$ & $0.67$\\
      $\alpha_{\rm bkg}$ & $-0.67$ & $0.67$ & $1.00$\\
      \hline\end{tabular}
  }
  \end{minipage}
  \caption{(Left) the parameters used in the generation of the pseudoexperiments.
    Only $N_{\rm sig}$, $N_{\rm bkg}$, and the background slope $\alpha_{\rm bkg}$ are varied in the mass fit. 
    The background slope $\alpha_{\rm bkg}$ is then fixed for the determination of the {\it sWeights}. 
    (Right) the mean correlation matrix from the mass fit 
    when both the yields and $\alpha_{\rm bkg}$ are allowed to float.}
\end{table}

\begin{figure}
  \centering
  \subfloat[Exponential background model]{
  \includegraphics[width=0.49\textwidth]{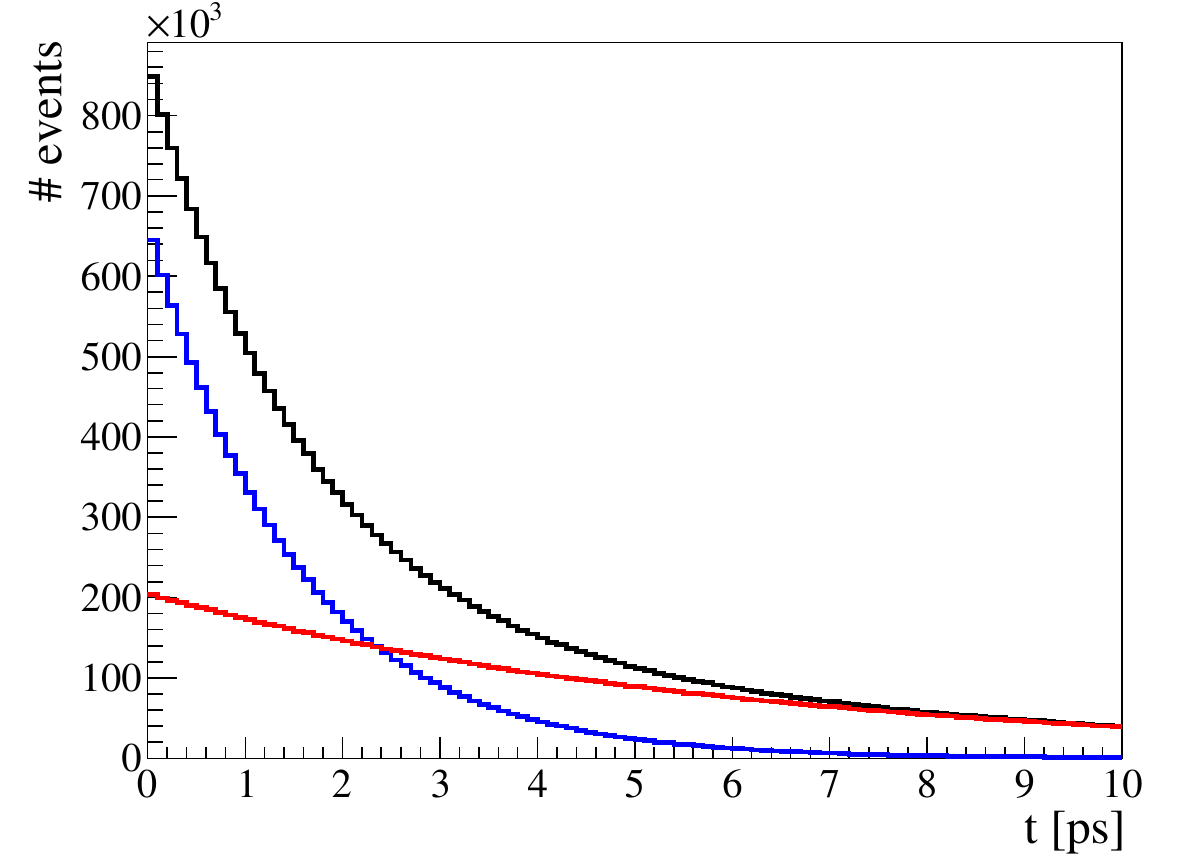}}
  \subfloat[Gaussian background model]{
  \includegraphics[width=0.49\textwidth]{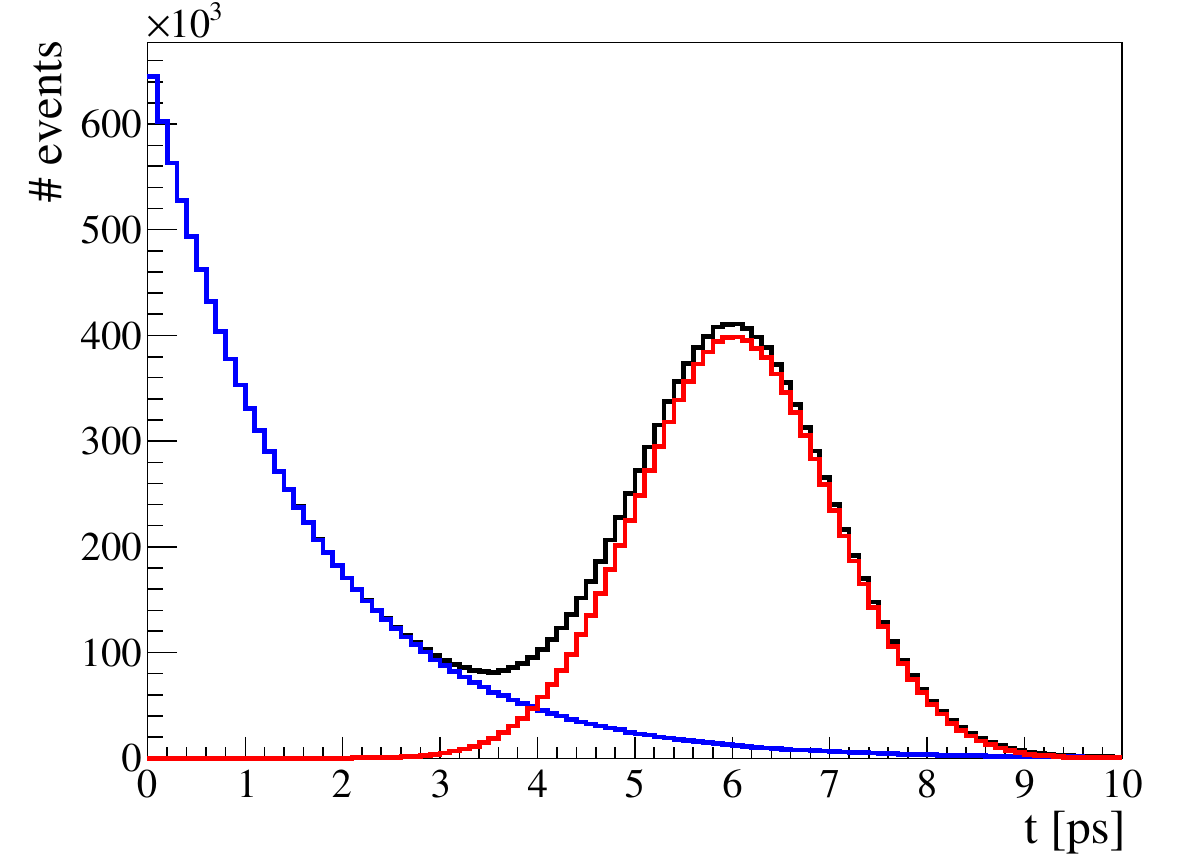}}\\
  \subfloat[Triangular background model]{
  \includegraphics[width=0.49\textwidth]{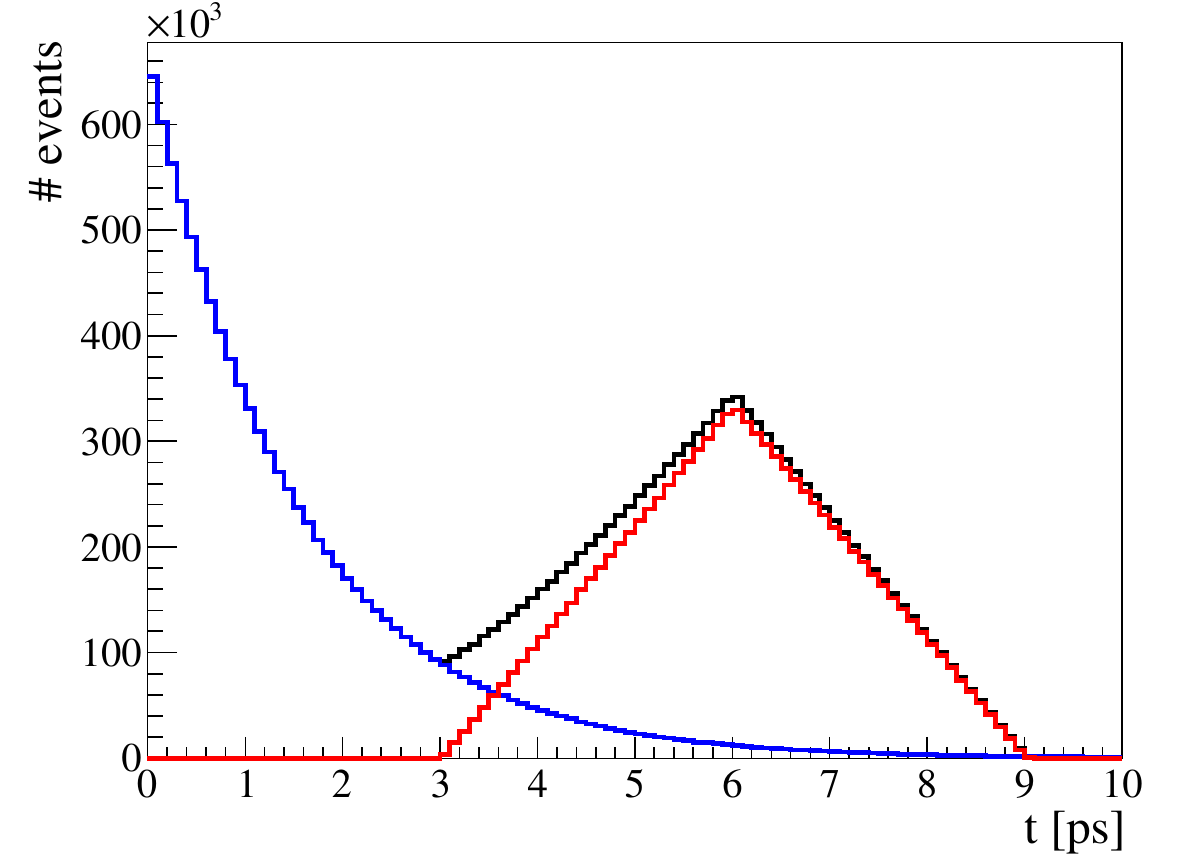}}
  \subfloat[Flat background model]{
  \includegraphics[width=0.49\textwidth]{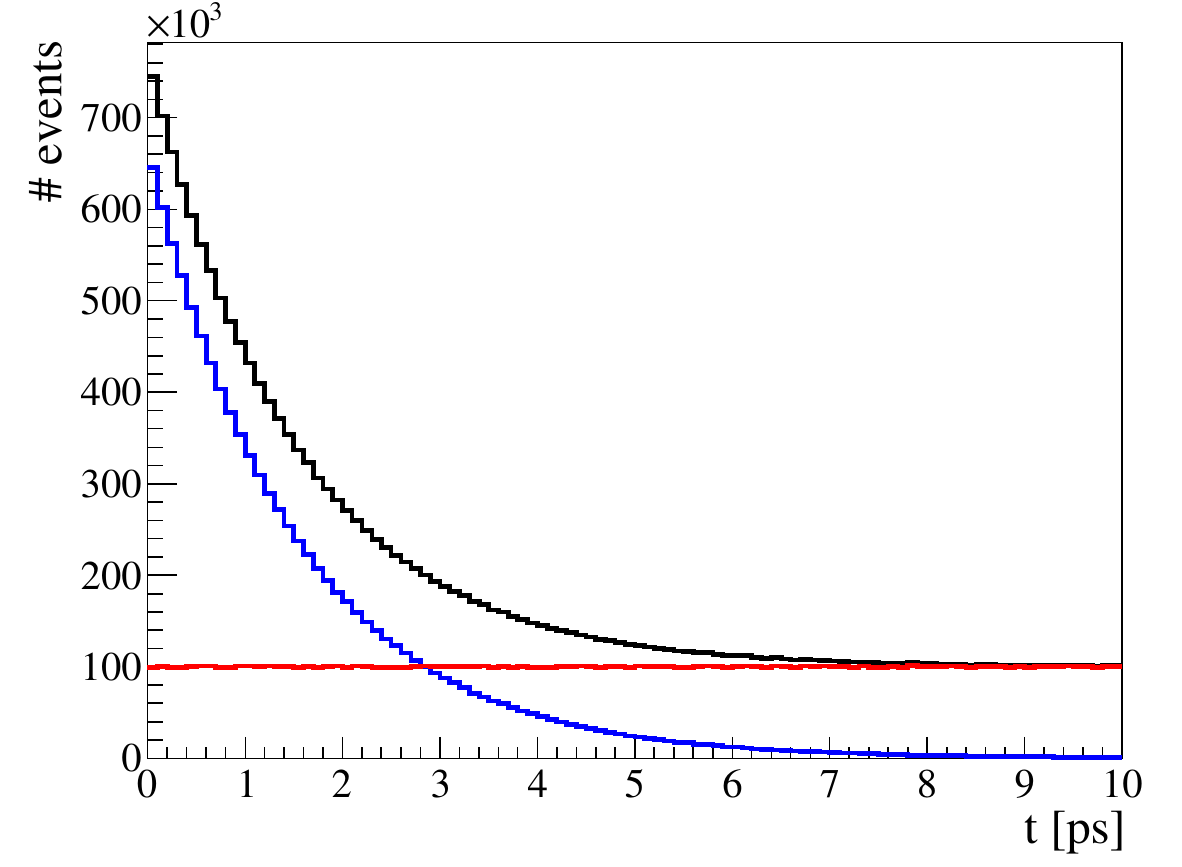}}  
  \caption{Decay time distributions for the four different background options for (black) the full data, (blue) signal and (red) background.
    The signal and background components are obtained using {\it sWeights}.\label{fig:time}}
\end{figure}

The decay time distribution (the {\it control variable}) that is used to determine the lifetime is a single Exponential for the signal.
For the background component, several different shapes were tested:
(a) A single Exponential with long lifetime, (b) a Gaussian distribution, (c) a triangular distribution, and (d) a flat distribution in the decay time. 
Figure~\ref{fig:time} shows the decay time distribution for the different options. 
The decay time distributions for signal and background components shown are obtained using the {\it sPlot} formalism~\cite{Pivk:2004ty} described in Sec.~\ref{sec:splots}. 

The parameter $\tau$ is determined using a weighted unbinned maximum likelihood fit solving the maximum likelihood condition Eq.~\ref{eq:mlweighted} numerically.
Its uncertainty $\sigma(\tau)$ is determined using the different methods
for weighted unbinned maximum likelihood fits discussed in Sec.~\ref{sec:maximumlikelihood}.
The following approaches are studied:
\begin{enumerate}[label=(\alph*)]
\item A weighted fit determining the uncertainties according to Eq.~\ref{eq:covprime} without any correction.
  This method is denoted as {\it sFit} in the following. 
\item Scaling the weights according to Eq.~\ref{eq:scaling}.
  The approach is denoted as {\it scaled weights}. 
\item Determining the covariance matrix using Eq.~\ref{eq:approximate}. 
  This method is referred to as {\it squared correction}. 
\item
  Bootstrapping the data (using 1000 bootstraps) with replacement, without re\-de\-ri\-ving the {\it sWeights} (\ie\ keeping the original {\it sWeights} for each event). 
  Denoted as {\it bootstrapping} in the following.
\item Bootstrapping the data (again using 1000 bootstraps) and re\-de\-ri\-ving the {\it sWeights} for every bootstrapped sample, in the following denoted as {\it full bootstrapping}.
\item The asymptotic method to determine the covariance according to Eq.~\ref{eq:sweightsvariance} as discussed in Sec.~\ref{sec:splots}, 
  but not accounting for the impact of nuisance parameters in the determination of the {\it sWeights}.
  This approach is referred to as {\it asymptotic} method. 
\item The method to determine the covariance according to Eq.~\ref{eq:correctvarnuisance}, 
  which includes the effect of nuisance parameters in the \textit{sWeight} determination. 
  This method is denoted as {\it full asymptotic}. 
\item A conventional fit ({\it cFit}) modelling both signal and background components in two dimensions (mass and decay time) for comparison.
  As the main point of using {\it sWeights} is to remove the need to model the background contribution in the fit, this method is given purely for comparison.  
\end{enumerate}
The performance of the different methods is evaluated using pseudoexperiments.
Every study consists of 10\,000 data samples generated and then fit for an initial determination of the {\it sWeights}.
For every method, the same data samples are used. 

The performance of the different methods is compared using the distribution of the pull,
defined as $p_i(\tau)=(\tau_i-\tau^{\rm gen}_{\rm sig})/\sigma_i(\tau)$.
Here, $\tau_i$ is the central value determined by the weighted maximum likelihood fit and $\sigma_i(\tau)$ the uncertainty determined by the above methods. 
The lifetime used in the generation is denoted as $\tau^{\rm gen}_{\rm sig}$. 
To study the influence of statistics, pseudoexperiments are performed for different numbers of events. 
The total yields $N_{\rm tot}=N_{\rm sig}+N_{\rm bkg}$ generated correspond to 400, 1\,000, 2\,000, 4\,000, 10\,000 and 20\,000 events.
The signal fraction used in the generation is $f_{\rm sig}=N_{\rm sig}/(N_{\rm sig}+N_{\rm bkg})=0.5$. 

\begin{figure}
\centering
  \subfloat[Exponential background model\label{fig:pullsexp}]{
    \includegraphics[width=0.7\textwidth]{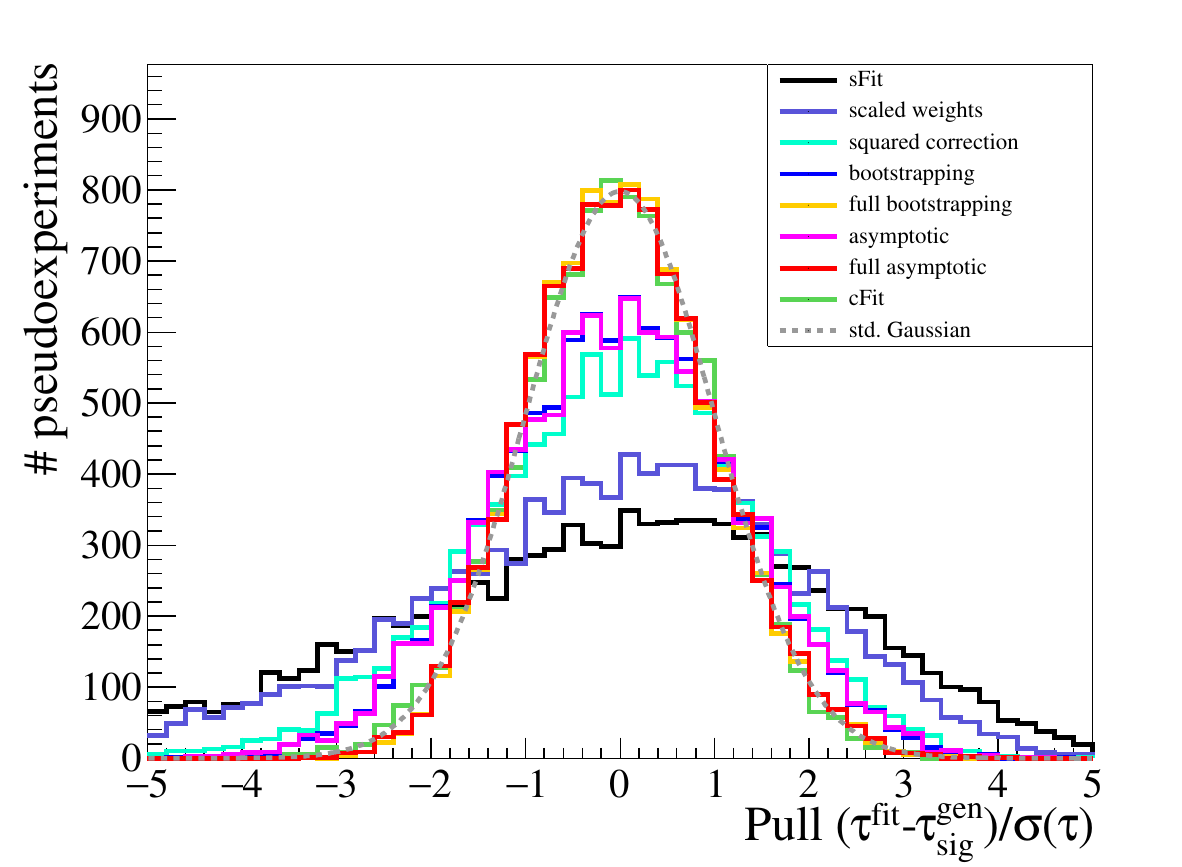}    
  }\\
  \subfloat[Gaussian background model\label{fig:pullsgauss}]{
    \includegraphics[width=0.7\textwidth]{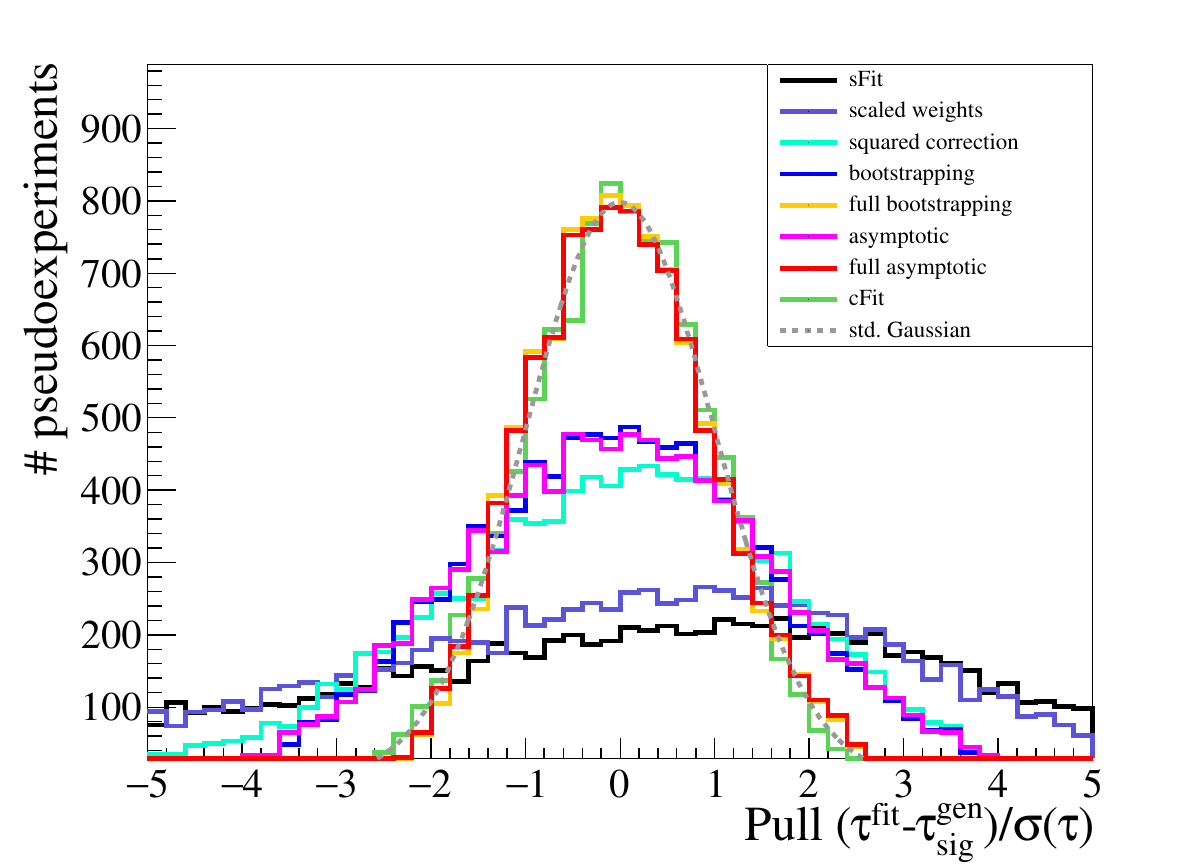}
  }
  \caption{Pull distributions from 10\,000 pseudoexperiments for the different approaches to the uncertainty estimation for a total yield of $2\,000$ events in each pseudoexperiment. 
    The different figures shown correspond to the different background models as specified in Sec.~\ref{sec:examples}.\label{fig:pullsa}}  
\end{figure}

\begin{figure}
\centering
  \subfloat[Triangular background model\label{fig:pullstriangle}]{
    \includegraphics[width=0.7\textwidth]{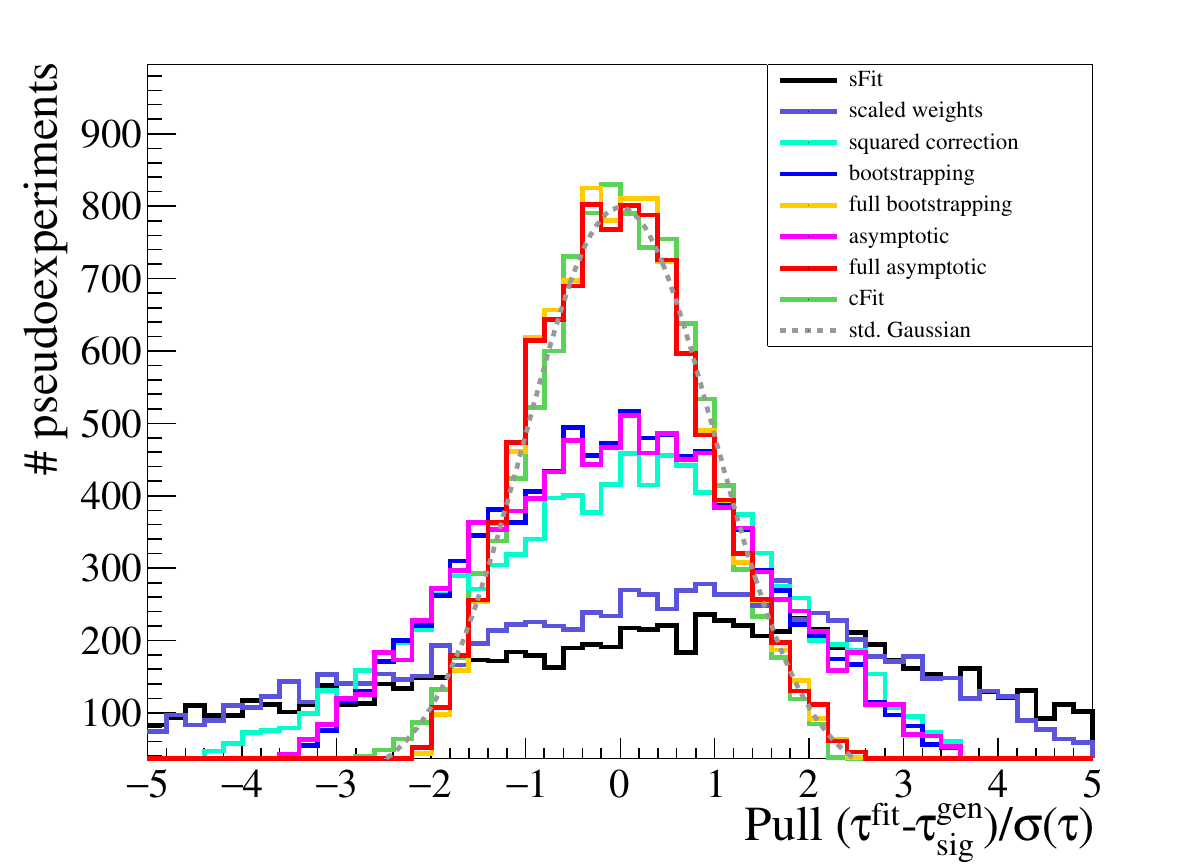}
  }\\
  \subfloat[Flat background model\label{fig:pullsflat}]{
    \includegraphics[width=0.7\textwidth]{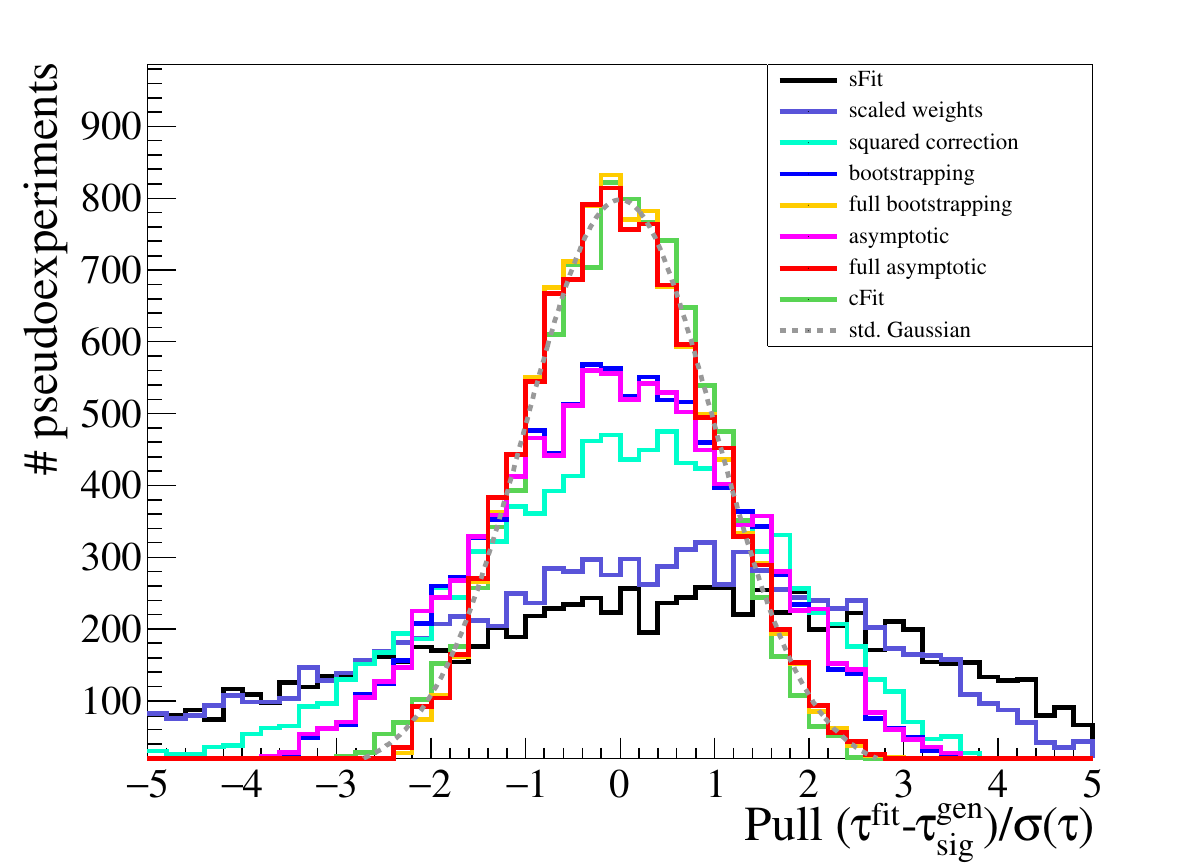}
  }
  \caption{Pull distributions from 10\,000 pseudoexperiments for the different approaches to the uncertainty estimation for a total yield of $2\,000$ events in each pseudoexperiment. 
    The different figures shown correspond to the different background models as specified in Sec.~\ref{sec:examples}.\label{fig:pullsb}}  
\end{figure}

\begin{figure}
\centering
  \subfloat[Exponential background model\label{fig:exp}]{
    \includegraphics[width=0.49\textwidth]{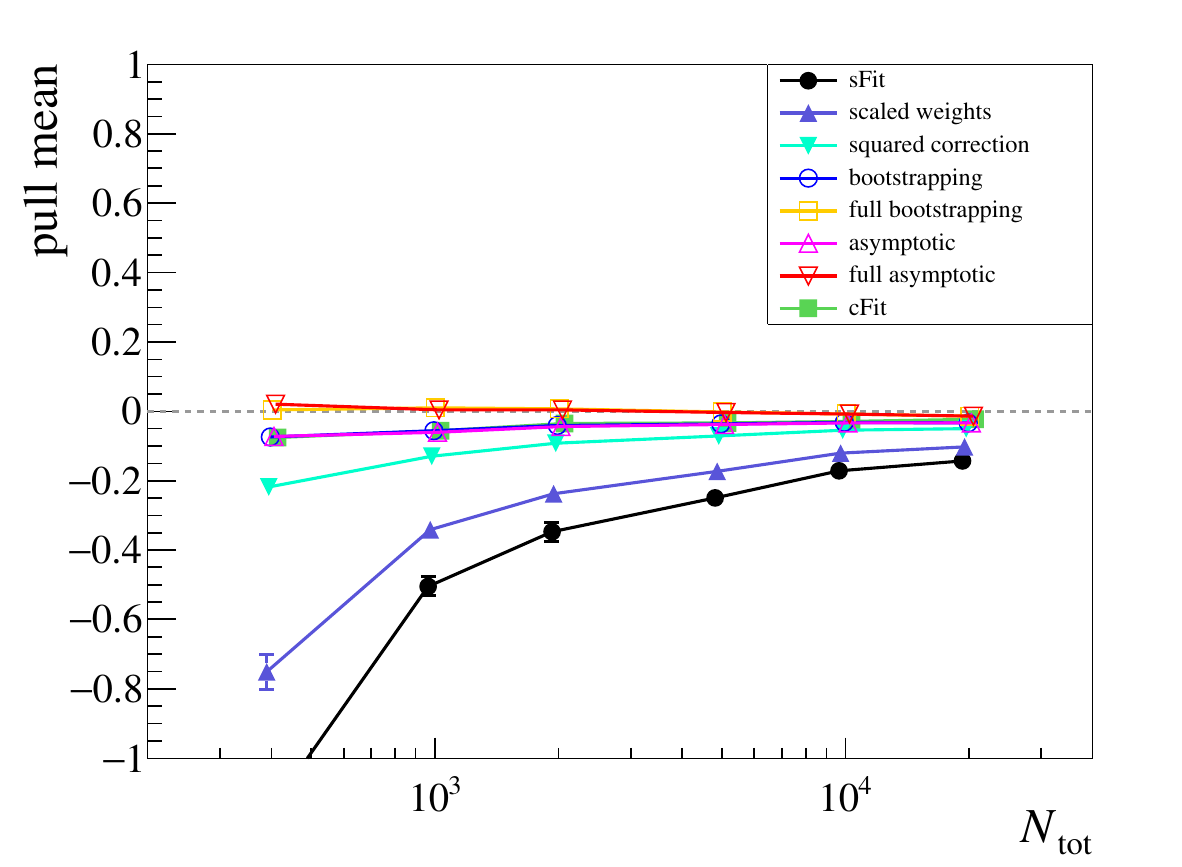}
    \includegraphics[width=0.49\textwidth]{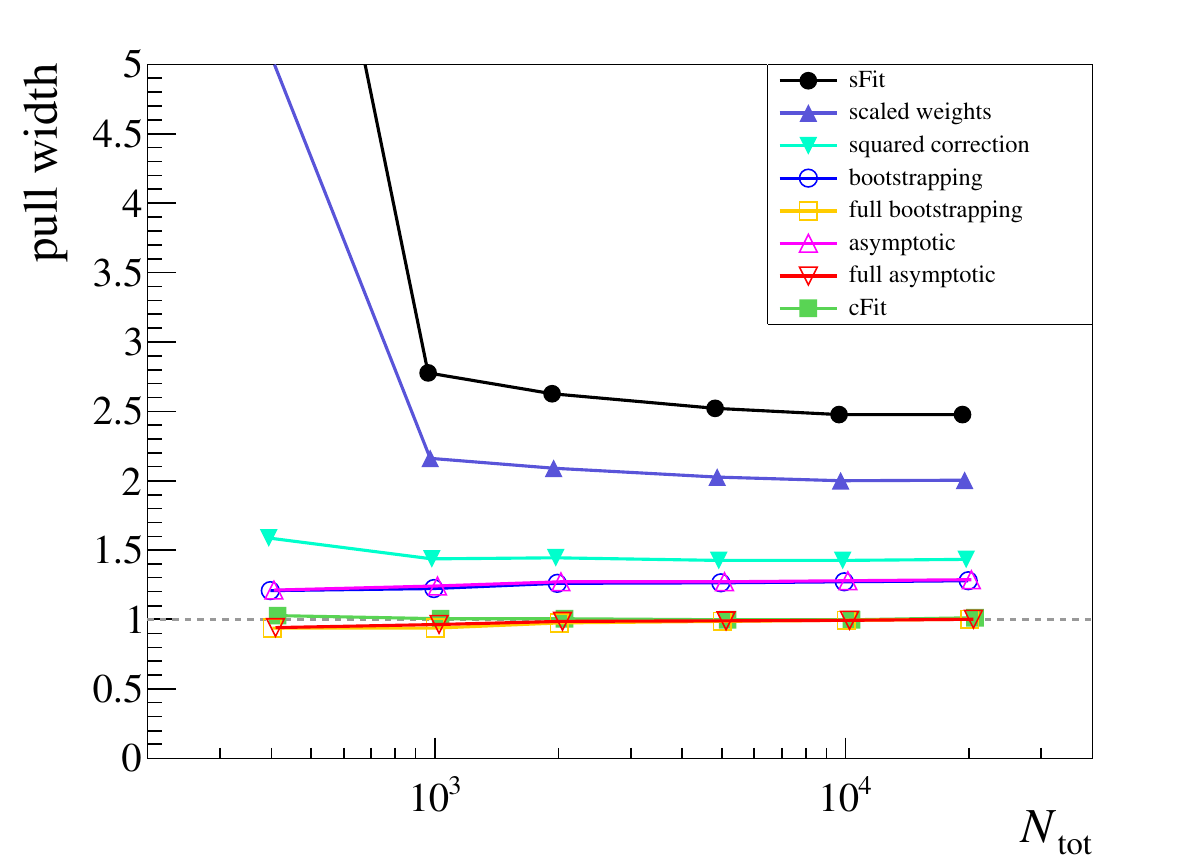}
  }\\
  \subfloat[Gaussian background model\label{fig:gauss}]{  
    \includegraphics[width=0.49\textwidth]{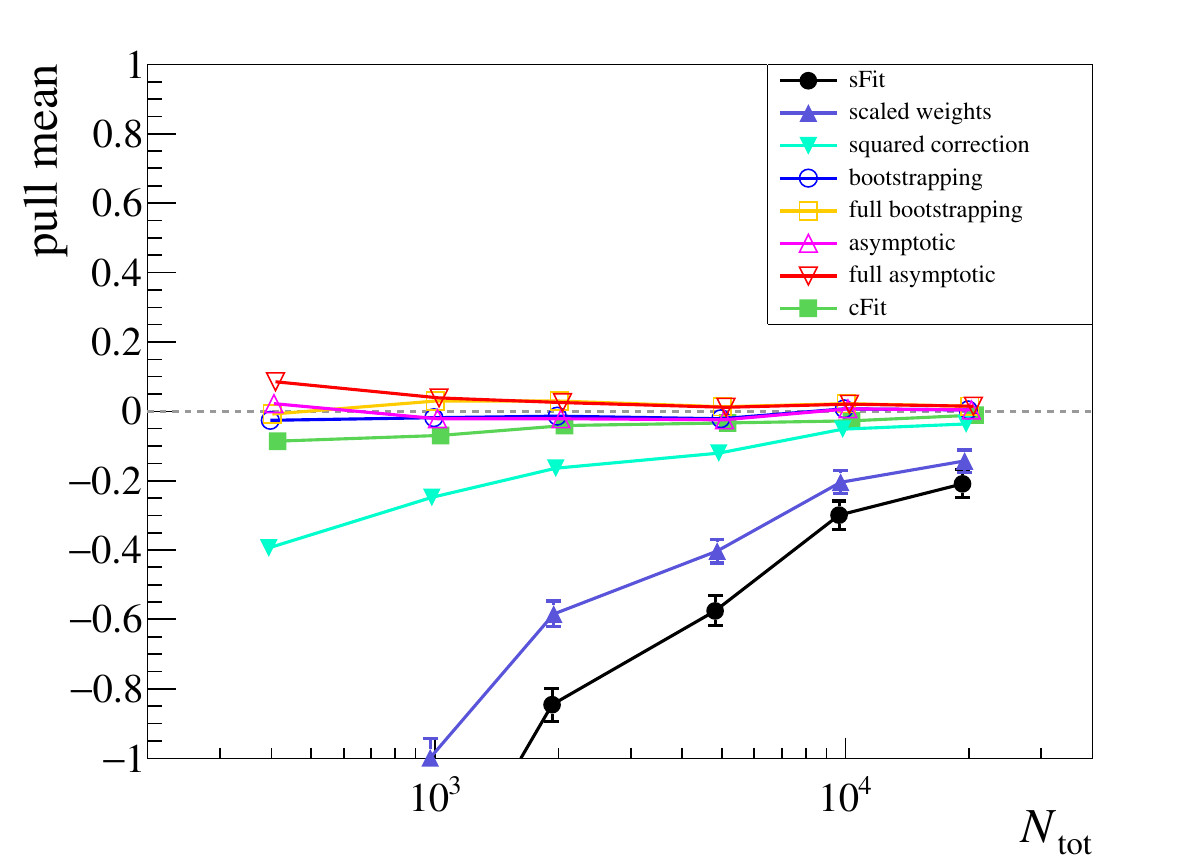}
    \includegraphics[width=0.49\textwidth]{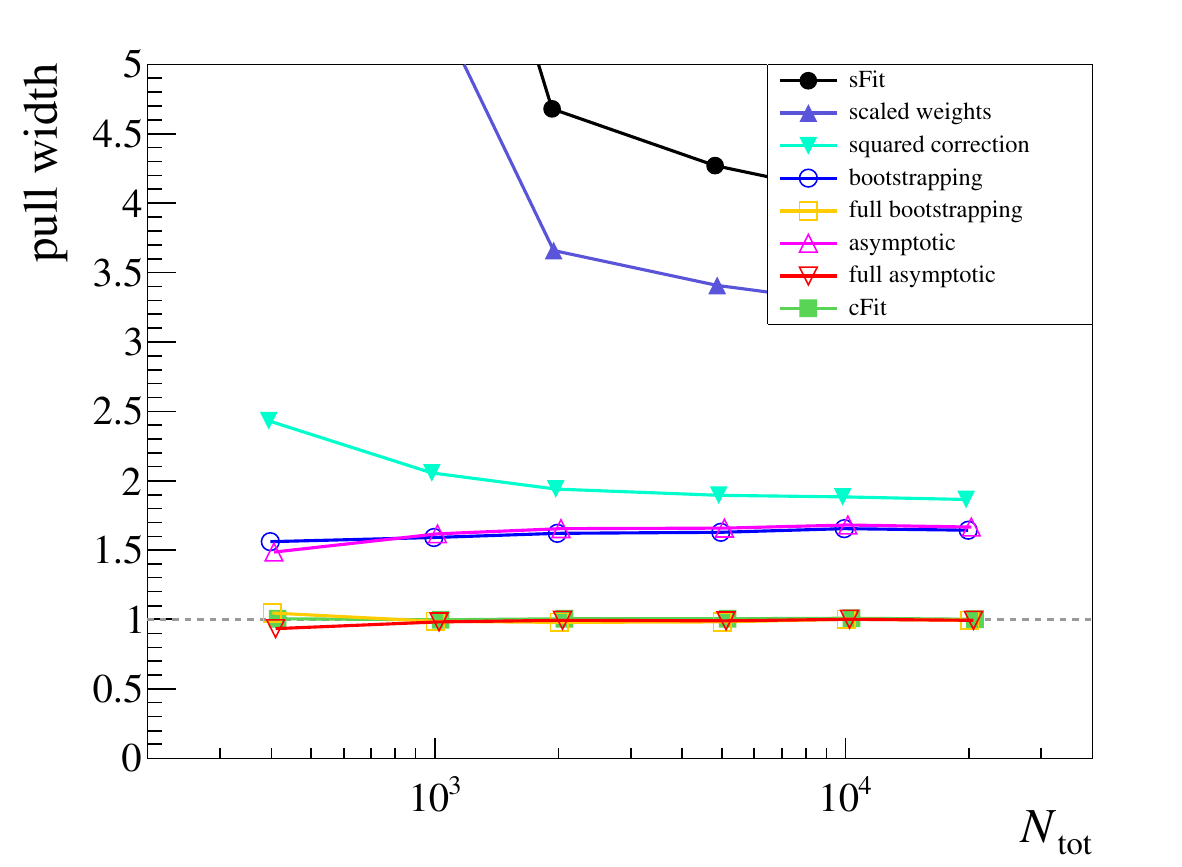}    
  }
  \caption{(Left) pull means and (right) pull widths depending on total event yield $N_{\rm tot}$.
    The markers are slightly horizontally staggered to improve readability.\label{fig:pullsc}}  
\end{figure}

\begin{figure}
\centering
  \subfloat[Triangular background model\label{fig:triangle}]{
    \includegraphics[width=0.49\textwidth]{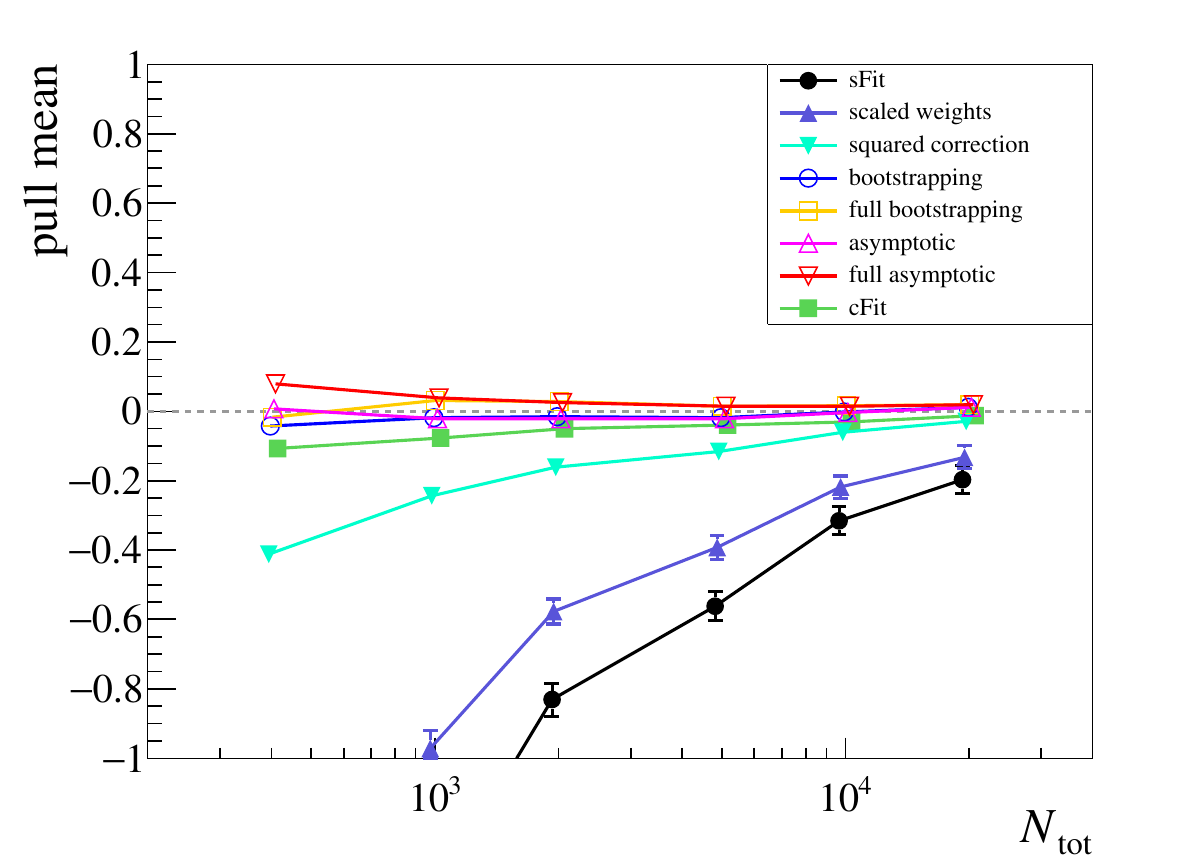}
    \includegraphics[width=0.49\textwidth]{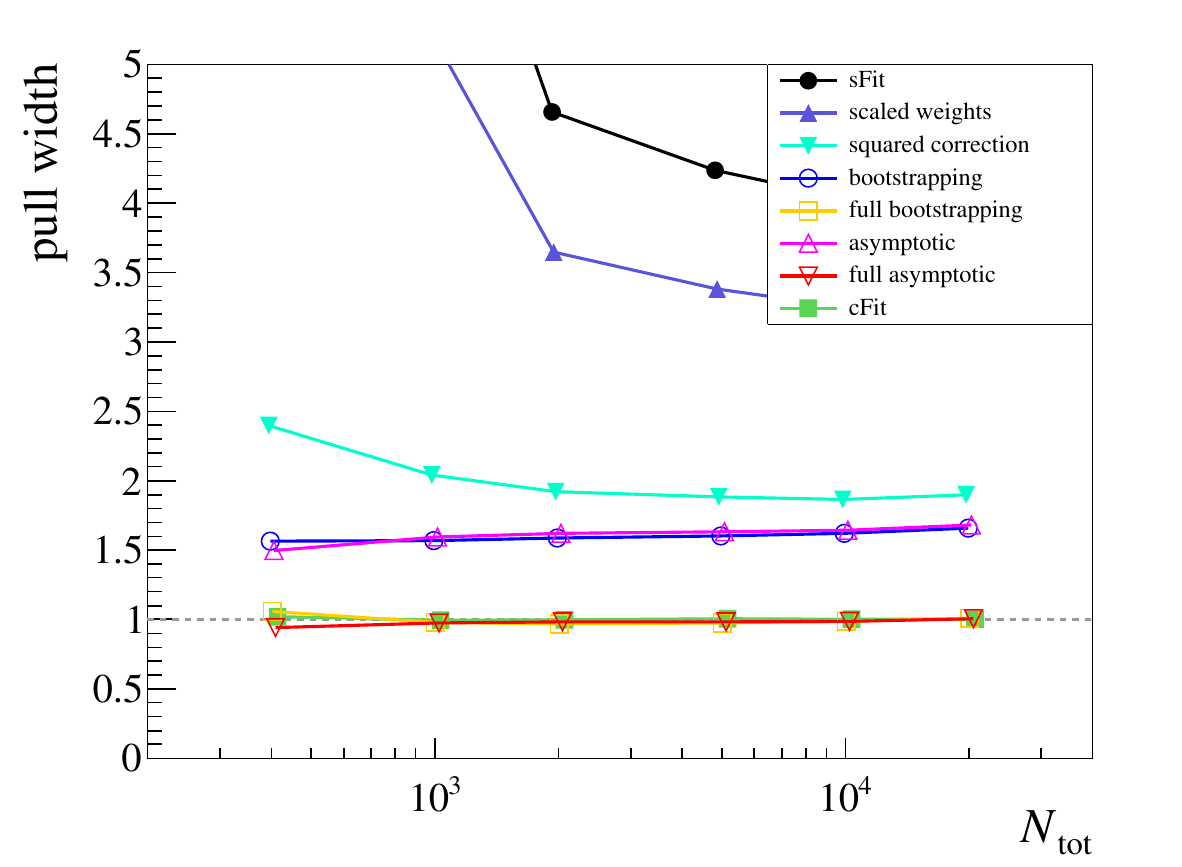}
  }\\
  \subfloat[Flat background model\label{fig:flat}]{
    \includegraphics[width=0.49\textwidth]{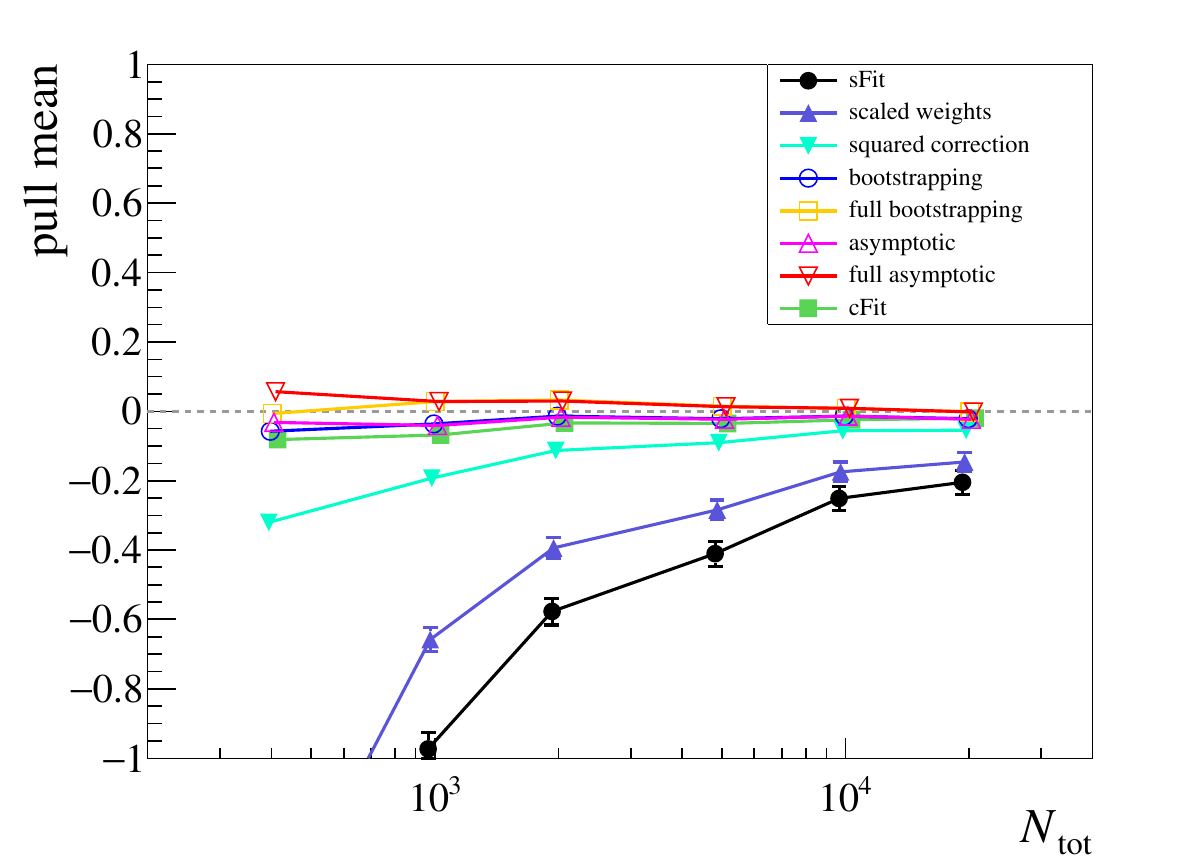}
    \includegraphics[width=0.49\textwidth]{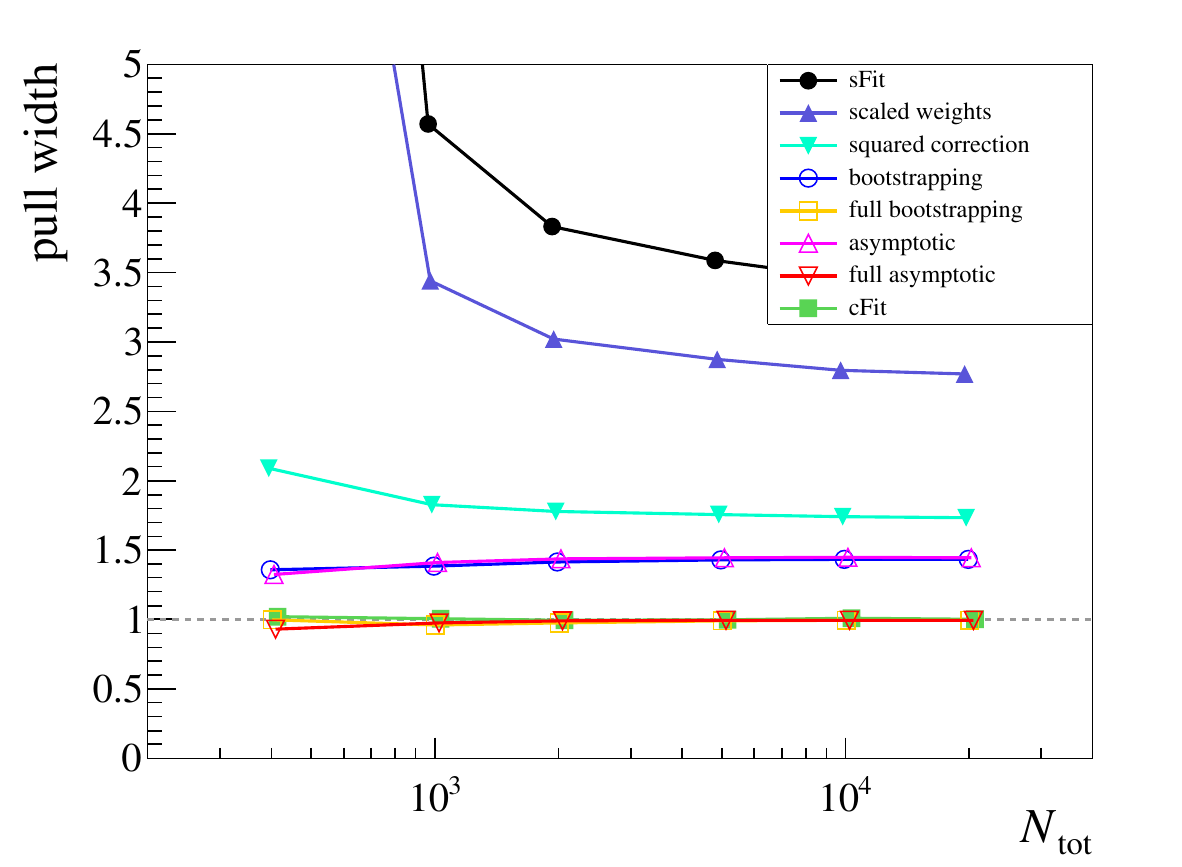}
  }
  \caption{(Left) pull means and (right) pull widths depending on total event yield $N_{\rm tot}$.
    The markers are slightly horizontally staggered to improve readability.\label{fig:pullsd}}  
\end{figure}

The pull distributions from 10\,000 pseudoexperiments, each with a total yield of 2000 events, are shown in Fig.~\ref{fig:pullsa} and~\ref{fig:pullsb}. 
The pull means and widths are shown in Figs.~\ref{fig:pullsc} and~\ref{fig:pullsd}. 
Numerical values for the different configurations are given in Tabs.~\ref{tab:pullsa} and~\ref{tab:pullsb} in App.~\ref{sec:appsweights}. 

As expected, both the {\it sFit} as well as the approach using {\it scaled weights} perform quite poorly, as they show large undercoverage, both for low statistics as well as for high statistics.
Furthermore, they exhibit significant bias at low statistics (which reduces at large statistics) due to a strong correlation of the uncertainty with the parameter $\tau$. 
This strongly disfavours the use of these methods for these {\it sWeighted} examples.
The {\it squared correction} method shows better performance, nevertheless also exhibits significant bias (which reduces for higher statistics) and undercoverage.
It should be stressed that significant undercoverage is still present at large statistics. 
This shows that the {\it squared correction} method 
in general does not provide asymptotically correct confidence intervals. 
Both {\it bootstrapping} as well as the {\it asymptotic} methods perform better for the examples studied here. 
However, both methods show some remaining undercoverage even at high statistics. 
It is instructive that {\it bootstrapping} the data without redetermining the {\it sWeights} performs identically to the {\it asymptotic method}
without accounting for the uncertainty due to nuisance parameters. 
However, 
when performing a {\it full bootstrapping} including re\-de\-riving the {\it sWeights} for the bootstrapped samples 
or when using the {\it full asymptotic} method
the confidence intervals generally cover correctly and 
no significant biases are observed.
Only at low statistics some slight overcoverage can be observed. 
This paper therefore advocates the use of the {\it full asymptotic} method, or alternatively, if computationally possible, the {\it full bootstrapping} approach  
for the determination of uncertainties in unbinned maximum likelihood fits using {\it sWeights}. 
If nuisance parameters have no large impact on the the {\it sWeights},
the {\it asymptotic} method can also be appropriate, as shown in App.~\ref{sec:appsweightsnocorr}. 

The conventional fit describing the background component in the decay time explicitly instead of using {\it sWeights} also shows good behaviour, as expected. 
When the background distribution is known, a conventional fit is generally advantageous as it has improved sensitivity due to the additional available information.
For this example, where the background pollution and parameter correlations are large,
the parameter sensitivity is significantly improved when using the conventional (unweighted) fit,
as shown by the relative efficiencies given in Tabs.~\ref{tab:pullsa} and~\ref{tab:pullsb}. 

\section{Conclusions}
\label{sec:conclusions} 
This paper derives the asymptotically correct method to determine parameter uncertainties in the presence of event weights for acceptance corrections, 
which was previously discussed in Ref.~\cite{doi:10.1146/annurev.ns.14.120164.002111} but does not currently see widespread use in high energy particle physics. 
The performance of this approach is validated on pseudoexperiments and compared with several commonly used methods.  
The asymptotically correct approach performs well, while several of the commonly used methods are shown to not generally result in correct coverage, 
even for large statistics. 
In addition, the effect of weight uncertainties for acceptance corrections is discussed. 
The paper furthermore derives asymptotically correct expressions for parameter uncertainties in fits that use event weights 
to statistically subtract background events using the \textit{sPlot} formalism~\cite{Pivk:2004ty}.
The asymptotically correct expression accounting for the presence of nuisance parameters in the determination of \textit{sWeights} is also given. 
On pseudoexperiments the asymptotically correct methods perform well,
whereas several commonly used methods show incorrect coverage also for this application.
Finally, the (co)variance for the sum of \textit{sWeights} in bins of the control variable is calculated,
which is a prerequisite for binned $\chi^2$ fits of \textit{sWeighted} data. 
If statistics are sufficiently large this paper 
advocates the use of the asymptotically correct expressions in weighted unbinned maximum likelihood fits, 
in particular over the current nominal method used in the {\scshape Roofit}~\cite{Verkerke:2003ir} fitting framework, which was proposed in Refs.~\cite{Eadie:100342,James:2006zz},
and is shown to not generally result in asymptotically correct uncertainties. 
If computationally feasible, the bootstrapping approach~\cite{10.2307/2958830} can be a useful alternative. 
A patch for {\scshape Roofit} to allow the determination of the covariance matrix according to Eq.~\ref{eq:correctcovariancemultidim} has been provided by the author and is available starting from {\scshape Root v6.20}. 

\section{Acknowledgements} 
C.\,L.\ gratefully acknowledges support by the Emmy Noether programme of the Deutsche Forschungsgemeinschaft (DFG), grant identifier LA 3937/1-1. 
Furthermore, C.\,L.\ would like to thank Roger Barlow for helpful comments and questions on an early version of this paper. 
Finally, C.\,L.\ would like to acknowledge useful communication from Michael~Schmelling, Hans~Dembinski and Matt~Kenzie. 

\setboolean{inbibliography}{true}
\bibliographystyle{JHEP}
\bibliography{main}

\clearpage

\appendix

\section{Expectation value Eq.~\ref{eq:expectation} for examples correcting for acceptance effects}  
\label{app:nonzeroacceptance}
As mentioned in Sec.~\ref{sec:commonapproaches}, Eq.~\ref{eq:squaredinequality} is not generally asymptotically valid.
To demonstrate this with an example, 
the expectation value in Eq.~\ref{eq:expectation} is explicitly calculated below for 
the angular fit in Sec.~\ref{sec:angularfit}.
Using the probability density function
\begin{align}
{\cal P}(\cos\theta;c_0,c_1) &= \frac{1+c_0\cos\theta+c_1\cos^2\theta}{2+\frac{2}{3}c_1}
\end{align}
and the efficiency correction (b)
\begin{align}
  \epsilon(\cos\theta)&=\frac{3}{10}+\frac{7}{10}\cos^2\theta
\end{align}
we derive the expectation value in Eq.~\ref{eq:expectation} according to (showing the asymptotic behaviour for the double partial derivative to $c_1$ is sufficient):
\begin{align}
  & E\biggl(\frac{1}{\epsilon^2(\cos\theta_e) {\cal P}(\cos\theta_e;c_0,c_1)}\frac{\partial ^2 {\cal P}(\cos\theta_e;c_0,c_1)}{\partial c_1^2}\biggr)\nonumber\\
  &= \frac{\int_{-1}^{+1} \frac{1}{\epsilon^2(\cos\theta) {\cal P}(\cos\theta_e;c_0,c_1)} \frac{\partial ^2 {\cal P}(\cos\theta_e;c_0,c_1)}{\partial c_1^2} {\cal P}(\cos\theta_e;c_0,c_1) \epsilon(\cos\theta) {\rm dcos}\theta}{ \int_{-1}^{+1} \epsilon(\cos\theta) {\cal P}(\cos\theta_e;c_0,c_1) {\rm dcos}\theta}\nonumber\\
  &= \frac{\frac{\partial^2}{\partial c_1^2}\int_{-1}^{+1} \frac{1}{\frac{3}{10}+\frac{7}{10}\cos^2\theta} \frac{1+c_0\cos\theta+c_1\cos^2\theta}{2+\frac{2}{3}c_1} {\rm dcos}\theta}{\int_{-1}^{+1} \left(\frac{3}{10}+\frac{7}{10}\cos^2\theta\right)\frac{1+c_0\cos\theta+c_1\cos^2\theta}{2+\frac{2}{3}c_1} {\rm dcos}\theta}.
\end{align}
Using computer algebra, these integrals can be easily evaluated analytically.
For the denominator we obtain
\begin{align}
  \int_{-1}^{+1} \left(\frac{3}{10}+\frac{7}{10}\cos^2\theta\right)\frac{1+c_0\cos\theta+c_1\cos^2\theta}{2+\frac{2}{3}c_1} {\rm dcos}\theta &= \frac{36}{50}\frac{c_1}{c_1+3} + \frac{48}{30}\frac{1}{c_1+3}\nonumber\\
  &\overset{c_1=0}{=} \frac{8}{15}.
\end{align}
The expression for the numerator is slightly more complicated, it results in
\begin{align}
  & \frac{\partial^2}{\partial c_1^2}\int_{-1}^{+1} \frac{1}{\frac{3}{10}+\frac{7}{10}\cos^2\theta} \frac{1+c_0\cos\theta+c_1\cos^2\theta}{2+\frac{2}{3}c_1} {\rm dcos}\theta\nonumber\\
  &= \frac{\partial^2}{\partial c_1^2}\left(
  \frac{1}{\frac{2}{3}c_1+2}\left[
    -\frac{20\sqrt{21}}{49}\tan^{-1}\left(\frac{\sqrt{21}}{3}\right)c_1
    +\frac{20}{7}c_1
    +\frac{20}{\sqrt{21}}\tan^{-1}\left(\frac{\sqrt{21}}{3}\right)
    \right]\right)\nonumber\\  
  &\overset{c_1=0}{=} \frac{20}{7\sqrt{21}}\tan^{-1}\left(\frac{\sqrt{21}}{3}\right)
  + \frac{140}{3\cdot 21^{3/2}} \tan^{-1}\left(\frac{\sqrt{21}}{3}\right)
  - \frac{20}{21}\nonumber\\
  &\approx 0.146.
\end{align}
We thus find
\begin{align}
  E\biggl(\frac{1}{\epsilon^2(\cos\theta_e) {\cal P}(\cos\theta_e;c_0,c_1)}\frac{\partial ^2 {\cal P}(\cos\theta_e;c_0,c_1)}{\partial c_1^2}\biggr)
&\approx 0.274 \neq 0.\label{eq:asymptotic}
\end{align}
This shows clearly, that Eq.~\ref{eq:squaredinequality} is not generally asymptotically correct.
The result in Eq.~\ref{eq:asymptotic} has been crosschecked using the pseudoexperiments described in Sec.~\ref{sec:angularfit} and indeed the additional term fluctuates around $N\times 0.2742$ as derived above. 

For acceptance~(a) we find similarly
\begin{align}
  E\biggl(\frac{1}{\epsilon^2(\cos\theta_e) {\cal P}(\cos\theta_e;c_0,c_1)}\frac{\partial ^2 {\cal P}(\cos\theta_e;c_0,c_1)}{\partial c_1^2}\biggr) &\approx -0.135 \neq 0,
\end{align}
which is also confirmed using the pseudoexperiments. 
The fact that the expectation value is negative for acceptance~(a) and positive for acceptance~(b) indicates that, as observed using the pseudoexperiments,
the {\it squared correction} method overcovers for acceptance~(a) and undercovers for acceptance~(b).

It is instructive to note that for the double partial derivative to $c_0$, 
$E( \epsilon^{-2} {\cal P}^{-1}\partial^2/\partial c_0^2 {\cal P})$, 
we find an expectation value of zero, as the integration of the numerator removes the $c_0$ dependence in that case.
This is the reason why the {\it squared correction} method results in compatible results with the {\it asymptotic} method for $c_0$, 
but shows incorrect coverage for $c_1$. 

\clearpage

\section{Covariance determination for unbinned \textit{sWeighted} fits}
\label{app:mestimationunbinned}
The asymptotic covariance 
for \textit{sWeighted} unbinned maximum likelihood fits is given by 
Eq.~\ref{eq:correctvar}~\cite{van2000asymptotic,davison_2003} which is reproduced below for convenience. 
\begin{align}
\bm{C}_{\bm{\theta}} &=
  {\underbrace{E\left(\frac{\partial \bm{g}(\bm{x},\bm{y};\bm{\theta})}{\partial \bm{\theta}^T}\right)}_{\textit{``denominator''}}}^{-1}
  \times \underbrace{E\bigl(\bm{g}(\bm{x},\bm{y};\bm{\theta})\bm{g}(\bm{x},\bm{y};\bm{\theta})^T\bigr)}_{\textit{``numerator''}}
  \times E\left(\frac{\partial \bm{g}(\bm{x},\bm{y};\bm{\theta})}{\partial \bm{\theta}^T}\right)^{-T}\label{eq:covreplica}
\end{align}
Here, the vector of estimating equations $\bm{g}$ is given by Eq.~\ref{eq:vectorg}, which is also repeated below
{\small
\begin{align}
  {\bm{g}}(\bm{x},\bm{y};\bm{\theta}) &= \left(
  \renewcommand{\arraystretch}{1.3}
  \begin{array}{c}
    \varphi_s(\bm{y};\ns,\nb)\\
    \varphi_b(\bm{y};\ns,\nb)\\
    \psi_{ss}(\bm{y};\vssi,\ns,\nb)\\
    \psi_{sb}(\bm{y};\vsbi,\ns,\nb)\\
    \psi_{bb}(\bm{y};\vbbi,\ns,\nb)\\
    \xi_i(\bm{x},\bm{y};\bm{\lambda},\vssi,\vsbi,\vbbi)
  \end{array}
  \right)  
  = \left(
  \begin{array}{c}
\sum_e\frac{\partial}{\partial\ns}\bigl[\ln(\ns\ps(y_e)+\nb\pb(y_e)) - \frac{\ns+\nb}{N} \bigr]\\
\sum_e\frac{\partial}{\partial\nb}\bigl[\ln(\ns\ps(y_e)+\nb\pb(y_e)) - \frac{\ns+\nb}{N} \bigr]\\
\sum_e \bigl[\frac{\ps(y_e)\ps(y_e)}{(\ns\ps(y_e)+\nb\pb(y_e))^2} - \frac{\vssi}{N}\bigr]\\    
\sum_e \bigl[\frac{\ps(y_e)\pb(y_e)}{(\ns\ps(y_e)+\nb\pb(y_e))^2} - \frac{\vsbi}{N}\bigr]\\    
\sum_e \bigl[\frac{\pb(y_e)\pb(y_e)}{(\ns\ps(y_e)+\nb\pb(y_e))^2} - \frac{\vbbi}{N}\bigr]\\    
\sum_e \ws(y_e;\vssi,\vsbi,\vbbi) \frac{\partial\ln\calp(x_e;\bm{\lambda})}{\partial\lambda_i}
    \end{array}
  \right). 
\end{align}}
The elements of the \textit{denominator}\footnote{In the following, the arguments of the estimating functions $\bm{\varphi}$, $\bm{\psi}$ and $\bm{\xi}$ will be omitted for brevity.}
$E\bigl(\partial \bm{g}(\bm{x},\bm{y};\bm{\theta})/\partial \bm{\theta}^T\bigr)$ given by 
\begin{align}
  E\left(\frac{\partial \bm{g}(\bm{x},\bm{y};\bm{\theta})}{\partial \bm{\theta}^T}\right) &=
  E\left(\begin{array}{cccccc}
   \frac{\partial\varphi_s}{\partial\ns} & \frac{\partial\varphi_s}{\partial\nb} & \frac{\partial\varphi_s}{\partial\vssi} & \frac{\partial\varphi_s}{\partial\vsbi} & \frac{\partial\varphi_s}{\partial\vbbi} & \frac{\partial\varphi_s}{\partial\bm{\lambda}^T}\\
   \frac{\partial\varphi_b}{\partial\ns} & \frac{\partial\varphi_b}{\partial\nb} & \frac{\partial\varphi_b}{\partial\vssi} & \frac{\partial\varphi_b}{\partial\vsbi} & \frac{\partial\varphi_b}{\partial\vbbi} & \frac{\partial\varphi_b}{\partial\bm{\lambda}^T}\\
   \frac{\partial\psi_{ss}}{\partial\ns} & \frac{\partial\psi_{ss}}{\partial\nb} & \frac{\partial\psi_{ss}}{\partial\vssi} & \frac{\partial\psi_{ss}}{\partial\vsbi} & \frac{\partial\psi_{ss}}{\partial\vbbi} & \frac{\partial\psi_{ss}}{\partial\bm{\lambda}^T}\\
   \frac{\partial\psi_{sb}}{\partial\ns} & \frac{\partial\psi_{sb}}{\partial\nb} & \frac{\partial\psi_{sb}}{\partial\vssi} & \frac{\partial\psi_{sb}}{\partial\vsbi} & \frac{\partial\psi_{sb}}{\partial\vbbi} & \frac{\partial\psi_{sb}}{\partial\bm{\lambda}^T}\\
   \frac{\partial\psi_{bb}}{\partial\ns} & \frac{\partial\psi_{bb}}{\partial\nb} & \frac{\partial\psi_{bb}}{\partial\vssi} & \frac{\partial\psi_{bb}}{\partial\vsbi} & \frac{\partial\psi_{bb}}{\partial\vbbi} & \frac{\partial\psi_{bb}}{\partial\bm{\lambda}^T}\\
   \frac{\partial\bm{\xi}}{\partial\ns} & \frac{\partial\bm{\xi}}{\partial\nb} & \frac{\partial\bm{\xi}}{\partial\vssi} & \frac{\partial\bm{\xi}}{\partial\vsbi} & \frac{\partial\bm{\xi}}{\partial\vbbi} & \frac{\partial\bm{\xi}}{\partial\bm{\lambda}^T}
  \end{array}\right)
\end{align}
are determined in Sec.~\ref{app:unbinneddenom} and the elements of the (symmetric) \textit{numerator}
\begin{align}
E\bigl(\bm{g}(\bm{x},\bm{y};\bm{\theta})\bm{g}(\bm{x},\bm{y};\bm{\theta})^T\bigr) &= 
  E\left(\begin{array}{cccccc}
   \varphi_s\varphi_s & \varphi_s\varphi_b & \varphi_s\psi_{ss} & \varphi_s\psi_{sb} & \varphi_s\psi_{bb} & \varphi_s\bm{\xi}^T\\
   \varphi_b\varphi_s & \varphi_b\varphi_b & \varphi_b\psi_{ss} & \varphi_b\psi_{sb} & \varphi_b\psi_{bb} & \varphi_b\bm{\xi}^T\\
   \psi_{ss}\varphi_s & \psi_{ss}\varphi_b & \psi_{ss}\psi_{ss} & \psi_{ss}\psi_{sb} & \psi_{ss}\psi_{bb} & \psi_{ss}\bm{\xi}^T\\
   \psi_{sb}\varphi_s & \psi_{sb}\varphi_b & \psi_{sb}\psi_{ss} & \psi_{sb}\psi_{sb} & \psi_{sb}\psi_{bb} & \psi_{sb}\bm{\xi}^T\\
   \psi_{bb}\varphi_s & \psi_{bb}\varphi_b & \psi_{bb}\psi_{ss} & \psi_{bb}\psi_{sb} & \psi_{bb}\psi_{bb} & \psi_{bb}\bm{\xi}^T\\
   \bm{\xi}\varphi_s & \bm{\xi}\varphi_b & \bm{\xi}\psi_{ss} & \bm{\xi}\psi_{sb} & \bm{\xi}\psi_{bb} & \bm{\xi}\bm{\xi}^T
  \end{array}\right)
\end{align}
are determined in Sec.~\ref{app:unbinnednum}. 
For both numerator and denominator the sample estimates are derived first and their expectation values are given afterwards. 
The resulting covariance is derived in Sec~\ref{app:unbinnedresult}. 

\subsection{Determination of the denominator}
\label{app:unbinneddenom}
We first determine the expectations $E(\partial \bm{g}/\partial \bm{\theta}^T)$, the denominator in Eq.~\ref{eq:covreplica}, which would be corresponding to the Hessian matrix if we were doing purely maximum likelihood estimation. 
We find for the two $\bm{\varphi}$ components
\begin{align}
  E\biggl(\frac{\partial}{\partial N_j} \varphi_i\biggr) &= 
  E\biggl(\sum \frac{\partial}{\partial N_j} \frac{\calpi(y_e)}{\ns\ps(y_e)+\nb\pb(y_e)} \biggr) \nonumber\\
  &= E\biggl(-\sum \frac{\calpi(y_e)\calpj(y_e)}{\bigl(\ns\ps(y_e)+\nb\pb(y_e)\bigr)^2} \biggr) \nonumber\\
  &= -\int \frac{\calpi(y)\calpj(y)}{\ns\ps(y)+\nb\pb(y)}\deriv y\equiv A_{ij}\label{eq:defA}\\
  E\biggl(\frac{\partial}{\partial\vkli} \varphi_i\biggr) &= 0\\
  E\biggl(\frac{\partial}{\partial\lambda_j} \varphi_i\biggr) &= 0,
\end{align}
for the $\bm{\psi}$ components
\begin{align}
  E\biggl(\frac{\partial}{\partial N_k} \psi_{ij}\biggr) &= E\biggl(\sum_e\frac{\partial}{\partial N_k}
  \biggl[\frac{\calpi(y_e)\calpj(y_e)}{\bigl(\ns\ps(y_e)+\nb\pb(y_e)\bigr)^2} - \frac{\viji}{N}\biggr]\biggr)\nonumber\\
  &= E\biggl(-2 \sum_e\frac{\calpi(y_e)\calpj(y_e)\calpk(y_e)}{\bigl(\ns\ps(y_e)+\nb\pb(y_e)\bigr)^3}\biggr)\nonumber\\
  &= -2 \int \frac{\calpi(y)\calpj(y)\calpk(y)}{\bigl(\ns\ps(y)+\nb\pb(y)\bigr)^2}\deriv y \equiv B_{(ij)k}, 
  \label{eq:defB}\\
  E\biggl(\frac{\partial}{\partial\vkli} \psi_{ij}\biggr) &= E\biggl(\frac{\partial}{\partial\vkli}
  \sum_e \biggl[\frac{\calpi(y_e)\calpj(y_e)}{\bigl(\ns\ps(y_e)+\nb\pb(y_e)\bigr)^2} -\frac{\viji}{N}\biggr]\biggr)\nonumber\\
  &= -\delta_{(kl)(ij)}\\
  E\biggl(\frac{\partial}{\partial\lambda_i} \psi_{kl}\biggr) &= 0,
\end{align}
and finally for the $\bm{\xi}$ components
\begin{align}
  E\biggl(\frac{\partial}{\partial N_k} \xi_l\biggr) &= 0\\
  E\biggl(\frac{\partial}{\partial\vssi} \xi_l\biggr) &= E\biggl(\frac{\partial}{\partial\vssi} \sum_e \ws(y_e)\frac{\partial}{\partial\lambda_l}\ln\ps(x_e;\bm{\lambda})\biggr)\nonumber\\
  &= -\int \frac{\vbbi\ps(y)\pb(y)-\vsbi\pb^2(y)}{\bigl((\vbbi-\vsbi)\ps(y)+(\vssi-\vsbi)\pb(y)\bigr)^2}\pb(y)\deriv y \times \kappa_l\nonumber\\
  &\equiv E_{l1} = \kappa_l e_1\label{eq:defEa}\\
  E\biggl(\frac{\partial}{\partial\vsbi} \xi_l\biggr) &= E\biggl(\frac{\partial}{\partial\vsbi} \sum_e \ws(y_e)\frac{\partial}{\partial\lambda_l}\ln\ps(x_e;\bm{\lambda})\biggr)\nonumber\\
  &= +\int \frac{\vbbi\ps^2(y)-\vssi\pb^2(y)}{\bigl((\vbbi-\vsbi)\ps(y)+(\vssi-\vsbi)\pb(y)\bigr)^2}\pb(y)\deriv y \times \kappa_l\nonumber\\
  &\equiv E_{l2} = \kappa_l e_2\\
  E\biggl(\frac{\partial}{\partial\vbbi} \xi_l\biggr) &= E\biggl(\frac{\partial}{\partial\vbbi} \sum_e \ws(y_e)\frac{\partial}{\partial\lambda_l}\ln\ps(x_e;\bm{\lambda})\biggr)\nonumber\\
  &= +\int \frac{-\vsbi\ps^2(y)+\vssi\ps(y)\pb(y)}{\bigl((\vbbi-\vsbi)\ps(y)+(\vssi-\vsbi)\pb(y)\bigr)^2}\pb(y)\deriv y \times \kappa_l\nonumber\\
  &\equiv E_{l3} = \kappa_l e_3\label{eq:defEb}\\
  E\biggl(\frac{\partial}{\partial\lambda_l} \xi_k\biggr) &= E\biggl(\sum_e \ws(y_e) \frac{\partial^2\ln\ps(x_e;\bm{\lambda})}{\partial\lambda_l\partial\lambda_k} \biggr)\nonumber\\
  &= \int \frac{\vbbi\ps(y)-\vsbi\pb(y)}{(\vbbi-\vsbi)\ps(y)+(\vssi-\vsbi)\pb(y)} \frac{\partial^2\ln\ps(x;\bm{\lambda})}{\partial\lambda_l\partial\lambda_k}\nonumber\\
  &\hphantom{=} \times \bigl(\ns\ps(x;\lambda)\ps(y)+\nb\pb(x)\pb(y)\bigr)\deriv x\deriv y\nonumber\\
  &= \int \ns\ps(x;\bm{\lambda})\frac{\partial^2\ln\ps(x;\bm{\lambda})}{\partial\lambda_l\partial\lambda_k} \deriv x \equiv H_{kl},\label{eq:defH}
\end{align}
where $\kappa_i=\int \nb\frac{\pb(x)}{\ps(x;\bm{\lambda})}\frac{\partial\ps(x;\bm{\lambda})}{\partial\lambda_i}\deriv x$ was defined in Eq.~\ref{eq:sweightsunbiasedthree}. 
In summary, the matrix in the denominator is
\begin{align}
  E\left(+\frac{\partial \bm{g}(\bm{x},\bm{y};\bm{\theta})}{\partial \bm{\theta}^T}\right) &=
  \left(\begin{array}{ccc}
    \bm{A} & \bm{0} & \bm{0}\\
    \bm{B} & -\bm{1} & \bm{0}\\
    \bm{0} & \bm{E} & \bm{H}\\
  \end{array}\right)
.\label{eq:denom}
\end{align}
As a reminder, the matrix $\bm{A}$ is the $2\times 2$ matrix for the signal and background yields defined by Eq.~\ref{eq:defA}, 
the matrix $\bm{B}$ is a $3\times 2$ matrix defined by Eq.~\ref{eq:defB},
$\bm{E}$ is the $\dim(\bm{\lambda})\times 3$ matrix defined by Eqs.~\ref{eq:defEa}--\ref{eq:defEb},
and $\bm{H}$ the $\dim(\bm{\lambda})\times \dim(\bm{\lambda})$ matrix given by Eq.~\ref{eq:defH}. 
The upper right corner of Eq.~\ref{eq:denom} is filled with zero matrices, as the estimation of the yields does not depend on the parameters \viji,
and the estimators for the \viji\ and the yields do not depend on $\bm{\lambda}$. 
This simplifies the inversion of the matrix in Eq.~\ref{eq:denom}, which results in 
{
\begin{align}
  E\left(+\frac{\partial \bm{g}(\bm{x},\bm{y};\bm{\theta})}{\partial \bm{\theta}^T}\right)^{-1} 
   &= \left(\begin{array}{ccc}
    \bm{A}^{-1} & \bm{0} & \bm{0}\\
    \bm{B}\bm{A}^{-1} & -\bm{1} & \bm{0}\\
    -\bm{H}^{-1}\bm{E}\bm{B}\bm{A}^{-1} & \bm{H}^{-1}\bm{E} & \bm{H}^{-1}\\
  \end{array}\right)
\end{align}}
This matrix can be further simplified, as
{\small
\begin{align}
 &  (B_{11}E_{l1}+B_{12}E_{l2}+B_{22}E_{l3}) \nonumber\\
=&\int \frac{\ps^3}{(\ldots)^2}\deriv y \int \frac{-\vbbi\ps\pb^2+\vsbi\pb^3}{(\ldots)^2}\times\kappa_l 
+\int \frac{\ps^2\pb}{(\ldots)^2}\deriv y \int \frac{\vbbi\ps^2\pb-\vssi\pb^3}{(\ldots)^2}\times\kappa_l\nonumber\\
&+\int \frac{\ps\pb^2}{(\ldots)^2}\deriv y \int \frac{-\vsbi\ps^2\pb+\vssi\ps\pb^2}{(\ldots)^2}\times\kappa_l\nonumber ~~~\text{expanding}~\viji=\int \frac{\calpi\calpj}{(\ldots)}\frac{\ns\ps+\nb\pb}{\ns\ps+\nb\pb}\deriv y\\
=& \biggl(\ns\int\frac{\ps\pb^2}{(\ldots)^2}\deriv y +\nb\int\frac{\pb^3}{(\ldots)^2}\deriv y\biggr)\times\biggl(
\int\frac{\ps^2\pb}{(\ldots)^2}\deriv y\int\frac{\ps^2\pb}{(\ldots)^2}\deriv y - \int\frac{\ps^3}{(\ldots)^2}\deriv y\int\frac{\ps\pb^2}{(\ldots)^2}\deriv y
\biggr)\times\kappa_l\nonumber\\
&+\biggl(\ns\int\frac{\ps^2\pb}{(\ldots)^2}\deriv y +\nb\int\frac{\ps\pb^2}{(\ldots)^2}\deriv y\biggr)\times\biggl(
\int\frac{\ps^3}{(\ldots)^2}\deriv y\int\frac{\pb^3}{(\ldots)^2}\deriv y - \int\frac{\ps\pb^2}{(\ldots)^2}\deriv y\int\frac{\ps^2\pb}{(\ldots)^2}\deriv y
\biggr)\times\kappa_l\nonumber\\
&+\biggl(\ns\int\frac{\ps^3}{(\ldots)^2}\deriv y +\nb\int\frac{\ps^2\pb}{(\ldots)^2}\deriv y\biggr)\times\biggl(
\int\frac{\ps\pb^2}{(\ldots)^2}\deriv y\int\frac{\ps\pb^2}{(\ldots)^2}\deriv y - \int\frac{\ps^2\pb}{(\ldots)^2}\deriv y\int\frac{\pb^3}{(\ldots)^2}\deriv y
\biggr)\times\kappa_l\nonumber\\
=& 0 \label{eq:zeroone}  
\end{align}
and
\begin{align}
  & (B_{12}E_{l1}+B_{22}E_{l2}+B_{32}E_{l3})\nonumber\\
=&\int \frac{\ps^2\pb}{(\ldots)^2}\deriv y \int \frac{-\vbbi\ps\pb^2+\vsbi\pb^3}{(\ldots)^2}\times\kappa_l 
+\int \frac{\ps\pb^2}{(\ldots)^2}\deriv y \int \frac{\vbbi\ps^2\pb-\vssi\pb^3}{(\ldots)^2}\times\kappa_l\nonumber\\
&+\int \frac{\pb^3}{(\ldots)^2}\deriv y \int \frac{-\vsbi\ps^2\pb+\vssi\ps\pb^2}{(\ldots)^2}\times\kappa_l\nonumber\\
=& \biggl(\ns\int\frac{\ps\pb^2}{(\ldots)^2}\deriv y +\nb\int\frac{\pb^3}{(\ldots)^2}\deriv y\biggr)\times\biggl(
\int\frac{\ps^2\pb}{(\ldots)^2}\deriv y\int\frac{\ps\pb^2}{(\ldots)^2}\deriv y - \int\frac{\ps^2\pb}{(\ldots)^2}\deriv y\int\frac{\ps\pb^2}{(\ldots)^2}\deriv y
\biggr)\times\kappa_l\nonumber\\
&+\biggl(\ns\int\frac{\ps^2\pb}{(\ldots)^2}\deriv y +\nb\int\frac{\ps\pb^2}{(\ldots)^2}\deriv y\biggr)\times\biggl(
\int\frac{\ps^2\pb}{(\ldots)^2}\deriv y\int\frac{\pb^3}{(\ldots)^2}\deriv y - \int\frac{\ps^2\pb}{(\ldots)^2}\deriv y\int\frac{\pb^3}{(\ldots)^2}\deriv y
\biggr)\times\kappa_l\nonumber\\
&+\biggl(\ns\int\frac{\ps^3}{(\ldots)^2}\deriv y +\nb\int\frac{\ps^2\pb}{(\ldots)^2}\deriv y\biggr)\times\biggl(
\int\frac{\ps\pb^2}{(\ldots)^2}\deriv y\int\frac{\pb^3}{(\ldots)^2}\deriv y - \int\frac{\ps\pb^2}{(\ldots)^2}\deriv y\int\frac{\pb^3}{(\ldots)^2}\deriv y
\biggr)\times\kappa_l\nonumber\\
  =& 0.  \label{eq:zerotwo}  
\end{align}}
We therefore find $\bm{E}\bm{B}=\bm{0}$ and the inverse of Eq.~\ref{eq:denom} is given by
\begin{align}
  E\left(+\frac{\partial \bm{g}(\bm{x},\bm{y};\bm{\theta})}{\partial \bm{\theta}^T}\right)^{-1}  
   &= \left(\begin{array}{ccc}
    \bm{A}^{-1} & \bm{0} & \bm{0}\\
    \bm{B}\bm{A}^{-1} & -\bm{1} & \bm{0}\\
    \bm{0} & \bm{H}^{-1}\bm{E} & \bm{H}^{-1}\\
  \end{array}\right)
\end{align}

\clearpage
\subsection{Determination of numerator}
\label{app:unbinnednum}
We now have to determine the elements of the numerator $E\bigl(\bm{g}(\bm{x},\bm{y};\bm{\theta})\bm{g}(\bm{x},\bm{y};\bm{\theta})^T\bigr)$ in Eq.~\ref{eq:covreplica}. 
For completeness we determine all sub-matrices, but it will be shown that we only need the components $E\bigl(\psi_{ij}\psi_{kl}\bigr)\equiv C_{(ij)(kl)}^\prime$, $E\bigl(\psi_{ij}\xi_l\bigr)\equiv E_{l(ij)}^\prime$ and $E\bigl(\xi_k\xi_l\bigr)\equiv H^\prime_{kl}$ to determine the covariance for the parameters of interest $\bm{\lambda}$. 
We find
{\small
  \begin{align}
    E\bigl(\varphi_i\varphi_j \bigr) &= E\biggl( \sum_e \biggl[\frac{\calpi(y_e)}{\ns\ps(y_e)+\nb\pb(y_e)}-\frac{1}{N}\biggr]\nonumber\\
    &\hphantom{=} \times \sum_f \biggl[\frac{\calpj(y_f)}{\ns\ps(y_f)+\nb\pb(y_f)}-\frac{1}{N}\biggr]\biggr)\nonumber\\
    &= E\biggl(\sum_e\frac{\calpi(y_e)\calpj(y_e)}{\big(\ns\ps+\nb\pb\bigr)^2} + \sum_{e\neq f}\frac{\calpi(y_e)}{\ns\ps+\nb\pb}\frac{\calpj(y_f)}{\ns\ps+\nb\pb} \nonumber\\
    &\hphantom{=} -\sum_e \frac{\calpi(y_e)}{\ns\ps+\nb\pb} - \sum_f\frac{\calpj(y_f)}{\ns\ps+\nb\pb} + 1\biggr)\nonumber\\
    &= E\biggl(\sum_e\frac{\calpi(y_e)\calpj(y_e)}{\bigl(\ns\ps(y_e)+\nb\pb(y_e)\bigr)^2}\biggr)\nonumber\\
    &= \int \frac{\calpi(y)\calpj(y)}{\ns\ps(y)+\nb\pb(y)}\deriv y \equiv A_{ij}^\prime\label{eq:firstelement}\\
    E\bigl(\varphi_k\psi_{ij} \bigr) &=
    E\biggl( \sum_e \biggl[\frac{\calpk(y_e)}{\ns\ps(y_e)+\nb\pb(y_e)}-\frac{1}{N}\biggr]\nonumber\\
    &\hphantom{=}\times \sum_f \biggl[\frac{\calpi(y_f)\calpj(y_f)}{\bigl(\ns\ps(y_f)+\nb\pb(y_f)\bigr)^2} - \frac{\viji}{N}\biggr]\biggr)\nonumber\\
    &= E\biggl(\sum_e\frac{\calpi(y_e)\calpj(y_e)\calpk(y_e)}{\bigl(\ns\ps(y_e)+\nb\pb(y_e)\bigr)^3}\biggr)\nonumber\\
    &= \int \frac{\calpi(y)\calpj(y)\calpk(y)}{\bigl(\ns\ps(y)+\nb\pb(y)\bigr)^2}\deriv y \equiv B_{(ij)k}^\prime\\ 
    E\bigl(\psi_{ij}\psi_{kl}\bigr) &= E\biggl(\sum_e \biggl[\frac{\calpi(y_e)\calpj(y_e)}{\bigl(\ns\ps(y_e)+\nb\pb(y_e)\bigr)^2} -\frac{\viji}{N}\biggr]\nonumber\\
    &\hphantom{=}\times \sum_f \biggl[\frac{\calp_k(y_f)\calp_l(y_f)}{\bigl(\ns\ps(y_f)+\nb\pb(y_f)\bigr)^2} -\frac{\vkli}{N}\biggr]\biggr)\nonumber\\
    &= E\biggl(\sum_e\frac{\calpi(y_e)\calpj(y_e)\calp_k(y_e)\calp_l(y_e)}{\bigl(\ns\ps(y_e)+\nb\pb(y_e)\bigr)^4}\biggr)\nonumber\\
    &= \int\frac{\calpi(y)\calpj(y)\calp_k(y)\calp_l(y)}{\bigl(\ns\ps(y)+\nb\pb(y)\bigr)^3}\deriv y \equiv C_{(ij)(kl)}^\prime, 
\end{align}}
and
{\small
\begin{align}
  E\bigl(\varphi_k\xi_l\bigr) &= 
  E\biggl( \sum_e \biggl[\frac{\calpk(y_e)}{\ns\ps(y_e)+\nb\pb(y_e)}-\frac{1}{N}\biggr]\nonumber\\
  & \hphantom{=}\times \sum_f \ws(y_f)\frac{\partial}{\partial\lambda_l}\ln \ps(x_f;\bm{\lambda})\biggr)\nonumber\\
  &= E\biggl(\sum_e \frac{\calpk(y_e)}{\ns\ps(y_e)+\nb\pb(y_e)} \ws(y_e)\frac{\partial}{\partial\lambda_l}\ln \ps(x_e;\bm{\lambda})\biggr)\nonumber\\
  &= \int \frac{\calpk(y_e)}{\ns\ps(y_e)+\nb\pb(y_e)}\nonumber\\
  &\hphantom{=} \times\frac{\vbbi\ps(y)-\vsbi\pb(y)}{(\vbbi-\vsbi)\ps(y)+(\vssi-\vsbi)\pb(y)}\pb(y)
  \deriv y \times\kappa_l \nonumber\\
  &\equiv D_{lk}^\prime = \kappa_l d_{k}^\prime\\
  E\bigl(\psi_{ij}\xi_l \bigr) &= E\biggl(\sum_e \biggl[\frac{\calpi(y_e)\calpj(y_e)}{\bigl(\ns\ps(y_e)+\nb\pb(y_e)\bigr)^2} -\frac{\viji}{N}\biggr]\nonumber\\
  & \hphantom{=}\times \sum_f \ws(y_f)\frac{\partial}{\partial\lambda_l}\ln \ps(x_f;\bm{\lambda})\biggr)\nonumber\\
  &= E\biggl(\sum_e \frac{\calpi(y_e)\calpj(y_e)}{\bigl(\ns\ps(y_e)+\nb\pb(y_e)\bigr)^2} \ws(y_e)\frac{\partial}{\partial\lambda_l}\ln \ps(x_e;\bm{\lambda}) \biggr)\nonumber\\
  &= \int \frac{\calpi(y)\calpj(y)}{\bigl(\ns\ps(y)+\nb\pb(y)\bigr)^2}\nonumber\\
  &\hphantom{=} \times\frac{\vbbi\ps(y)-\vsbi\pb(y)}{(\vbbi-\vsbi)\ps(y)+(\vssi-\vsbi)\pb(y)}\pb(y)
  \deriv y \times\kappa_l\nonumber\\
  &\equiv E_{l(ij)}^\prime = \kappa_l e_{(ij)}^\prime\\
  E\bigl(\xi_k\xi_l \bigr) &= E\biggl(
  \sum_e \ws(y_e)\frac{\partial}{\partial\lambda_k}\ln \ps(x_e;\bm{\lambda}) 
  \times \sum_f \ws(y_f)\frac{\partial}{\partial\lambda_l}\ln \ps(x_f;\bm{\lambda})\biggr)\nonumber\\
  &= E\biggl(
  \sum_e \ws^2(y_e) \frac{\partial\ln \ps(x_e;\bm{\lambda})}{\partial\lambda_k}\frac{\partial\ln \ps(x_e;\bm{\lambda})}{\partial\lambda_l} \biggr)\nonumber\\
  &= \int \ws^2(y) \frac{\partial\ln \ps(x;\bm{\lambda})}{\partial\lambda_k}\frac{\partial\ln \ps(x;\bm{\lambda})}{\partial\lambda_l}\nonumber\\ 
  &\hphantom{=}\times\bigl(\ns\ps(x)\ps(y)+\nb\pb(x)\pb(y)\bigr)\deriv x\deriv y\nonumber\\
  &= \ns \int \ws^2(y)\ps(y)\deriv y \int \frac{\partial\ln \ps(x;\bm{\lambda})}{\partial\lambda_k}\frac{\partial\ln \ps(x;\bm{\lambda})}{\partial\lambda_l} \ps(x) \deriv x\nonumber\\
  &\hphantom{=}+ \nb \int \ws^2(y)\pb(y)\deriv y \int \frac{\partial\ln \ps(x;\bm{\lambda})}{\partial\lambda_k}\frac{\partial\ln \ps(x;\bm{\lambda})}{\partial\lambda_l} \pb(x) \deriv x\nonumber\\
  &\equiv H^\prime_{kl}. 
  \label{eq:lastelement}
\end{align}}
The matrix in the numerator is therefore given by
\begin{align}
  {E\bigl(\bm{g}(\bm{x},\bm{y};\bm{\theta})\bm{g}(\bm{x},\bm{y};\bm{\theta})^T\bigr)} &= 
  \left(\begin{array}{ccc}
    \bm{A}^\prime & \bm{B}^{\prime T} & \bm{D}^{\prime T}\\
    \bm{B}^\prime & \bm{C}^{\prime} & \bm{E}^{\prime T}\\
    \bm{D}^\prime & \bm{E}^{\prime} & \bm{H}^{\prime}\\    
  \end{array}\right)
  ,
\end{align}
with the matrix elements defined in Eqs.~\ref{eq:firstelement}--\ref{eq:lastelement}.

\subsection{Resulting covariance}
\label{app:unbinnedresult}
The full covariance matrix is then given by the following matrix multiplication
\begin{align}
  & E\left(\frac{\partial \bm{g}(\bm{x},\bm{y};\bm{\theta})}{\partial\bm{\theta}^T}\right)^{-1}\times
  E\bigl(\bm{g}(\bm{x},\bm{y};\bm{\theta})\bm{g}(\bm{x},\bm{y};\bm{\theta})^T\bigr)
  \times E\left(\frac{\partial \bm{g}(\bm{x},\bm{y};\bm{\theta})}{\partial\bm{\theta}^T}\right)^{-T}\nonumber\\
 =& \left(\begin{array}{ccc}
    \bm{A}^{-1} & \bm{0} & \bm{0}\\
    \bm{B}\bm{A}^{-1} & -\bm{1} & \bm{0}\\
    \bm{0} & \bm{H}^{-1}\bm{E} & \bm{H}^{-1}\\
\end{array}\right)
\times
\left(\begin{array}{ccc}
    \bm{A}^\prime & \bm{B}^{\prime T} & \bm{D}^{\prime T}\\
    \bm{B}^\prime & \bm{C}^{\prime} & \bm{E}^{\prime T}\\
    \bm{D}^\prime & \bm{E}^{\prime} & \bm{H}^{\prime}\\    
\end{array}\right)
\times
\left(\begin{array}{ccc}
    \bm{A}^{-T} & \bm{A}^{-T}\bm{B}^T & \bm{0}\\
    \bm{0} & -\bm{1} & \bm{E}^T\bm{H}^{-T}\\
    \bm{0} & \bm{0} & \bm{H}^{-T}\\
\end{array}\right).
\end{align}
We are interested in the bottom right elements for the covariance matrix for the parameters of interest $\bm{\lambda}$.
The matrix multiplication yields
\begin{align}
\bm{C}_{\bm{\lambda}}
  &= \bigl(\bm{H}^{-1}\bm{E}\bm{C}^\prime + \bm{H}^{-1}\bm{E}^\prime\bigr)\bm{E}^T\bm{H}^{-T} + \bigl(\bm{H}^{-1}\bm{E}\bm{E}^{\prime T} + \bm{H}^{-1}\bm{H}^\prime\bigr)\bm{H}^{-T}\nonumber\\
  &= \bm{H}^{-1} \bigl[\bm{E}\bm{C}^\prime\bm{E}^T + \bm{E}^\prime\bm{E}^T + \bm{E}\bm{E}^{\prime T} + \bm{H}^\prime\bigr] \bm{H}^{-T}\nonumber\\
  &= \bm{H}^{-1}\bm{H}^\prime \bm{H}^{-T}  + \bm{H}^{-1} \bigl[\bm{E}\bm{C}^\prime\bm{E}^T + \bm{E}^\prime\bm{E}^T + \bm{E}\bm{E}^{\prime T} \bigr] \bm{H}^{-T}\nonumber\\
  &= \bm{H}^{-1}\bm{H}^\prime \bm{H}^{-T}  + \bm{H}^{-1} \bigl[\bm{E}\bm{C}^\prime\bm{E}^T + 2\bm{E}\bm{E}^{\prime T} \bigr] \bm{H}^{-T} ~~~\textrm{because}\label{eq:fullcovariance}\\
 \bigl(\bm{E}^\prime\bm{E}^T\bigr)_{ik} &= \sum_{j} {E}_{ij}^\prime {E}_{kj} = \sum_j e_j^\prime\kappa_i e_j \kappa_k = \kappa_i\kappa_k \sum_j e_j^\prime e_j = \sum_j e_j\kappa_i e_j^\prime \kappa_k = \sum_j {E}_{ij}{E}_{kj}^\prime = \bigl(\bm{E}\bm{E}^{\prime T}\bigr)_{ik} \nonumber 
\end{align}
The additional covariance beyond the first term can be further simplified.
To this end we use our definitions
\begin{align}
  {E}_{l1} &= \frac{\kappa_l}{\bigl(\det\vi\bigr)^2}\biggl(-\vbbi \int \frac{\ps\pb^2}{\bigl(\ns\ps+\nb\pb\bigr)^2}\deriv y + \vsbi \int \frac{\pb^3}{\bigl(\ns\ps+\nb\pb\bigr)^2}\deriv y\biggr)\nonumber\\
  &= \frac{\kappa_l}{\bigl(\det\vi\bigr)^2}\bigl(-\vbbi (\ns {C}_{13}^\prime + \nb {C}_{23}^\prime) + \vsbi (\ns {C}_{23}^\prime + \nb {C}_{33}^\prime)\bigr)\nonumber\\
  {E}_{12} &= \frac{\kappa_l}{\bigl(\det\vi\bigr)^2}\biggl(\vbbi \int \frac{\ps^2\pb}{\bigl(\ns\ps+\nb\pb\bigr)^2}\deriv y - \vssi \int \frac{\pb^3}{\bigl(\ns\ps+\nb\pb\bigr)^2}\deriv y\biggr)\nonumber\\
  &= \frac{\kappa_l}{\bigl(\det\vi\bigr)^2}\bigl(\vbbi (\ns {C}_{12}^\prime + \nb {C}_{13}^\prime) - \vssi (\ns {C}_{23}^\prime + \nb {C}_{33}^\prime)\bigr)\nonumber\\
  {E}_{l3} &= \frac{\kappa_l}{\bigl(\det\vi\bigr)^2}\biggl(-\vsbi \int \frac{\ps^2\pb}{\bigl(\ns\ps+\nb\pb\bigr)^2}\deriv y + \vssi \int \frac{\ps\pb^2}{\bigl(\ns\ps+\nb\pb\bigr)^2}\deriv y\biggr)\nonumber\\  
  &= \frac{\kappa_l}{\bigl(\det\vi\bigr)^2}\bigl(-\vsbi (\ns {C}_{12}^\prime + \nb {C}_{13}^\prime) + \vssi (\ns {C}_{13}^\prime + \nb {C}_{23}^\prime)\bigr)
\end{align}
and 
\begin{align}
  {{E}}_{l1}^\prime &= \frac{\kappa_l}{\det\vi}\bigl(\vbbi {C}_{12}^\prime -\vsbi {C}_{13}^\prime\bigr)\nonumber\\
  {{E}}_{l2}^\prime &= \frac{\kappa_l}{\det\vi}\bigl(\vbbi {C}_{13}^\prime -\vsbi {C}_{23}^\prime\bigr)\nonumber\\
  {{E}}_{l3}^\prime &= \frac{\kappa_l}{\det\vi}\bigl(\vbbi {C}_{23}^\prime -\vsbi {C}_{33}^\prime\bigr)
\end{align}
and furthermore replace the \viji\ to only retain dependencies on ${C}_{ij}^\prime$ and the yields
\begin{align}
  \vssi &= \int \frac{\ps^2}{\ns\ps+\nb\pb} \frac{(\ns\ps+\nb\pb)^2}{(\ns\ps+\nb\pb)^2}\deriv y = \ns^2 {C}_{11}^\prime + 2\ns\nb {C}_{12}^\prime + \nb^2 {C}_{13}^\prime\nonumber\\
  \vsbi &= \ns^2 {C}_{12}^\prime + 2\ns\nb {C}_{13}^\prime + \nb^2 {C}_{23}^\prime\nonumber\\
  \vbbi &= \ns^2 {C}_{13}^\prime + 2\ns\nb {C}_{23}^\prime + \nb^2 {C}_{33}^\prime.
\end{align}
We can then calculate (easiest to verify using computer algebra)
{\small
\begin{align}
& \bigl(\bm{E}\bm{C}^\prime\bm{E}^T + \bm{E}\bm{E}^{\prime T}\bigr)_{ik}\nonumber\\
  =& \sum_{j,l} {E}_{ij}{C}_{jl}^\prime {E}_{kl} + \sum_j {E}_{ij} {E}_{kj}^\prime\nonumber\\
  =& \biggl[ 
    -\Bigl({C}^{\prime}_{11}{C}^{\prime}_{13}{C}^{\prime}_{33}-{C}^{\prime 2}_{12} {C}^{\prime}_{33}-{C}^{\prime}_{11}{C}^{\prime 2}_{23}+2{C}^{\prime}_{12}{C}^{\prime}_{13}{C}^{\prime}_{23}-{C}^{\prime 3}_{13}\Bigr)\nonumber\\
   & \times\ns^2\underbrace{\Bigl({C}^{\prime}_{12}\ns^3+3{C}^{\prime}_{13}\nb\ns^2+3{C}^{\prime}_{23}\nb^2\ns+{C}^{\prime}_{33}\nb^3-1\Bigr)}_{=0,~\text{as}~1=\int \pb\deriv y = \int \frac{(\ns\ps+\nb\pb)^3}{(\ns\ps+\nb\pb)^3}\pb\deriv y}\nonumber\\
    &\times  \Bigl({C}^{\prime}_{11}{C}^{\prime}_{13}\ns^4-{C}^{\prime 2}_{12}\ns^4+2{C}^{\prime}_{11}{C}^{\prime}_{23}\nb\ns^3-2{C}^{\prime}_{12}{C}^{\prime}_{13}\nb\ns^3+{C}^{\prime}_{11}{C}^{\prime}_{33}\nb^2\ns^2 +2{C}^{\prime}_{12}{C}^{\prime}_{23}\nb^2\ns^2\nonumber\\
    &~ -3{C}^{\prime 2}_{13}\nb^2\ns^2+2{C}^{\prime}_{12}{C}^{\prime}_{33}\nb^3\ns-2{C}^{\prime}_{13}{C}^{\prime}_{23}\nb^3\ns+{C}^{\prime}_{13}{C}^{\prime}_{33}\nb^4-{C}^{\prime 2}_{23}\nb^4\Bigr)\biggr]\times \frac{\kappa_i\kappa_k}{\bigl(\det\vi\bigr)^4}\nonumber\\
  =& 0.
\end{align}}
Equation~\ref{eq:fullcovariance} therefore simplifies to
\begin{align}
  \bm{C}_{\bm{\lambda}} &= \bm{H}^{-1}\bm{H}^\prime\bm{H}^{-T} + \bm{H}^{-1}\bm{E}\bm{E}^{\prime T}\bm{H}^{-T}\nonumber\\
  &= \bm{H}^{-1}\bm{H}^\prime\bm{H}^{-T} - \bm{H}^{-1}\bm{E}\bm{C}^{\prime}\bm{E}^T\bm{H}^{-T}.\label{eq:simplifiedcovariance}
\end{align}
The additional variance for the parameters of interest is therefore $\leq 0$, 
as the matrix $\bm{C}^\prime$ (and also $\bm{E}\bm{C}^\prime\bm{E}^T$) is positive definite. 
Using only the first term of Eq.~\ref{eq:simplifiedcovariance} (the \textit{naive} sandwich estimate) is therefore conservative for \textit{sWeights}, but can overestimate the variances. 
To guarantee asymptotically correct coverage, Eq.~\ref{eq:simplifiedcovariance} should be used. 

\clearpage

\section{Covariance determination for binned \textit{sWeighted} quantities}
\label{app:mestimationbinned}
To determine the variance for the sum of \textit{sWeights} in a bin $i$ in the control variable\footnote{Ref.~\cite{Pivk:2004ty} gives this variance as $\sum_{e\in\,\text{bin}\,i}\ws^2(y_e)$ in Eq.~22.}, $\sum_{e\in\,\text{bin}\,i}\ws(y_e)$, 
we use the same technique of systematic error propagation as detailed above in App.~\ref{app:mestimationunbinned}. 
We have as estimating equations
\begin{align}
  \varphi_i(\bm{y};\ns,\nb) &= 
  \sum \frac{\partial}{\partial N_i}\biggl[\ln(\ns\ps(y_e)+\nb\pb(y_e)) -\frac{\ns+\nb}{N}\biggr]\nonumber\\
  &= \sum \biggl[\frac{\calpi(y_e)}{\ns\ps(y_e)+\nb\pb(y_e)}  -\frac{1}{N}\biggr]\\
  \psi_{ij}(\bm{y};\viji,\ns,\nb) &= \sum_e \biggl[\frac{\calpi(y_e)\calpj(y_e)}{\bigl(\ns\ps(y_e)+\nb\pb(y_e)\bigr)^2}-\frac{\viji}{N}\biggr]\\
  \xi_i(\bm{x},\bm{y};\vssi,\vsbi,\vbbi,S_i) &= \sum_{e} \biggl[\theta(e\in\,\text{bin}\,i)\ws(y_e;\vssi,\vsbi,\vbbi)- \frac{S_i}{N}\biggr],\nonumber\\
  &= \biggl[\sum_{e\in\,\text{bin}\,i}\ws(y_e;\vssi,\vsbi,\vbbi)\biggr] - S_i
\end{align}
where $S_i$ denotes the expected signal yield in bin $i$.
We now want to determine the variance for the expected signal yield in bin $i$, and, in addition, 
its covariance with a different, non-overlapping bin $j$ with expected signal yield $S_j$. 

\subsection{Determination of the denominator}
The elements of the denominator $E(\partial\bm{g}/\partial\bm{\theta}^T)$ are calculated in full analogy to the unbinned case. 
We have
\begin{align}
  E\biggl(\frac{\partial}{\partial N_j} \varphi_i\biggr) &= 
  E\biggl(-\sum \frac{\calpi(y_e)\calpj(y_e)}{\bigl(\ns\ps(y_e)+\nb\pb(y_e)\bigr)^2} \biggr) \nonumber\\
  &= -\int \frac{\calpi(y)\calpj(y)}{\ns\ps(y)+\nb\pb(y)}\deriv y \equiv A_{ij}\\
E\biggl(\frac{\partial}{\partial\vkli} \varphi_i\biggr) &= 0\\
E\biggl(\frac{\partial}{\partial S_j} \varphi_i\biggr) &= 0,
\end{align}
for the $\bm{\varphi}$ components. 
For the $\bm{\psi}$ components we find
\begin{align}
  E\biggl(\frac{\partial}{\partial N_k} \psi_{ij}\biggr) &=
  E\biggl(-2 \sum_e\frac{\calpi(y_e)\calpj(y_e)\calpk(y_e)}{\bigl(\ns\ps(y_e)+\nb\pb(y_e)\bigr)^3}\biggr)\nonumber\\
&= -2 \int \frac{\calpi(y)\calpj(y)\calpk(y)}{\bigl(\ns\ps(y)+\nb\pb(y)\bigr)^2}\deriv y \equiv B_{(ij)k}\\
E\biggl(\frac{\partial}{\partial\vkli} \psi_{ij}\biggr) &=  -\delta_{(ij)(kl)}\\
E\biggl(\frac{\partial}{\partial S_k} \psi_{ij}\biggr) &= 0,
\end{align}
and finally, using the shorthand
\begin{align}
  \beta_1 &= \int \frac{\ps^3(y)}{\bigl(\ns\ps(y)+\nb\pb(y)\bigr)^2}\deriv y \Bigl(=-\frac{B_{11}}{2}\Bigr)\\
  \beta_2 &= \int \frac{\ps^2(y)\pb(y)}{\bigl(\ns\ps(y)+\nb\pb(y)\bigr)^2}\deriv y \Bigl(=-\frac{B_{12}}{2}\Bigr)\nonumber\\
  \beta_3 &= \int \frac{\ps(y)\pb^2(y)}{\bigl(\ns\ps(y)+\nb\pb(y)\bigr)^2}\deriv y \Bigl(=-\frac{B_{22}}{2}\Bigr)\nonumber\\
  \beta_4 &= \int \frac{\pb^3(y)}{\bigl(\ns\ps(y)+\nb\pb(y)\bigr)^2}\deriv y \Bigl(=-\frac{B_{32}}{2}\Bigr)\nonumber
\end{align}
we find
{\small
\begin{align}
E\biggl(\frac{\partial}{\partial N_i} \xi_k\biggr) &= 0\\
E\biggl(\frac{\partial}{\partial \vssi} \xi_k\biggr) &= E\biggl(\sum_{e\in\text{bin}\,k} \frac{-\vbbi\ps(y_e)\pb(y_e) +\vsbi\pb^2(y_e)}{\bigl((\vbbi-\vsbi)\ps(y_e)+(\vssi-\vsbi)\pb(y_e)\bigr)^2}\biggr)\nonumber\\
&= \frac{1}{\bigl(\det\vi\bigr)^2} 
\int_y \int_{\text{bin}\,k}\frac{-\vbbi\ps(y)\pb(y) +\vsbi\pb^2(y)}{\bigl(\ns\ps(y)+\nb\pb(y)\bigr)^2}\nonumber\\
&\hphantom{=} \times \bigl[\ns\ps(y)\ps(x)+\nb\pb(y)\pb(x)\bigr]\deriv y\deriv x\nonumber\\
&= \frac{1}{\bigl(\det\vi\bigr)^2}\biggl[\ns \int_{\text{bin}\,k} \ps(x)\deriv x\int \frac{-\vbbi\ps^2(y)\pb(y) + \vsbi\ps(y)\pb^2(y)}{\bigl(\ns\ps(y)+\nb\pb(y)\bigr)^2} \deriv y\nonumber\\
&\hphantom{=} +\nb \int_{\text{bin}\,k} \pb(x)\deriv x\int \frac{-\vbbi\ps(y)\pb^2(y) + \vsbi\pb^3(y)}{\bigl(\ns\ps(y)+\nb\pb(y)\bigr)^2} \deriv y\biggr]\nonumber\\
&= \frac{1}{\bigl(\det\vi\bigr)^2}\biggl[\ns\int\ps(x)\deriv x\bigl(-\vbbi \beta_2 +\vsbi \beta_3\bigr)\nonumber\\
&\hphantom{=} +\nb\int\pb(x)\deriv x \bigl(-\vbbi \beta_3+\vsbi \beta_4\bigr)\biggr]\nonumber\\
&= \frac{\beta_3^2-\beta_2\beta_4}{\bigl(\det\vi\bigr)^2}\ns\nb 
\biggl(\underbrace{\int_{\text{bin}\,k} \ps(x)\deriv x}_{=\delta_k} - \underbrace{\int_{\text{bin}\,k}\pb(x)\deriv x}_{=\epsilon_k}\biggr)\nonumber\\
&\equiv E_{k1} = (\delta_k-\epsilon_k) e_1,
\end{align}}
{\small
\begin{align}
E\biggl(\frac{\partial}{\partial \vsbi} \xi_k\biggr) &= E\biggl(\sum_{e\in\text{bin}\,k} \frac{+\vbbi\ps^2(y_e) -\vssi\pb^2(y_e)}{\bigl((\vbbi-\vsbi)\ps(y_e)+(\vssi-\vsbi)\pb(y_e)\bigr)^2}\biggr)\nonumber\\
&= \frac{1}{\bigl(\det\vi\bigr)^2}\biggl[\ns \int_{\text{bin}\,k} \ps(x)\deriv x\int \frac{\vbbi\ps^3(y) - \vssi\ps(y)\pb^2(y)}{\bigl(\ns\ps(y)+\nb\pb(y)\bigr)^2} \deriv y\nonumber\\
&\hphantom{=} +\nb \int_{\text{bin}\,k} \pb(x)\deriv x\int \frac{\vbbi\ps^2(y)\pb(y) - \vssi\pb^3(y)}{\bigl(\ns\ps(y)+\nb\pb(y)\bigr)^2} \deriv y\biggr]\nonumber\\
&= \frac{1}{\bigl(\det\vi\bigr)^2}\biggl[\ns\int\ps(x)\deriv x\bigl(\vbbi \beta_1 -\vssi \beta_3\bigr)\nonumber\\
  & \hphantom{=} +\nb\int\pb(x)\deriv x \bigl(\vbbi \beta_2-\vssi \beta_4\bigr)\biggr]\nonumber\\
&= \frac{\beta_1\beta_4-\beta_2\beta_3}{\bigl(\det\vi\bigr)^2}\ns\nb\int_{\text{bin}\,k}\bigl(\ps(x)-\pb(x)\bigr)\deriv x\nonumber\\
&\equiv E_{k2}=(\delta_k-\epsilon_k) e_2,
\end{align}}
and
{\small\begin{align}
E\biggl(\frac{\partial}{\partial \vbbi} \xi_k\biggr) &= E\biggl(\sum_{e\in\text{bin}\,k} \frac{-\vsbi\ps^2(y_e) +\vssi\ps(y_e)\pb(y_e)}{\bigl((\vbbi-\vsbi)\ps(y_e)+(\vssi-\vsbi)\pb(y_e)\bigr)^2}\biggr)\nonumber\\
&= \frac{1}{\bigl(\det\vi\bigr)^2}\biggl[\ns \int_{\text{bin}\,k} \ps(x)\deriv x\int \frac{-\vsbi\ps^3(y) + \vssi\ps^2(y)\pb(y)}{\bigl(\ns\ps(y)+\nb\pb(y)\bigr)^2} \deriv y\nonumber\\
& \hphantom{=} +\nb \int_{\text{bin}\,k} \pb(x)\deriv x\int \frac{-\vsbi\ps^2(y)\pb(y) + \vssi\ps\pb^2(y)}{\bigl(\ns\ps(y)+\nb\pb(y)\bigr)^2} \deriv y\biggr]\nonumber\\
&= \frac{1}{\bigl(\det\vi\bigr)^2}\biggl[\ns\int\ps(x)\deriv x\bigl(-\vsbi \beta_1 +\vssi \beta_2\bigr)\nonumber\\
& \hphantom{=} +\nb\int\pb(x)\deriv x \bigl(-\vsbi \beta_2+\vssi \beta_3\bigr)\biggr]\nonumber\\
&= \frac{\beta_2^2-\beta_1\beta_3}{\bigl(\det\vi\bigr)^2}\ns\nb\int_{\text{bin}\,k}\bigl(\ps(x)-\pb(x)\bigr)\deriv x\nonumber\\
&\equiv E_{k3}=(\delta_k-\epsilon_k) e_3,
\end{align}}
and finally
\begin{align}
E\biggl(\frac{\partial}{\partial S_k} \xi_l\biggr) &= -\delta_{kl}.
\end{align}
The  denominator matrix is then given by
\begin{align}
  E\left(\frac{\partial \bm{g}(\bm{x},\bm{y};\bm{\theta})}{\partial\bm{\theta}^T}\right) &=
  \left(\begin{array}{ccc}
    \bm{A} & \bm{0} & \bm{0}\\
    \bm{B} & -\bm{1} & \bm{0}\\
    \bm{0} & \bm{E} & -\bm{1}
  \end{array}\right)
\end{align}
with the inverse
\begin{align}
  E\left(\frac{\partial \bm{g}(\bm{x},\bm{y};\bm{\theta})}{\partial\bm{\theta}^T}\right)^{-1} &=
  \left(\begin{array}{ccc}
   \bm{A}^{-1} & \bm{0} & \bm{0}\\
   \bm{B}\bm{A}^{-1} & -\bm{1} & \bm{0}\\
   +\bm{E}\bm{B}\bm{A}^{-1} & -\bm{E} & -\bm{1}
  \end{array}\right)
  =\left(\begin{array}{ccc}
   \bm{A}^{-1} & \bm{0} & \bm{0}\\
   \bm{B}\bm{A}^{-1} & -\bm{1} & \bm{0}\\
   \bm{0} & -\bm{E} & -\bm{1}
  \end{array}\right)
\end{align}
for which we again needed to show $\bm{E}\bm{B}=\bm{0}$, \ie 
\begin{align}
  &B_{11}E_{k1} + B_{12}E_{k2} + B_{22}E_{k3}\nonumber\\
  =& -2\bigl(\beta_1(\beta_3^2-\beta_2\beta_4)+\beta_2(\beta_1\beta_4-\beta_2\beta_3)+\beta_3(\beta_2^2-\beta_1\beta_3)\bigr) \frac{\ns\nb}{ \bigl(\det\vi\bigr)^2} (\delta_k-\epsilon_k)\nonumber\\
  =& -2\bigl(\beta_1\beta_3^2- \beta_1\beta_2\beta_4 + \beta_1\beta_2\beta_4 - \beta_2^2\beta_3 + \beta_3\beta_2^2 -\beta_1\beta_3^2\bigr) \frac{\ns\nb}{ \bigl(\det\vi\bigr)^2} (\delta_k-\epsilon_k)=0 ~~~\textrm{and}\\
  & B_{12}E_{k1} + B_{22}E_{k2} + B_{32}E_{k3}\nonumber\\
  =& -2\bigl(\beta_2(\beta_3^2-\beta_2\beta_4)+\beta_3(\beta_1\beta_4-\beta_2\beta_3)+\beta_4(\beta_2^2-\beta_1\beta_3)\bigr) \frac{\ns\nb}{ \bigl(\det\vi\bigr)^2} (\delta_k-\epsilon_k)\nonumber\\
  =& -2\bigl(\beta_2\beta_3^2-\beta_2^2\beta_4+\beta_1\beta_3\beta_4-\beta_2\beta_3^2+\beta_2^2\beta_4-\beta_1\beta_3\beta_4\bigr) \frac{\ns\nb}{ \bigl(\det\vi\bigr)^2} (\delta_k-\epsilon_k)=0.
\end{align}

\subsection{Determination of the numerator}
The elements of the numerator $E\bigl(\bm{g}(\bm{x},\bm{y};\bm{\theta})\bm{g}(\bm{x},\bm{y};\bm{\theta})^T\bigr)$ are given by
{\small
\begin{align}
  E\bigl(\varphi_i\varphi_j \bigr) &= E\biggl(\sum_e\frac{\calpi(y_e)\calpj(y_e)}{\bigl(\ns\ps(y_e)+\nb\pb(y_e)\bigr)^2}\biggr)\nonumber\\
  &= \int \frac{\calpi(y)\calpj(y)}{\ns\ps(y)+\nb\pb(y)}\deriv y \equiv A_{ij}^\prime\\
  E\bigl(\varphi_k\psi_{ij} \bigr) &=
E\biggl(\sum_e\frac{\calpi(y_e)\calpj(y_e)\calpk(y_e)}{\bigl(\ns\ps(y_e)+\nb\pb(y_e)\bigr)^3}\biggr)\nonumber\\
&= \int \frac{\calpi(y)\calpj(y)\calpk(y)}{\bigl(\ns\ps(y)+\nb\pb(y)\bigr)^2}\deriv y \equiv B_{(ij)k}^\prime\\ 
E\bigl(\psi_{ij}\psi_{kl}\bigr) &=
E\biggl(\sum_e\frac{\calpi(y_e)\calpj(y_e)\calp_k(y_e)\calp_l(y_e)}{\bigl(\ns\ps(y_e)+\nb\pb(y_e)\bigr)^4}\biggr)\nonumber\\
  &= \int\frac{\calpi(y)\calpj(y)\calp_k(y)\calp_l(y)}{\bigl(\ns\ps(y)+\nb\pb(y)\bigr)^3}\deriv y \equiv C_{(ij)(kl)}^\prime. 
\end{align}}
For the combinations including the terms $\bm{\xi}$ we find
{\small
\begin{align}
  E\bigl(\xi_k\xi_l \bigr) &= E\biggl(\sum_e\biggl[\theta(x_e\in\text{bin}\,k)\ws(y_e)-\frac{S_k}{N}\biggr]\sum_f\biggl[\theta(x_f\in\text{bin}\,l)\ws(y_f)-\frac{S_l}{N}\biggr]\biggr)\nonumber\\
  &\overset{k=l}{=} E\biggl(\sum\theta(e\in\text{bin}\,k) \ws^2(y_e) + \sum_{e\neq f} \theta(e\in\text{bin}\,k)\theta(f\in\text{bin}\,l)\ws(y_e)\ws(y_f)\nonumber\\
  &\hphantom{=} - S_k\sum_f\theta(f\in\text{bin}\,l) \ws(y_f) - S_l\sum_e \theta(e\in\text{bin}\,k)\ws(y_e) + S_kS_l\biggr)\nonumber\\
  &= E\biggl(\sum_{e\in\text{bin}\,k} \ws^2(y_e)\biggr)\nonumber\\
  &= \ns \underbrace{\int_{\text{bin}\,k}\ps(x)\deriv x}_{\delta_k}\int \ws^2(y)\ps(y)\deriv y + \nb\underbrace{\int_{\text{bin}\,k}\pb(x)\deriv x}_{\epsilon_k}\int \ws^2(y)\pb(y)\deriv y \nonumber\\
  &\equiv H^\prime_{kk}\\
E\bigl(\xi_k\xi_l \bigr)  &\overset{k\neq l}{=} S_kS_l - S_l\sum_e\theta(e\in\text{bin}\,k)\ws(y_e) - S_k\sum_f\theta(f\in\text{bin}\,l)\ws(y_f) + S_kS_l\nonumber\\
  &=0,
\end{align}}
and furthermore
{\small
\begin{align}
  E\bigl(\xi_k\psi_{ij} \bigr) &= E\biggl(\sum_e\biggl[\theta(x_e\in\text{bin}\,k)\ws(y_e) -\frac{S_k}{N}\biggr] 
  \sum_f \biggl[\frac{\calpi(y_f)\calpj(y_f)}{\bigl(\ns\ps(y_f)+\nb\pb(y_f)\bigr)^2} -\frac{\viji}{N}\biggr]\biggr)\nonumber\\
  &= E\biggl( \sum_e \theta(x_e\in\text{bin}\,k)\ws(y_e)\frac{\calpi(y_e)\calpj(y_e)}{\bigl(\ns\ps(y_e)+\nb\pb(y_e)\bigr)^2}\nonumber\\
  &\hphantom{=} +\sum_{e\neq f} \theta(x_e\in\text{bin}\,k)\ws(y_e)\frac{\calpi(y_f)\calpj(y_f)}{\bigl(\ns\ps(y_f)+\nb\pb(y_f)\bigr)^2}+S_k\viji\nonumber\\
  &\hphantom{=} -\sum_e\theta(x_e\in\text{bin}\,k)\ws(y_e)\viji -S_k\sum_e\frac{\calpi(y_e)\calpj(y_e)}{\bigl(\ns\ps(y_e)+\nb\pb(y_e)\bigr)^2}\biggr) \nonumber\\
  &= E\biggl(\sum_{e\in\text{bin}\,k} \ws(y_e)\frac{\calpi(y_e)\calpj(y_e)}{\bigl(\ns\ps(y_e)+\nb\pb(y_e)\bigr)^2}\biggr)\nonumber\\
  &= \ns\int_{\text{bin}\,k}\ps(x)\deriv x \int \ws(y)\frac{\calpi(y)\calpj(y)\ps(y)}{\bigl(\ns\ps(y)+\nb\pb(y)\bigr)^2}\deriv y\nonumber\\
  &\hphantom{=} +\nb \int_{\text{bin}\,k}\pb(x)\deriv x \int \ws(y)\frac{\calpi(y)\calpj(y)\pb(y)}{\bigl(\ns\ps(y)+\nb\pb(y)\bigr)^2}\deriv y,\nonumber\\
  &= \frac{\ns}{\det\vi}\delta_k\biggl(\vbbi\int \frac{\calpi\calpj\ps^2}{\bigl(\ns\ps+\nb\pb\bigr)^3}\deriv y-\vsbi\int \frac{\calpi\calpj\ps\pb}{\bigl(\ns\ps+\nb\pb\bigr)^3}\deriv y\biggr)\nonumber\\
  &\hphantom{=} +\frac{\nb}{\det\vi}\epsilon_k\biggl(\vbbi\int \frac{\calpi\calpj\ps\pb}{\bigl(\ns\ps+\nb\pb\bigr)^3}\deriv y-\vsbi\int \frac{\calpi\calpj\pb^2}{\bigl(\ns\ps+\nb\pb\bigr)^3}\deriv y\biggr),\nonumber\\
  &\equiv E_{k(ij)}^\prime,
\end{align}}
\ie
{\small\begin{align}
  E\bigl(\xi_k\psi_{ss} \bigr) &= \frac{\ns}{N\det\vi}\delta_k\bigl(\vbbi C_{11}^\prime-\vsbi C_{12}^\prime\bigr) 
  +\frac{\nb}{N\det\vi}\epsilon_k\bigl(\vbbi C_{12}^\prime-\vsbi C_{13}^\prime\bigr) = E_{k1}^\prime\nonumber\\
  E\bigl(\xi_k\psi_{sb} \bigr) &= \frac{\ns}{N\det\vi}\delta_k\bigl(\vbbi C_{12}^\prime-\vsbi C_{13}^\prime\bigr) 
  +\frac{\nb}{N\det\vi}\epsilon_k\bigl(\vbbi C_{13}^\prime-\vsbi C_{23}^\prime\bigr) = E_{k2}^\prime\nonumber\\
  E\bigl(\xi_k\psi_{bb} \bigr) &= \frac{\ns}{N\det\vi}\delta_k\bigl(\vbbi C_{13}^\prime-\vsbi C_{23}^\prime\bigr) 
  +\frac{\nb}{N\det\vi}\epsilon_k\bigl(\vbbi C_{23}^\prime-\vsbi C_{33}^\prime\bigr) = E_{k3}^\prime\nonumber
\end{align}}
and finally
{\small\begin{align}
  E\bigl(\xi_k\varphi_i \bigr) &= E\biggl(\sum_e\biggl[\theta(x_e\in\text{bin}\,k)\ws(y_e) -\frac{S_k}{N}\biggr] 
  \biggl[\biggl(\sum_f\frac{\calpi(y_f)}{\ns\ps(y_f)+\nb\pb(y_f)}\biggr)-1\biggr]\nonumber\\
  &= E\biggl(
  \sum_e\theta(x_e\in\text{bin}\,k) \frac{\calpi(y_e)}{\ns\ps(y_e)+\nb\pb(y_e)}\nonumber\\
  &\hphantom{=} +\sum_{e\neq f}\theta(x_e\in\text{bin}\,k) \frac{\calpi(y_f)}{\ns\ps(y_f)+\nb\pb(y_f)}\nonumber\\
  &\hphantom{=} -\sum_e\theta(x_e\in\text{bin}\,k)\ws(y_e)
  -S_k\sum_f\frac{\calpi(y_f)}{\ns\ps(y_f)+\nb\pb(y_f)} + S_k
  \biggr)\nonumber\\
  &= E\biggl(\sum_e \theta(x_e\in\text{bin}\,k)\frac{\calpi(y_e)}{\ns\ps(y_e)+\nb\pb(y_e)}\biggr)\nonumber\\
  &=  \int \theta(x\in\text{bin}\,k) \ws(y)\frac{\calpi(y)}{\ns\ps(y)+\nb\pb(y)}\nonumber\\
  &\hphantom{=} \times \bigl[\ns\ps(x)\ps(y)+\nb\pb(x)\pb(y)\bigr]\deriv x\deriv y\nonumber\\
  &= \ns \int_{\text{bin}\,k}\ps(x)\deriv x \int \ws(y)\frac{\calpi(y)\ps(y)}{\ns\ps(y)+\nb\pb(y)}\deriv y\nonumber\\
  &\hphantom{=} + \nb\int_{\text{bin}\,k}\pb(x)\deriv x \int \ws(y)\frac{\calpi(y)\pb(y)}{\ns\ps(y)+\nb\pb(y)}\deriv y\nonumber\\
  &\equiv D_{ki}^\prime.
\end{align}}
As before, the matrix in the numerator is given by
\begin{align}
E\bigl(\bm{g}(\bm{x},\bm{y};\bm{\theta})\bm{g}(\bm{x},\bm{y};\bm{\theta})^T\bigr)
  &= \left(\begin{array}{ccc}
    \bm{A}^\prime & \bm{B}^{\prime T} & \bm{D}^{\prime T} \\
    \bm{B}^\prime & \bm{C}^{\prime} & \bm{E}^{\prime T} \\
    \bm{D}^\prime & \bm{E}^{\prime} & \bm{H}^{\prime}
  \end{array}\right).
\end{align}

\subsection{Resulting covariance}
The full covariance matrix is given by the product
\begin{align}
  & E\left(\frac{\partial \bm{g}(\bm{x},\bm{y};\bm{\theta})}{\partial\bm{\theta}^T}\right)^{-1}\times
E\bigl(\bm{g}(\bm{x},\bm{y};\bm{\theta})\bm{g}(\bm{x},\bm{y};\bm{\theta})^T\bigr)
  \times E\left(\frac{\partial \bm{g}(\bm{x},\bm{y};\bm{\theta})}{\partial\bm{\theta}^T}\right)^{-T}\nonumber\\
 =& \left(\begin{array}{ccc}
    \bm{A}^{-1} & \bm{0} & \bm{0}\\
    \bm{B}\bm{A}^{-1} & -\bm{1} & \bm{0}\\
    \bm{0} & -\bm{E} & -\bm{1}\\
\end{array}\right)
\times
\left(\begin{array}{ccc}
    \bm{A}^\prime & \bm{B}^{\prime T} & \bm{D}^{\prime T}\\
    \bm{B}^\prime & \bm{C}^{\prime} & \bm{E}^{\prime T}\\
    \bm{D}^\prime & \bm{E}^{\prime} & \bm{H}^{\prime}\\    
\end{array}\right)
\times
\left(\begin{array}{ccc}
    \bm{A}^{-T} & \bm{A}^{-T}\bm{B}^T & \bm{0}\\
    \bm{0} & -\bm{1} & -\bm{E}^T\\
    \bm{0} & \bm{0} & -\bm{1}\\
\end{array}\right)
\end{align}
The covariance matrix for the parameters of interest, in this instance the \textit{sWeighted} bin contents, is given by
\begin{align}
\bm{C}_{\bm{S}}
  &=  \bm{E}\bm{C}^\prime\bm{E}^T + \bm{E}^\prime\bm{E}^T + \bm{E}\bm{E}^{\prime T} + \bm{H}^\prime\nonumber\\
&= \bm{H}^\prime - \bm{E}\bm{C}^\prime\bm{E}^T.\label{eq:correctvarbinned}
\end{align}
Equation \ref{eq:correctvarbinned} is shown by first realising
\begin{align}
  \bigl(\bm{E}^{\prime}\bm{E}^T\bigr)_{ik} &= \sum_j E_{ij}^\prime E_{jk}^{ T} = (\delta_k-\epsilon_k) \sum_{j} E_{ij}^\prime e_{j} = (\delta_k-\epsilon_k) \bigl(E_{i1}^\prime e_1+E_{i2}^\prime e_2+E_{i3}^\prime e_3\bigr)\nonumber\\
  &= -\ns^2\nb^2 (\delta_i-\epsilon_i)(\delta_k-\epsilon_k) \times \bigl(C_{11}^\prime C_{13}^\prime C_{33}^\prime -C_{12}^{\prime 2} C_{33}^\prime -C_{11}^\prime C_{23}^{\prime 2} +2C_{12}^\prime C_{13}^\prime C_{23}^\prime - C_{13}^{\prime 3} \bigr) \nonumber\\
  &\hphantom{=} \times \underbrace{\bigl(C_{12}^{\prime} \ns^3 + 3 C_{13}^{\prime} \ns^2\nb + 3 C_{23}^{\prime} \ns\nb^2 + C_{33}^{\prime}\nb^3\bigr)}_{=1}\times\bigl(\det\vi\bigr)^{-3}\nonumber\\
  &= (\delta_i-\epsilon_i) \sum_j e_{kj}^{\prime} e_j = \sum_j e_{ij} e_{kj}^{\prime} = \bigl(\bm{E}\bm{E}^{\prime T}\bigr)_{ik}
\end{align}
and
\begin{align}
  \bigl(\bm{E}\bm{C}^\prime\bm{E}^T\bigr)_{ik} &= \sum_{j,l}E_{ij} C_{jl}^\prime E_{lk}^T = (\delta_i-\epsilon_i)(\delta_k-\epsilon_k)\sum_{j,l} e_{j} C_{jl}^{\prime} e_{l}\nonumber\\
  &= \ns^2\nb^2 (\delta_i-\epsilon_i)(\delta_k-\epsilon_k)
  \times \bigl(C_{11}^{\prime}C_{13}^{\prime}C_{33}^{\prime} - C_{12}^{\prime 2}C_{33}^{\prime} - C_{11}^{\prime}C_{23}^{\prime 2} + 2C_{12}^{\prime}C_{13}^{\prime}C_{23}^{\prime} - C_{13}^{\prime 3}\bigr)\nonumber\\
  &  \hphantom{=}\times \bigl(\det\vi\bigr)^{-3}\nonumber\\
  &= -\bigl(\bm{E}^\prime\bm{E}^T\bigr)_{ik}.
\end{align}
The first term in Eq.~\ref{eq:correctvarbinned} can be estimated by $\sum_{e\in\mathrm{bin}\,k}\ws^2(y_e)$ from the sample. 
As the matrix $\bm{C}^\prime$ is positive definite, using only the first term can overestimate the variances and is therefore conservative. 
To guarantee asymptotically correct coverage for a $\chi^2$ fit the full expression in Eq.~\ref{eq:correctvarbinned} should be used, which also accounts for correlations between bins. 

\clearpage
\section{Results from pseudoexperiments correcting acceptance effects}
\label{sec:appeffs}
\begin{table}[hb!]
\centering
\subfloat[Parameter $c_0$\label{tab:pullsc0a}]{
\scalebox{0.7}{
\begin{tabular}{llrrrrrrr}\hline
method & pull & 500 & 1\,k & 2\,k & 5\,k & 10\,k & 20\,k & 50\,k \\ \hline\hline
wFit  & mean & $-0.01 \pm 0.01$ & $\hphantom{-}0.01 \pm 0.01$ & $\hphantom{-}0.01 \pm 0.01$ & $-0.01 \pm 0.01$ & $\hphantom{-}0.01 \pm 0.01$ & $\hphantom{-}0.02 \pm 0.01$ & $\hphantom{-}0.01 \pm 0.01$\\
      & width & $1.38 \pm 0.01$ & $1.40 \pm 0.01$ & $1.40 \pm 0.01$ & $1.37 \pm 0.01$ & $1.38 \pm 0.01$ & $1.38 \pm 0.01$ & $1.37 \pm 0.01$\\
scaled weights & mean & $-0.00 \pm 0.01$ & $0.01 \pm 0.01$ & $0.01 \pm 0.01$ & $-0.01 \pm 0.01$ & $0.01 \pm 0.01$ & $0.01 \pm 0.01$ & $0.01 \pm 0.01$\\
      & width & $1.15 \pm 0.01$ & $1.16 \pm 0.01$ & $1.17 \pm 0.01$ & $1.14 \pm 0.01$ & $1.15 \pm 0.01$ & $1.15 \pm 0.01$ & $1.14 \pm 0.01$\\
squared correction  & mean & $-0.00 \pm 0.01$ & $0.01 \pm 0.01$ & $0.01 \pm 0.01$ & $-0.01 \pm 0.01$ & $0.01 \pm 0.01$ & $0.01 \pm 0.01$ & $0.00 \pm 0.01$\\
      & width & $0.99 \pm 0.01$ & $1.01 \pm 0.01$ & $1.01 \pm 0.01$ & $0.99 \pm 0.01$ & $1.00 \pm 0.01$ & $1.00 \pm 0.01$ & $0.99 \pm 0.01$\\
bootstrapping  & mean & $-0.00 \pm 0.01$ & $0.01 \pm 0.01$ & $0.01 \pm 0.01$ & $-0.01 \pm 0.01$ & $0.01 \pm 0.01$ & $0.01 \pm 0.01$ & $0.00 \pm 0.01$\\
      & width & $0.97 \pm 0.01$ & $1.00 \pm 0.01$ & $1.01 \pm 0.01$ & $0.99 \pm 0.01$ & $1.00 \pm 0.01$ & $1.00 \pm 0.01$ & $0.99 \pm 0.01$\\
asymptotic & mean & $-0.00 \pm 0.01$ & $0.01 \pm 0.01$ & $0.01 \pm 0.01$ & $-0.01 \pm 0.01$ & $0.01 \pm 0.01$ & $0.01 \pm 0.01$ & $0.00 \pm 0.01$\\
      & width & $1.00 \pm 0.01$ & $1.01 \pm 0.01$ & $1.01 \pm 0.01$ & $0.99 \pm 0.01$ & $1.00 \pm 0.01$ & $1.00 \pm 0.01$ & $0.99 \pm 0.01$\\
cFit  & mean & $0.00 \pm 0.01$ & $0.01 \pm 0.01$ & $0.01 \pm 0.01$ & $-0.00 \pm 0.01$ & $0.01 \pm 0.01$ & $0.01 \pm 0.01$ & $0.00 \pm 0.01$\\
      & width & $1.00 \pm 0.01$ & $1.01 \pm 0.01$ & $1.01 \pm 0.01$ & $0.99 \pm 0.01$ & $1.00 \pm 0.01$ & $1.00 \pm 0.01$ & $0.99 \pm 0.01$\\
\hline \multicolumn{2}{l}{rel. efficiency cFit/weighted} & $0.90 \pm 0.02$ & $0.91 \pm 0.02$ & $0.90 \pm 0.02$ & $0.90 \pm 0.02$ & $0.90 \pm 0.02$ & $0.90 \pm 0.02$ & $0.90 \pm 0.02$\\
\hline\end{tabular}
}}\\
\subfloat[Parameter $c_1$\label{tab:pullsc1a}]{
\scalebox{0.7}{
\begin{tabular}{llrrrrrrr}\hline
method & pull & 500 & 1\,k & 2\,k & 5\,k & 10\,k & 20\,k & 50\,k \\ \hline\hline
wFit  & mean & $-0.17 \pm 0.01$ & $-0.11 \pm 0.01$ & $-0.07 \pm 0.01$ & $-0.06 \pm 0.01$ & $-0.05 \pm 0.01$ & $-0.02 \pm 0.01$ & $-0.03 \pm 0.01$\\
      & width & $1.39 \pm 0.01$ & $1.40 \pm 0.01$ & $1.39 \pm 0.01$ & $1.38 \pm 0.01$ & $1.38 \pm 0.01$ & $1.38 \pm 0.01$ & $1.39 \pm 0.01$\\
scaled weights & mean & $-0.15 \pm 0.01$ & $-0.10 \pm 0.01$ & $-0.06 \pm 0.01$ & $-0.05 \pm 0.01$ & $-0.04 \pm 0.01$ & $-0.02 \pm 0.01$ & $-0.02 \pm 0.01$\\
      & width & $1.16 \pm 0.01$ & $1.17 \pm 0.01$ & $1.16 \pm 0.01$ & $1.15 \pm 0.01$ & $1.15 \pm 0.01$ & $1.15 \pm 0.01$ & $1.16 \pm 0.01$\\
squared correction  & mean & $-0.10 \pm 0.01$ & $-0.07 \pm 0.01$ & $-0.04 \pm 0.01$ & $-0.04 \pm 0.01$ & $-0.03 \pm 0.01$ & $-0.01 \pm 0.01$ & $-0.02 \pm 0.01$\\
      & width & $0.79 \pm 0.01$ & $0.80 \pm 0.01$ & $0.79 \pm 0.01$ & $0.79 \pm 0.01$ & $0.79 \pm 0.01$ & $0.79 \pm 0.01$ & $0.79 \pm 0.01$\\
bootstrapping  & mean & $-0.13 \pm 0.01$ & $-0.07 \pm 0.01$ & $-0.04 \pm 0.01$ & $-0.04 \pm 0.01$ & $-0.03 \pm 0.01$ & $-0.01 \pm 0.01$ & $-0.02 \pm 0.01$\\
      & width & $1.01 \pm 0.01$ & $0.99 \pm 0.01$ & $1.00 \pm 0.01$ & $1.00 \pm 0.01$ & $1.00 \pm 0.01$ & $1.00 \pm 0.01$ & $1.01 \pm 0.01$\\
asymptotic & mean & $-0.11 \pm 0.01$ & $-0.07 \pm 0.01$ & $-0.05 \pm 0.01$ & $-0.04 \pm 0.01$ & $-0.03 \pm 0.01$ & $-0.01 \pm 0.01$ & $-0.02 \pm 0.01$\\
      & width & $1.00 \pm 0.01$ & $1.01 \pm 0.01$ & $1.00 \pm 0.01$ & $1.00 \pm 0.01$ & $1.00 \pm 0.01$ & $1.00 \pm 0.01$ & $1.00 \pm 0.01$\\
cFit  & mean & $-0.08 \pm 0.01$ & $-0.05 \pm 0.01$ & $-0.03 \pm 0.01$ & $-0.03 \pm 0.01$ & $-0.02 \pm 0.01$ & $-0.01 \pm 0.01$ & $-0.02 \pm 0.01$\\
      & width & $1.00 \pm 0.01$ & $1.01 \pm 0.01$ & $1.00 \pm 0.01$ & $1.00 \pm 0.01$ & $1.00 \pm 0.01$ & $1.00 \pm 0.01$ & $1.00 \pm 0.01$\\
\hline \multicolumn{2}{l}{rel. efficiency cFit/weighted} & $0.91 \pm 0.02$ & $0.92 \pm 0.02$ & $0.90 \pm 0.02$ & $0.91 \pm 0.02$ & $0.91 \pm 0.02$ & $0.90 \pm 0.02$ & $0.90 \pm 0.02$\\
\hline\end{tabular}
}}
\caption{Means and widths of the pull distribution for the different approaches to the uncertainty estimation for the efficiency correction $\epsilon(\cos\theta)=1.0-0.7\cos^2\theta$, depending on the number of events.
  In addition, the relative efficiencies (\ie\ the ratios of variances) of the cFit estimator and the weighted estimator defined by Eq.~\ref{eq:mlweighted} are given.
  \label{tab:effpullsa}}
\end{table}

\begin{table}
\centering
\subfloat[Parameter $c_0$\label{tab:pullsc0b}]{
\scalebox{0.7}{
\begin{tabular}{llrrrrrrr}\hline
method & pull & 500 & 1\,k & 2\,k & 5\,k & 10\,k & 20\,k & 50\,k \\ \hline\hline
wFit  & mean & $\hphantom{-}0.01 \pm 0.01$ & $\hphantom{-}0.00 \pm 0.01$ & $-0.00 \pm 0.01$ & $-0.01 \pm 0.01$ & $\hphantom{-}0.01 \pm 0.01$ & $-0.01 \pm 0.01$ & $-0.00 \pm 0.01$\\
      & width & $1.22 \pm 0.01$ & $1.22 \pm 0.01$ & $1.23 \pm 0.01$ & $1.24 \pm 0.01$ & $1.24 \pm 0.01$ & $1.22 \pm 0.01$ & $1.21 \pm 0.01$\\
scaled weights & mean & $0.00 \pm 0.01$ & $0.00 \pm 0.01$ & $-0.00 \pm 0.01$ & $-0.00 \pm 0.01$ & $0.00 \pm 0.01$ & $-0.01 \pm 0.01$ & $-0.00 \pm 0.01$\\
      & width & $0.83 \pm 0.01$ & $0.83 \pm 0.01$ & $0.83 \pm 0.01$ & $0.84 \pm 0.01$ & $0.84 \pm 0.01$ & $0.83 \pm 0.01$ & $0.82 \pm 0.01$\\
squared correction  & mean & $0.01 \pm 0.01$ & $0.00 \pm 0.01$ & $-0.00 \pm 0.01$ & $-0.00 \pm 0.01$ & $0.00 \pm 0.01$ & $-0.01 \pm 0.01$ & $-0.00 \pm 0.01$\\
      & width & $1.01 \pm 0.01$ & $1.00 \pm 0.01$ & $1.00 \pm 0.01$ & $1.01 \pm 0.01$ & $1.01 \pm 0.01$ & $1.00 \pm 0.01$ & $0.99 \pm 0.01$\\
bootstrapping  & mean & $0.01 \pm 0.01$ & $0.00 \pm 0.01$ & $-0.00 \pm 0.01$ & $-0.00 \pm 0.01$ & $0.00 \pm 0.01$ & $-0.01 \pm 0.01$ & $-0.00 \pm 0.01$\\
      & width & $0.97 \pm 0.01$ & $0.99 \pm 0.01$ & $1.00 \pm 0.01$ & $1.01 \pm 0.01$ & $1.01 \pm 0.01$ & $1.00 \pm 0.01$ & $0.99 \pm 0.01$\\
asymptotic & mean & $0.01 \pm 0.01$ & $0.00 \pm 0.01$ & $-0.00 \pm 0.01$ & $-0.00 \pm 0.01$ & $0.00 \pm 0.01$ & $-0.01 \pm 0.01$ & $-0.00 \pm 0.01$\\
      & width & $0.99 \pm 0.01$ & $0.99 \pm 0.01$ & $1.00 \pm 0.01$ & $1.01 \pm 0.01$ & $1.01 \pm 0.01$ & $1.00 \pm 0.01$ & $0.99 \pm 0.01$\\
cFit  & mean & $0.01 \pm 0.01$ & $0.00 \pm 0.01$ & $-0.00 \pm 0.01$ & $-0.01 \pm 0.01$ & $0.01 \pm 0.01$ & $-0.01 \pm 0.01$ & $-0.01 \pm 0.01$\\
      & width & $0.99 \pm 0.01$ & $0.99 \pm 0.01$ & $1.00 \pm 0.01$ & $1.01 \pm 0.01$ & $1.00 \pm 0.01$ & $1.00 \pm 0.01$ & $0.99 \pm 0.01$\\
\hline \multicolumn{2}{l}{rel. efficiency cFit/weighted} & $0.91 \pm 0.02$ & $0.92 \pm 0.02$ & $0.92 \pm 0.02$ & $0.92 \pm 0.02$ & $0.92 \pm 0.02$ & $0.92 \pm 0.02$ & $0.92 \pm 0.02$\\
\hline\end{tabular}
}}\\
\subfloat[Parameter $c_1$\label{tab:pullsc1b}]{
  \scalebox{0.7}{
\begin{tabular}{llrrrrrrr}\hline
method & pull & 500 & 1\,k & 2\,k & 5\,k & 10\,k & 20\,k & 50\,k \\ \hline\hline
wFit  & mean & $-0.06 \pm 0.01$ & $-0.03 \pm 0.01$ & $-0.01 \pm 0.01$ & $-0.01 \pm 0.01$ & $0.00 \pm 0.01$ & $-0.02 \pm 0.01$ & $-0.00 \pm 0.01$\\
      & width & $1.37 \pm 0.01$ & $1.36 \pm 0.01$ & $1.37 \pm 0.01$ & $1.40 \pm 0.01$ & $1.38 \pm 0.01$ & $1.37 \pm 0.01$ & $1.37 \pm 0.01$\\
scaled weights & mean & $-0.04 \pm 0.01$ & $-0.02 \pm 0.01$ & $-0.00 \pm 0.01$ & $-0.01 \pm 0.01$ & $0.01 \pm 0.01$ & $-0.01 \pm 0.01$ & $-0.00 \pm 0.01$\\
      & width & $0.93 \pm 0.01$ & $0.93 \pm 0.01$ & $0.93 \pm 0.01$ & $0.95 \pm 0.01$ & $0.94 \pm 0.01$ & $0.93 \pm 0.01$ & $0.93 \pm 0.01$\\
squared correction  & mean & $0.53 \pm 0.04$ & $0.29 \pm 0.03$ & $0.21 \pm 0.03$ & $0.12 \pm 0.03$ & $0.11 \pm 0.03$ & $0.04 \pm 0.03$ & $0.04 \pm 0.03$\\
      & width & $3.93 \pm 0.03$ & $3.05 \pm 0.02$ & $2.92 \pm 0.02$ & $2.91 \pm 0.02$ & $2.86 \pm 0.02$ & $2.84 \pm 0.02$ & $2.82 \pm 0.02$\\
bootstrapping  & mean & $-0.06 \pm 0.01$ & $-0.03 \pm 0.01$ & $-0.02 \pm 0.01$ & $-0.01 \pm 0.01$ & $-0.00 \pm 0.01$ & $-0.02 \pm 0.01$ & $-0.00 \pm 0.01$\\
      & width & $0.96 \pm 0.01$ & $0.98 \pm 0.01$ & $0.99 \pm 0.01$ & $1.02 \pm 0.01$ & $1.01 \pm 0.01$ & $1.00 \pm 0.01$ & $1.00 \pm 0.01$\\
asymptotic & mean & $-0.06 \pm 0.01$ & $-0.03 \pm 0.01$ & $-0.02 \pm 0.01$ & $-0.01 \pm 0.01$ & $0.00 \pm 0.01$ & $-0.01 \pm 0.01$ & $-0.00 \pm 0.01$\\
      & width & $1.00 \pm 0.01$ & $0.99 \pm 0.01$ & $1.00 \pm 0.01$ & $1.02 \pm 0.01$ & $1.00 \pm 0.01$ & $1.00 \pm 0.01$ & $1.00 \pm 0.01$\\
cFit  & mean & $-0.07 \pm 0.01$ & $-0.04 \pm 0.01$ & $-0.02 \pm 0.01$ & $-0.02 \pm 0.01$ & $-0.00 \pm 0.01$ & $-0.02 \pm 0.01$ & $-0.01 \pm 0.01$\\
      & width & $1.00 \pm 0.01$ & $1.00 \pm 0.01$ & $1.00 \pm 0.01$ & $1.02 \pm 0.01$ & $1.01 \pm 0.01$ & $1.00 \pm 0.01$ & $1.00 \pm 0.01$\\
\hline \multicolumn{2}{l}{rel. efficiency cFit/weighted} & $0.89 \pm 0.02$ & $0.91 \pm 0.02$ & $0.91 \pm 0.02$ & $0.90 \pm 0.02$ & $0.92 \pm 0.02$ & $0.91 \pm 0.02$ & $0.91 \pm 0.02$\\
\hline\end{tabular}
}}
\caption{Means and widths of the pull distribution for the different approaches to the uncertainty estimation for the efficiency correction $\epsilon(\cos\theta)=0.3+0.7\cos^2\theta$, depending on the number of events. 
  In addition, the relative efficiencies (\ie\ the ratios of variances) of the cFit estimator and the weighted estimator defined by Eq.~\ref{eq:mlweighted} are given. 
  \label{tab:effpullsb}}
\end{table}

\clearpage

\section{Results from pseudoexperiments using \textit{\textbf{sWeights}}}
\label{sec:appsweights}
\begin{table}[hb!]
\centering
  \subfloat[Exponential background model\label{tab:pullsexp}]{
  \scalebox{0.75}{
\begin{tabular}{llrrrrrr}\hline
method & pull & 400 & 1\,k & 2\,k & 5\,k & 10\,k & 20\,k \\ \hline\hline
sFit  & mean & $-1.17 \pm 0.09$ & $-0.50 \pm 0.03$ & $-0.35 \pm 0.03$ & $-0.25 \pm 0.03$ & $-0.17 \pm 0.02$ & $-0.14 \pm 0.02$\\
      & width & $8.53 \pm 0.06$ & $2.78 \pm 0.02$ & $2.63 \pm 0.02$ & $2.52 \pm 0.02$ & $2.48 \pm 0.02$ & $2.48 \pm 0.02$\\
scaled weights & mean & $-0.75 \pm 0.05$ & $-0.34 \pm 0.02$ & $-0.24 \pm 0.02$ & $-0.17 \pm 0.02$ & $-0.12 \pm 0.02$ & $-0.10 \pm 0.02$\\
      & width & $5.14 \pm 0.04$ & $2.16 \pm 0.02$ & $2.09 \pm 0.01$ & $2.03 \pm 0.01$ & $2.00 \pm 0.01$ & $2.00 \pm 0.01$\\
squared correction  & mean & $-0.22 \pm 0.02$ & $-0.13 \pm 0.01$ & $-0.09 \pm 0.01$ & $-0.07 \pm 0.01$ & $-0.05 \pm 0.01$ & $-0.05 \pm 0.01$\\
      & width & $1.59 \pm 0.01$ & $1.44 \pm 0.01$ & $1.44 \pm 0.01$ & $1.43 \pm 0.01$ & $1.43 \pm 0.01$ & $1.43 \pm 0.01$\\
bootstrapping  & mean & $-0.07 \pm 0.01$ & $-0.06 \pm 0.01$ & $-0.04 \pm 0.01$ & $-0.04 \pm 0.01$ & $-0.03 \pm 0.01$ & $-0.03 \pm 0.01$\\
      & width & $1.21 \pm 0.01$ & $1.22 \pm 0.01$ & $1.26 \pm 0.01$ & $1.26 \pm 0.01$ & $1.27 \pm 0.01$ & $1.28 \pm 0.01$\\
full bootstrapping & mean & $0.00 \pm 0.01$ & $0.01 \pm 0.01$ & $0.01 \pm 0.01$ & $-0.00 \pm 0.01$ & $-0.01 \pm 0.01$ & $-0.01 \pm 0.01$\\
      & width & $0.94 \pm 0.01$ & $0.94 \pm 0.01$ & $0.98 \pm 0.01$ & $0.98 \pm 0.01$ & $0.99 \pm 0.01$ & $1.00 \pm 0.01$\\
asymptotic & mean & $-0.07 \pm 0.01$ & $-0.06 \pm 0.01$ & $-0.04 \pm 0.01$ & $-0.04 \pm 0.01$ & $-0.03 \pm 0.01$ & $-0.03 \pm 0.01$\\
      & width & $1.21 \pm 0.01$ & $1.24 \pm 0.01$ & $1.27 \pm 0.01$ & $1.27 \pm 0.01$ & $1.28 \pm 0.01$ & $1.29 \pm 0.01$\\
full asymptotic & mean & $0.02 \pm 0.01$ & $0.00 \pm 0.01$ & $0.00 \pm 0.01$ & $-0.00 \pm 0.01$ & $-0.01 \pm 0.01$ & $-0.01 \pm 0.01$\\
      & width & $0.94 \pm 0.01$ & $0.96 \pm 0.01$ & $0.99 \pm 0.01$ & $0.99 \pm 0.01$ & $0.99 \pm 0.01$ & $1.00 \pm 0.01$\\
cFit  & mean & $-0.08 \pm 0.01$ & $-0.06 \pm 0.01$ & $-0.03 \pm 0.01$ & $-0.03 \pm 0.01$ & $-0.03 \pm 0.01$ & $-0.02 \pm 0.01$\\
      & width & $1.03 \pm 0.01$ & $1.00 \pm 0.01$ & $1.01 \pm 0.01$ & $1.00 \pm 0.01$ & $1.00 \pm 0.01$ & $1.01 \pm 0.01$\\
\hline \multicolumn{2}{l}{rel. efficiency cFit/weighted} & $0.393 \pm 0.008$ & $0.429 \pm 0.009$ & $0.433 \pm 0.009$ & $0.440 \pm 0.009$ & $0.444 \pm 0.009$ & $0.451 \pm 0.009$\\
\hline\end{tabular}
  }}\\
    \subfloat[Gaussian background model\label{tab:pullsgauss}]{
\scalebox{0.75}{
\begin{tabular}{llrrrrrr}\hline
method & pull & 400 & 1\,k & 2\,k & 5\,k & 10\,k & 20\,k \\ \hline\hline
sFit  & mean & $-3.75 \pm 0.25$ & $-1.46 \pm 0.08$ & $-0.85 \pm 0.05$ & $-0.58 \pm 0.04$ & $-0.30 \pm 0.04$ & $-0.21 \pm 0.04$\\
      & width & $25.33 \pm 0.18$ & $7.69 \pm 0.05$ & $4.68 \pm 0.03$ & $4.27 \pm 0.03$ & $4.08 \pm 0.03$ & $4.00 \pm 0.03$\\
scaled weights & mean & $-2.47 \pm 0.17$ & $-1.00 \pm 0.06$ & $-0.58 \pm 0.04$ & $-0.40 \pm 0.03$ & $-0.20 \pm 0.03$ & $-0.14 \pm 0.03$\\
      & width & $16.58 \pm 0.12$ & $5.51 \pm 0.04$ & $3.66 \pm 0.03$ & $3.41 \pm 0.02$ & $3.29 \pm 0.02$ & $3.24 \pm 0.02$\\
squared correction  & mean & $-0.39 \pm 0.02$ & $-0.25 \pm 0.02$ & $-0.16 \pm 0.02$ & $-0.12 \pm 0.02$ & $-0.05 \pm 0.02$ & $-0.04 \pm 0.02$\\
      & width & $2.43 \pm 0.02$ & $2.06 \pm 0.01$ & $1.94 \pm 0.01$ & $1.89 \pm 0.01$ & $1.89 \pm 0.01$ & $1.87 \pm 0.01$\\
bootstrapping  & mean & $-0.03 \pm 0.02$ & $-0.02 \pm 0.02$ & $-0.01 \pm 0.02$ & $-0.02 \pm 0.02$ & $0.01 \pm 0.02$ & $0.00 \pm 0.02$\\
      & width & $1.56 \pm 0.01$ & $1.59 \pm 0.01$ & $1.62 \pm 0.01$ & $1.63 \pm 0.01$ & $1.65 \pm 0.01$ & $1.64 \pm 0.01$\\
full bootstrapping & mean & $-0.01 \pm 0.01$ & $0.03 \pm 0.01$ & $0.03 \pm 0.01$ & $0.01 \pm 0.01$ & $0.02 \pm 0.01$ & $0.01 \pm 0.01$\\
      & width & $1.05 \pm 0.01$ & $0.99 \pm 0.01$ & $0.98 \pm 0.01$ & $0.98 \pm 0.01$ & $1.00 \pm 0.01$ & $0.99 \pm 0.01$\\
asymptotic & mean & $0.02 \pm 0.01$ & $-0.02 \pm 0.02$ & $-0.02 \pm 0.02$ & $-0.02 \pm 0.02$ & $0.01 \pm 0.02$ & $0.00 \pm 0.02$\\
      & width & $1.49 \pm 0.01$ & $1.62 \pm 0.01$ & $1.66 \pm 0.01$ & $1.66 \pm 0.01$ & $1.68 \pm 0.01$ & $1.67 \pm 0.01$\\
full asymptotic & mean & $0.09 \pm 0.01$ & $0.04 \pm 0.01$ & $0.03 \pm 0.01$ & $0.01 \pm 0.01$ & $0.02 \pm 0.01$ & $0.01 \pm 0.01$\\
      & width & $0.93 \pm 0.01$ & $0.98 \pm 0.01$ & $1.00 \pm 0.01$ & $0.99 \pm 0.01$ & $1.00 \pm 0.01$ & $0.99 \pm 0.01$\\
cFit  & mean & $-0.09 \pm 0.01$ & $-0.07 \pm 0.01$ & $-0.04 \pm 0.01$ & $-0.03 \pm 0.01$ & $-0.03 \pm 0.01$ & $-0.01 \pm 0.01$\\
      & width & $1.00 \pm 0.01$ & $1.00 \pm 0.01$ & $1.00 \pm 0.01$ & $1.00 \pm 0.01$ & $1.01 \pm 0.01$ & $1.00 \pm 0.01$\\
\hline \multicolumn{2}{l}{rel. efficiency cFit/weighted} & $0.118 \pm 0.002$ & $0.113 \pm 0.002$ & $0.119 \pm 0.002$ & $0.124 \pm 0.002$ & $0.126 \pm 0.003$ & $0.127 \pm 0.003$\\
\hline\end{tabular}
    }}
  \caption{Means and widths of the pull distribution for the different approaches to the uncertainty estimation, depending on the total yield $N_{\rm tot}$.
     In addition, the relative efficiencies (\ie\ the ratios of variances) of the cFit estimator and the weighted estimator defined by Eq.~\ref{eq:mlweighted} are given.
    The different tables shown correspond to the different background models as specified in Sec.~\ref{sec:examples}.\label{tab:pullsa}}  
\end{table}

\begin{table}
\centering
\subfloat[Triangular background model\label{tab:pullstriangle}]{
\scalebox{0.75}{
\begin{tabular}{llrrrrrr}\hline
method & pull & 400 & 1\,k & 2\,k & 5\,k & 10\,k & 20\,k \\ \hline\hline
sFit  & mean & $-3.55 \pm 0.21$ & $-1.42 \pm 0.07$ & $-0.83 \pm 0.05$ & $-0.56 \pm 0.04$ & $-0.32 \pm 0.04$ & $-0.20 \pm 0.04$\\
      & width & $20.61 \pm 0.15$ & $7.25 \pm 0.05$ & $4.66 \pm 0.03$ & $4.24 \pm 0.03$ & $4.04 \pm 0.03$ & $4.08 \pm 0.03$\\
scaled weights & mean & $-2.37 \pm 0.14$ & $-0.97 \pm 0.05$ & $-0.58 \pm 0.04$ & $-0.39 \pm 0.03$ & $-0.22 \pm 0.03$ & $-0.13 \pm 0.03$\\
      & width & $13.93 \pm 0.10$ & $5.24 \pm 0.04$ & $3.65 \pm 0.03$ & $3.38 \pm 0.02$ & $3.26 \pm 0.02$ & $3.29 \pm 0.02$\\
squared correction  & mean & $-0.41 \pm 0.02$ & $-0.24 \pm 0.02$ & $-0.16 \pm 0.02$ & $-0.12 \pm 0.02$ & $-0.06 \pm 0.02$ & $-0.03 \pm 0.02$\\
      & width & $2.40 \pm 0.02$ & $2.04 \pm 0.01$ & $1.92 \pm 0.01$ & $1.88 \pm 0.01$ & $1.86 \pm 0.01$ & $1.90 \pm 0.01$\\
bootstrapping  & mean & $-0.04 \pm 0.02$ & $-0.02 \pm 0.02$ & $-0.01 \pm 0.02$ & $-0.02 \pm 0.02$ & $-0.00 \pm 0.02$ & $0.01 \pm 0.02$\\
      & width & $1.57 \pm 0.01$ & $1.57 \pm 0.01$ & $1.59 \pm 0.01$ & $1.60 \pm 0.01$ & $1.62 \pm 0.01$ & $1.66 \pm 0.01$\\
full bootstrapping & mean & $-0.02 \pm 0.01$ & $0.03 \pm 0.01$ & $0.03 \pm 0.01$ & $0.02 \pm 0.01$ & $0.02 \pm 0.01$ & $0.02 \pm 0.01$\\
      & width & $1.06 \pm 0.01$ & $0.98 \pm 0.01$ & $0.97 \pm 0.01$ & $0.98 \pm 0.01$ & $0.98 \pm 0.01$ & $1.01 \pm 0.01$\\
asymptotic & mean & $0.01 \pm 0.02$ & $-0.02 \pm 0.02$ & $-0.02 \pm 0.02$ & $-0.02 \pm 0.02$ & $-0.00 \pm 0.02$ & $0.01 \pm 0.02$\\
      & width & $1.50 \pm 0.01$ & $1.60 \pm 0.01$ & $1.62 \pm 0.01$ & $1.63 \pm 0.01$ & $1.64 \pm 0.01$ & $1.68 \pm 0.01$\\
full asymptotic & mean & $0.08 \pm 0.01$ & $0.04 \pm 0.01$ & $0.03 \pm 0.01$ & $0.01 \pm 0.01$ & $0.02 \pm 0.01$ & $0.02 \pm 0.01$\\
      & width & $0.94 \pm 0.01$ & $0.98 \pm 0.01$ & $0.98 \pm 0.01$ & $0.98 \pm 0.01$ & $0.99 \pm 0.01$ & $1.01 \pm 0.01$\\
cFit  & mean & $-0.11 \pm 0.01$ & $-0.08 \pm 0.01$ & $-0.05 \pm 0.01$ & $-0.04 \pm 0.01$ & $-0.03 \pm 0.01$ & $-0.01 \pm 0.01$\\
      & width & $1.02 \pm 0.01$ & $1.00 \pm 0.01$ & $1.00 \pm 0.01$ & $1.01 \pm 0.01$ & $1.00 \pm 0.01$ & $1.00 \pm 0.01$\\
\hline \multicolumn{2}{l}{rel. efficiency cFit/weighted} & $0.128 \pm 0.003$ & $0.122 \pm 0.002$ & $0.130 \pm 0.003$ & $0.136 \pm 0.003$ & $0.137 \pm 0.003$ & $0.132 \pm 0.003$\\
\hline\end{tabular}
  }}\\
    \subfloat[Flat background model\label{tab:pullsflat}]{
\scalebox{0.75}{
\begin{tabular}{llrrrrrr}\hline
method & pull & 400 & 1\,k & 2\,k & 5\,k & 10\,k & 20\,k \\ \hline\hline
sFit  & mean & $-2.38 \pm 0.16$ & $-0.97 \pm 0.05$ & $-0.58 \pm 0.04$ & $-0.41 \pm 0.04$ & $-0.25 \pm 0.03$ & $-0.20 \pm 0.03$\\
      & width & $16.02 \pm 0.11$ & $4.57 \pm 0.03$ & $3.83 \pm 0.03$ & $3.59 \pm 0.03$ & $3.47 \pm 0.02$ & $3.42 \pm 0.02$\\
scaled weights & mean & $-1.56 \pm 0.10$ & $-0.66 \pm 0.03$ & $-0.39 \pm 0.03$ & $-0.28 \pm 0.03$ & $-0.17 \pm 0.03$ & $-0.15 \pm 0.03$\\
      & width & $10.43 \pm 0.07$ & $3.44 \pm 0.02$ & $3.02 \pm 0.02$ & $2.87 \pm 0.02$ & $2.80 \pm 0.02$ & $2.77 \pm 0.02$\\
squared correction  & mean & $-0.32 \pm 0.02$ & $-0.19 \pm 0.02$ & $-0.11 \pm 0.02$ & $-0.09 \pm 0.02$ & $-0.06 \pm 0.02$ & $-0.05 \pm 0.02$\\
      & width & $2.09 \pm 0.01$ & $1.83 \pm 0.01$ & $1.78 \pm 0.01$ & $1.76 \pm 0.01$ & $1.74 \pm 0.01$ & $1.73 \pm 0.01$\\
bootstrapping  & mean & $-0.06 \pm 0.01$ & $-0.04 \pm 0.01$ & $-0.01 \pm 0.01$ & $-0.02 \pm 0.01$ & $-0.01 \pm 0.01$ & $-0.02 \pm 0.01$\\
      & width & $1.36 \pm 0.01$ & $1.38 \pm 0.01$ & $1.41 \pm 0.01$ & $1.43 \pm 0.01$ & $1.43 \pm 0.01$ & $1.43 \pm 0.01$\\
full bootstrapping & mean & $-0.01 \pm 0.01$ & $0.03 \pm 0.01$ & $0.03 \pm 0.01$ & $0.01 \pm 0.01$ & $0.01 \pm 0.01$ & $-0.00 \pm 0.01$\\
      & width & $1.00 \pm 0.01$ & $0.96 \pm 0.01$ & $0.97 \pm 0.01$ & $0.99 \pm 0.01$ & $0.99 \pm 0.01$ & $0.99 \pm 0.01$\\
asymptotic & mean & $-0.03 \pm 0.01$ & $-0.04 \pm 0.01$ & $-0.02 \pm 0.01$ & $-0.02 \pm 0.01$ & $-0.01 \pm 0.01$ & $-0.02 \pm 0.01$\\
      & width & $1.32 \pm 0.01$ & $1.41 \pm 0.01$ & $1.44 \pm 0.01$ & $1.45 \pm 0.01$ & $1.45 \pm 0.01$ & $1.45 \pm 0.01$\\
full asymptotic & mean & $0.06 \pm 0.01$ & $0.03 \pm 0.01$ & $0.03 \pm 0.01$ & $0.01 \pm 0.01$ & $0.01 \pm 0.01$ & $-0.00 \pm 0.01$\\
      & width & $0.93 \pm 0.01$ & $0.98 \pm 0.01$ & $0.99 \pm 0.01$ & $1.00 \pm 0.01$ & $1.00 \pm 0.01$ & $0.99 \pm 0.01$\\
cFit  & mean & $-0.08 \pm 0.01$ & $-0.07 \pm 0.01$ & $-0.03 \pm 0.01$ & $-0.03 \pm 0.01$ & $-0.02 \pm 0.01$ & $-0.02 \pm 0.01$\\
      & width & $1.02 \pm 0.01$ & $1.01 \pm 0.01$ & $0.99 \pm 0.01$ & $1.00 \pm 0.01$ & $1.01 \pm 0.01$ & $1.00 \pm 0.01$\\
\hline \multicolumn{2}{l}{rel. efficiency cFit/weighted} & $0.207 \pm 0.004$ & $0.203 \pm 0.004$ & $0.208 \pm 0.004$ & $0.215 \pm 0.004$ & $0.224 \pm 0.004$ & $0.223 \pm 0.004$\\
\hline\end{tabular}
    }}
  \caption{Means and widths of the pull distribution for the different approaches to the uncertainty estimation, depending on the total yield $N_{\rm tot}$.
    In addition, the relative efficiencies (\ie\ the ratios of variances) of the cFit estimator and the weighted estimator defined by Eq.~\ref{eq:mlweighted} are given. 
    The different tables shown correspond to the different background models as specified in Sec.~\ref{sec:examples}.\label{tab:pullsb}}  
\end{table}

\clearpage

\section{\textit{\textbf{sWeights}} with negligible nuisance parameter correlations}
\label{sec:appsweightsnocorr}
\begin{figure}[h]
  \centering
  \includegraphics[width=0.49\textwidth]{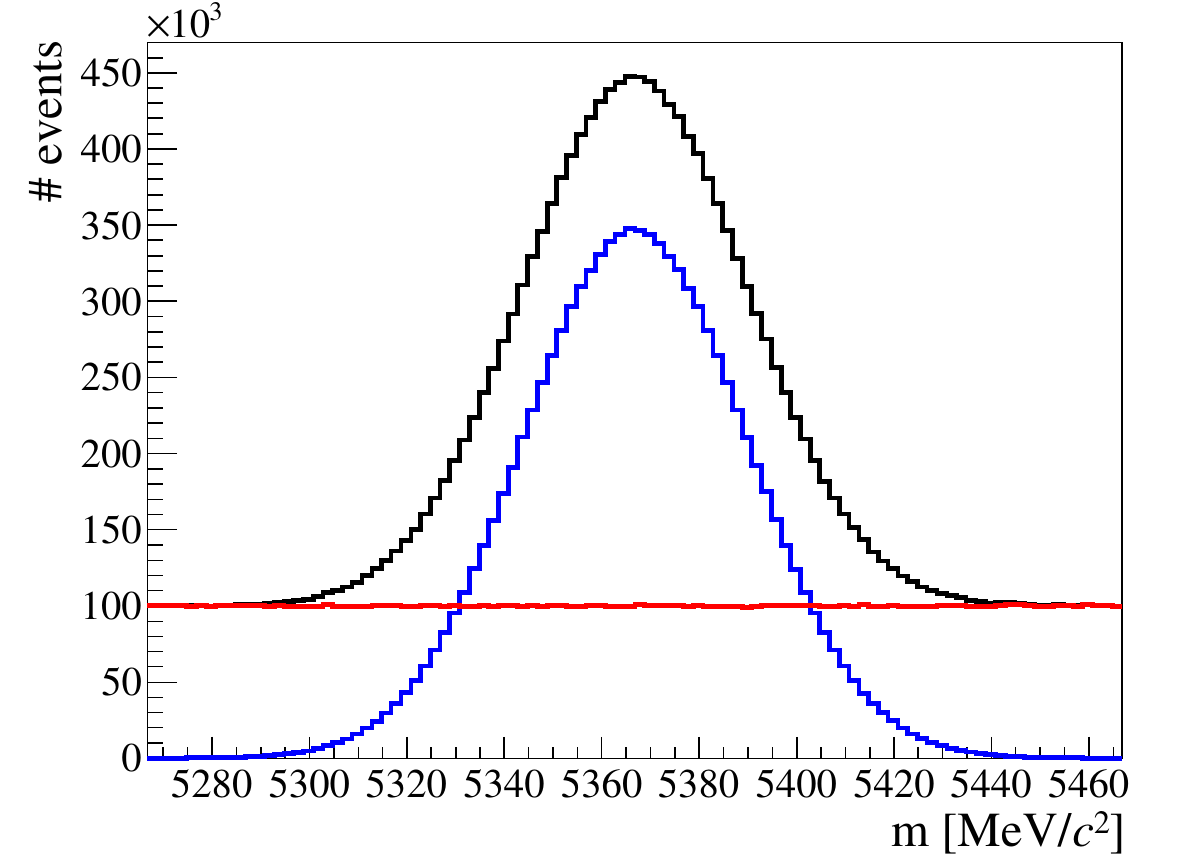}
  \caption{Discriminating mass distribution for (black) the full data, (blue) signal and (red) background using the mass range $[5\,267,5\,467]\mevcc$.\label{fig:massnocorr}}
\end{figure}

In Sec.~\ref{sec:sweights} the mass range which is used to determine the {\it sWeights} is chosen such that the background slope $\alpha_{\rm bkg}$ is significantly correlated with the event yields.
This results in an additional uncertainty on the {\it sWeights} that needs to be accounted for using Eq.~\ref{eq:correctvarnuisance}. 
For completeness, in this section the different methods are studied in the absence of any significant impact of nuisance parameters on the {\it sWeights}.
To this end, the mass range $[5\,267,5\,467]\mevcc$ is chosen, which is a symmetrical mass window around the peak at $5\,367\mevcc$, as shown in Fig.~\ref{fig:massnocorr}. 
This choice results in negligible correlation of $\alpha_{\rm bkg}$ with $N_{\rm sig}$ and $N_{\rm bkg}$.
All other settings are kept as in Sec.~\ref{sec:sweights}. 

The means and widths of the resulting pull distributions are given in Tabs.~\ref{tab:pullsanocorr} and~\ref{tab:pullsbnocorr}.
As for the case discussed in Sec.~\ref{sec:sweights},
the {\it sFit}, the {\it scaled weights} method and the {\it squared weights} method show significant undercoverage. 
All other methods, namely the ({\it full}) {\it bootstrapping}, the ({\it full}) {\it asymptotic} method and the {\it cFit} show correct coverage.
\begin{table}
\centering
    \subfloat[Exponential background model\label{tab:pullsexpnocorr}]{
      \scalebox{0.75}{
\begin{tabular}{llrrrrrr}\hline
method & pull & 400 & 1\,k & 2\,k & 5\,k & 10\,k & 20\,k \\ \hline\hline
sFit  & mean & $-0.46 \pm 0.03$ & $-0.29 \pm 0.02$ & $-0.18 \pm 0.02$ & $-0.14 \pm 0.02$ & $-0.09 \pm 0.02$ & $-0.06 \pm 0.02$\\
      & width & $2.82 \pm 0.02$ & $2.47 \pm 0.02$ & $2.40 \pm 0.02$ & $2.37 \pm 0.02$ & $2.32 \pm 0.02$ & $2.33 \pm 0.02$\\
scaled weights & mean & $-0.30 \pm 0.02$ & $-0.19 \pm 0.02$ & $-0.12 \pm 0.02$ & $-0.09 \pm 0.02$ & $-0.06 \pm 0.02$ & $-0.04 \pm 0.02$\\
      & width & $1.97 \pm 0.01$ & $1.76 \pm 0.01$ & $1.72 \pm 0.01$ & $1.69 \pm 0.01$ & $1.66 \pm 0.01$ & $1.67 \pm 0.01$\\
squared correction  & mean & $-0.08 \pm 0.01$ & $-0.06 \pm 0.01$ & $-0.03 \pm 0.01$ & $-0.03 \pm 0.01$ & $-0.02 \pm 0.01$ & $-0.01 \pm 0.01$\\
      & width & $1.17 \pm 0.01$ & $1.14 \pm 0.01$ & $1.13 \pm 0.01$ & $1.13 \pm 0.01$ & $1.11 \pm 0.01$ & $1.12 \pm 0.01$\\
bootstrapping  & mean & $0.01 \pm 0.01$ & $-0.00 \pm 0.01$ & $0.00 \pm 0.01$ & $-0.00 \pm 0.01$ & $-0.00 \pm 0.01$ & $-0.00 \pm 0.01$\\
      & width & $0.95 \pm 0.01$ & $0.97 \pm 0.01$ & $0.99 \pm 0.01$ & $0.99 \pm 0.01$ & $0.98 \pm 0.01$ & $0.99 \pm 0.01$\\
full bootstrapping & mean & $0.00 \pm 0.01$ & $-0.01 \pm 0.01$ & $0.00 \pm 0.01$ & $-0.00 \pm 0.01$ & $-0.00 \pm 0.01$ & $-0.00 \pm 0.01$\\
      & width & $0.96 \pm 0.01$ & $0.98 \pm 0.01$ & $1.00 \pm 0.01$ & $1.00 \pm 0.01$ & $0.99 \pm 0.01$ & $1.00 \pm 0.01$\\
asymptotic & mean & $0.00 \pm 0.01$ & $-0.01 \pm 0.01$ & $0.00 \pm 0.01$ & $-0.00 \pm 0.01$ & $-0.00 \pm 0.01$ & $-0.00 \pm 0.01$\\
      & width & $0.98 \pm 0.01$ & $0.99 \pm 0.01$ & $1.00 \pm 0.01$ & $1.00 \pm 0.01$ & $0.99 \pm 0.01$ & $1.00 \pm 0.01$\\
full asymptotic & mean & $0.00 \pm 0.01$ & $-0.01 \pm 0.01$ & $0.00 \pm 0.01$ & $-0.00 \pm 0.01$ & $-0.00 \pm 0.01$ & $-0.00 \pm 0.01$\\
      & width & $0.98 \pm 0.01$ & $0.99 \pm 0.01$ & $1.00 \pm 0.01$ & $1.00 \pm 0.01$ & $0.99 \pm 0.01$ & $1.00 \pm 0.01$\\
cFit  & mean & $-0.05 \pm 0.01$ & $-0.05 \pm 0.01$ & $-0.02 \pm 0.01$ & $-0.02 \pm 0.01$ & $-0.01 \pm 0.01$ & $-0.01 \pm 0.01$\\
      & width & $1.02 \pm 0.01$ & $1.01 \pm 0.01$ & $1.01 \pm 0.01$ & $1.01 \pm 0.01$ & $1.01 \pm 0.01$ & $1.00 \pm 0.01$\\
\hline \multicolumn{2}{l}{rel. efficiency cFit/weighted} & $0.466 \pm 0.009$ & $0.470 \pm 0.009$ & $0.468 \pm 0.009$ & $0.467 \pm 0.009$ & $0.478 \pm 0.010$ & $0.465 \pm 0.009$\\
\hline\end{tabular}        
  }}\\
    \subfloat[Gaussian background model\label{tab:pullsgaussnocorr}]{
\scalebox{0.75}{
\begin{tabular}{llrrrrrr}\hline
method & pull & 400 & 1\,k & 2\,k & 5\,k & 10\,k & 20\,k \\ \hline\hline
sFit  & mean & $-0.84 \pm 0.04$ & $-0.42 \pm 0.03$ & $-0.25 \pm 0.03$ & $-0.16 \pm 0.03$ & $-0.14 \pm 0.03$ & $-0.10 \pm 0.03$\\
      & width & $4.47 \pm 0.03$ & $3.29 \pm 0.02$ & $3.09 \pm 0.02$ & $3.01 \pm 0.02$ & $2.92 \pm 0.02$ & $2.92 \pm 0.02$\\
scaled weights & mean & $-0.54 \pm 0.03$ & $-0.27 \pm 0.02$ & $-0.16 \pm 0.02$ & $-0.10 \pm 0.02$ & $-0.09 \pm 0.02$ & $-0.07 \pm 0.02$\\
      & width & $3.05 \pm 0.02$ & $2.32 \pm 0.02$ & $2.20 \pm 0.02$ & $2.15 \pm 0.02$ & $2.09 \pm 0.01$ & $2.09 \pm 0.01$\\
squared correction  & mean & $-0.12 \pm 0.01$ & $-0.06 \pm 0.01$ & $-0.03 \pm 0.01$ & $-0.01 \pm 0.01$ & $-0.03 \pm 0.01$ & $-0.02 \pm 0.01$\\
      & width & $1.22 \pm 0.01$ & $1.15 \pm 0.01$ & $1.14 \pm 0.01$ & $1.13 \pm 0.01$ & $1.11 \pm 0.01$ & $1.11 \pm 0.01$\\
bootstrapping  & mean & $0.03 \pm 0.01$ & $0.03 \pm 0.01$ & $0.03 \pm 0.01$ & $0.03 \pm 0.01$ & $0.00 \pm 0.01$ & $-0.00 \pm 0.01$\\
      & width & $0.96 \pm 0.01$ & $0.96 \pm 0.01$ & $0.97 \pm 0.01$ & $0.98 \pm 0.01$ & $0.97 \pm 0.01$ & $0.97 \pm 0.01$\\
full bootstrapping & mean & $0.03 \pm 0.01$ & $0.03 \pm 0.01$ & $0.03 \pm 0.01$ & $0.03 \pm 0.01$ & $0.00 \pm 0.01$ & $-0.00 \pm 0.01$\\
      & width & $0.97 \pm 0.01$ & $0.98 \pm 0.01$ & $1.00 \pm 0.01$ & $1.00 \pm 0.01$ & $0.99 \pm 0.01$ & $1.00 \pm 0.01$\\
asymptotic & mean & $0.04 \pm 0.01$ & $0.03 \pm 0.01$ & $0.03 \pm 0.01$ & $0.03 \pm 0.01$ & $0.00 \pm 0.01$ & $-0.00 \pm 0.01$\\
      & width & $0.98 \pm 0.01$ & $0.99 \pm 0.01$ & $1.00 \pm 0.01$ & $1.00 \pm 0.01$ & $0.99 \pm 0.01$ & $0.99 \pm 0.01$\\
full asymptotic & mean & $0.03 \pm 0.01$ & $0.03 \pm 0.01$ & $0.03 \pm 0.01$ & $0.03 \pm 0.01$ & $0.00 \pm 0.01$ & $-0.00 \pm 0.01$\\
      & width & $0.97 \pm 0.01$ & $0.99 \pm 0.01$ & $1.00 \pm 0.01$ & $1.00 \pm 0.01$ & $0.99 \pm 0.01$ & $0.99 \pm 0.01$\\
cFit  & mean & $-0.09 \pm 0.01$ & $-0.06 \pm 0.01$ & $-0.04 \pm 0.01$ & $-0.03 \pm 0.01$ & $-0.03 \pm 0.01$ & $-0.02 \pm 0.01$\\
      & width & $1.02 \pm 0.01$ & $1.01 \pm 0.01$ & $1.02 \pm 0.01$ & $1.01 \pm 0.01$ & $1.00 \pm 0.01$ & $1.00 \pm 0.01$\\
\hline \multicolumn{2}{l}{rel. efficiency cFit/weighted} & $0.218 \pm 0.004$ & $0.228 \pm 0.005$ & $0.235 \pm 0.005$ & $0.235 \pm 0.005$ & $0.239 \pm 0.005$ & $0.236 \pm 0.005$\\
\hline\end{tabular}
    }}\\
  \caption{Means and widths of the pull distribution for the different approaches to the uncertainty estimation, depending on the total yield $N_{\rm tot}$.
    In addition, the relative efficiencies (\ie\ the ratios of variances) of the cFit estimator and the weighted estimator defined by Eq.~\ref{eq:mlweighted} are given. 
    The different tables shown correspond to the different background models as specified in Sec.~\ref{sec:examples}.
    In this case, the mass range $[5\,267,5\,467]\mevcc$ is chosen, such that the nuisance parameter $\alpha_{\rm bkg}$ has negligible correlation with the event yields.\label{tab:pullsanocorr}}  
\end{table}

\begin{table}
  \centering
  \subfloat[Triangular background model\label{tab:pullstrianglenocorr}]{
  \scalebox{0.75}{
\begin{tabular}{llrrrrrr}\hline
method & pull & 400 & 1\,k & 2\,k & 5\,k & 10\,k & 20\,k \\ \hline\hline
sFit  & mean & $-0.84 \pm 0.05$ & $-0.42 \pm 0.03$ & $-0.29 \pm 0.03$ & $-0.20 \pm 0.03$ & $-0.14 \pm 0.03$ & $-0.09 \pm 0.03$\\
      & width & $5.33 \pm 0.04$ & $3.24 \pm 0.02$ & $3.08 \pm 0.02$ & $3.02 \pm 0.02$ & $2.97 \pm 0.02$ & $2.97 \pm 0.02$\\
scaled weights & mean & $-0.54 \pm 0.04$ & $-0.27 \pm 0.02$ & $-0.19 \pm 0.02$ & $-0.13 \pm 0.02$ & $-0.09 \pm 0.02$ & $-0.06 \pm 0.02$\\
      & width & $3.55 \pm 0.03$ & $2.29 \pm 0.02$ & $2.20 \pm 0.02$ & $2.16 \pm 0.02$ & $2.13 \pm 0.02$ & $2.12 \pm 0.02$\\
squared correction  & mean & $-0.11 \pm 0.01$ & $-0.06 \pm 0.01$ & $-0.05 \pm 0.01$ & $-0.03 \pm 0.01$ & $-0.02 \pm 0.01$ & $-0.01 \pm 0.01$\\
      & width & $1.23 \pm 0.01$ & $1.15 \pm 0.01$ & $1.14 \pm 0.01$ & $1.13 \pm 0.01$ & $1.13 \pm 0.01$ & $1.13 \pm 0.01$\\
bootstrapping  & mean & $0.03 \pm 0.01$ & $0.03 \pm 0.01$ & $0.02 \pm 0.01$ & $0.01 \pm 0.01$ & $0.00 \pm 0.01$ & $0.01 \pm 0.01$\\
      & width & $0.95 \pm 0.01$ & $0.95 \pm 0.01$ & $0.96 \pm 0.01$ & $0.97 \pm 0.01$ & $0.97 \pm 0.01$ & $0.98 \pm 0.01$\\
full bootstrapping & mean & $0.03 \pm 0.01$ & $0.03 \pm 0.01$ & $0.02 \pm 0.01$ & $0.01 \pm 0.01$ & $0.00 \pm 0.01$ & $0.00 \pm 0.01$\\
      & width & $0.97 \pm 0.01$ & $0.97 \pm 0.01$ & $0.98 \pm 0.01$ & $1.00 \pm 0.01$ & $1.00 \pm 0.01$ & $1.00 \pm 0.01$\\
asymptotic & mean & $0.04 \pm 0.01$ & $0.02 \pm 0.01$ & $0.02 \pm 0.01$ & $0.01 \pm 0.01$ & $0.00 \pm 0.01$ & $0.00 \pm 0.01$\\
      & width & $0.97 \pm 0.01$ & $0.98 \pm 0.01$ & $0.99 \pm 0.01$ & $1.00 \pm 0.01$ & $1.00 \pm 0.01$ & $1.00 \pm 0.01$\\
full asymptotic & mean & $0.04 \pm 0.01$ & $0.02 \pm 0.01$ & $0.01 \pm 0.01$ & $0.01 \pm 0.01$ & $0.00 \pm 0.01$ & $0.00 \pm 0.01$\\
      & width & $0.97 \pm 0.01$ & $0.98 \pm 0.01$ & $0.99 \pm 0.01$ & $1.00 \pm 0.01$ & $1.00 \pm 0.01$ & $1.00 \pm 0.01$\\
cFit  & mean & $-0.11 \pm 0.01$ & $-0.07 \pm 0.01$ & $-0.05 \pm 0.01$ & $-0.04 \pm 0.01$ & $-0.03 \pm 0.01$ & $-0.02 \pm 0.01$\\
      & width & $1.02 \pm 0.01$ & $1.01 \pm 0.01$ & $1.01 \pm 0.01$ & $1.01 \pm 0.01$ & $1.00 \pm 0.01$ & $1.00 \pm 0.01$\\
\hline \multicolumn{2}{l}{rel. efficiency cFit/weighted} & $0.231 \pm 0.005$ & $0.239 \pm 0.005$ & $0.244 \pm 0.005$ & $0.246 \pm 0.005$ & $0.242 \pm 0.005$ & $0.242 \pm 0.005$\\
\hline\end{tabular}    
  }}\\
  \subfloat[Flat background model\label{tab:pullsflatnocorr}]{
\scalebox{0.75}{
\begin{tabular}{llrrrrrr}\hline
method & pull & 400 & 1\,k & 2\,k & 5\,k & 10\,k & 20\,k \\ \hline\hline
sFit  & mean & $-0.71 \pm 0.04$ & $-0.46 \pm 0.03$ & $-0.36 \pm 0.03$ & $-0.28 \pm 0.03$ & $-0.18 \pm 0.03$ & $-0.13 \pm 0.03$\\
      & width & $3.86 \pm 0.03$ & $3.14 \pm 0.02$ & $3.03 \pm 0.02$ & $2.98 \pm 0.02$ & $2.89 \pm 0.02$ & $2.88 \pm 0.02$\\
scaled weights & mean & $-0.46 \pm 0.03$ & $-0.30 \pm 0.02$ & $-0.24 \pm 0.02$ & $-0.19 \pm 0.02$ & $-0.12 \pm 0.02$ & $-0.09 \pm 0.02$\\
      & width & $2.67 \pm 0.02$ & $2.23 \pm 0.02$ & $2.16 \pm 0.02$ & $2.13 \pm 0.02$ & $2.07 \pm 0.01$ & $2.07 \pm 0.01$\\
squared correction  & mean & $-0.10 \pm 0.01$ & $-0.09 \pm 0.01$ & $-0.08 \pm 0.01$ & $-0.07 \pm 0.01$ & $-0.05 \pm 0.01$ & $-0.03 \pm 0.01$\\
      & width & $1.26 \pm 0.01$ & $1.22 \pm 0.01$ & $1.22 \pm 0.01$ & $1.22 \pm 0.01$ & $1.20 \pm 0.01$ & $1.20 \pm 0.01$\\
bootstrapping  & mean & $0.02 \pm 0.01$ & $-0.00 \pm 0.01$ & $-0.02 \pm 0.01$ & $-0.02 \pm 0.01$ & $-0.02 \pm 0.01$ & $-0.01 \pm 0.01$\\
      & width & $0.94 \pm 0.01$ & $0.95 \pm 0.01$ & $0.98 \pm 0.01$ & $0.98 \pm 0.01$ & $0.97 \pm 0.01$ & $0.98 \pm 0.01$\\
full bootstrapping & mean & $0.01 \pm 0.01$ & $-0.01 \pm 0.01$ & $-0.02 \pm 0.01$ & $-0.02 \pm 0.01$ & $-0.02 \pm 0.01$ & $-0.01 \pm 0.01$\\
      & width & $0.95 \pm 0.01$ & $0.97 \pm 0.01$ & $0.99 \pm 0.01$ & $0.99 \pm 0.01$ & $0.99 \pm 0.01$ & $0.99 \pm 0.01$\\
asymptotic & mean & $0.02 \pm 0.01$ & $-0.01 \pm 0.01$ & $-0.02 \pm 0.01$ & $-0.02 \pm 0.01$ & $-0.02 \pm 0.01$ & $-0.01 \pm 0.01$\\
      & width & $0.96 \pm 0.01$ & $0.98 \pm 0.01$ & $1.00 \pm 0.01$ & $1.00 \pm 0.01$ & $0.99 \pm 0.01$ & $0.99 \pm 0.01$\\
full asymptotic & mean & $0.02 \pm 0.01$ & $-0.01 \pm 0.01$ & $-0.02 \pm 0.01$ & $-0.02 \pm 0.01$ & $-0.02 \pm 0.01$ & $-0.01 \pm 0.01$\\
      & width & $0.96 \pm 0.01$ & $0.98 \pm 0.01$ & $1.00 \pm 0.01$ & $1.00 \pm 0.01$ & $0.99 \pm 0.01$ & $0.99 \pm 0.01$\\
cFit  & mean & $-0.06 \pm 0.01$ & $-0.04 \pm 0.01$ & $-0.03 \pm 0.01$ & $-0.03 \pm 0.01$ & $-0.02 \pm 0.01$ & $-0.02 \pm 0.01$\\
      & width & $1.02 \pm 0.01$ & $1.01 \pm 0.01$ & $1.00 \pm 0.01$ & $1.00 \pm 0.01$ & $1.00 \pm 0.01$ & $1.00 \pm 0.01$\\
\hline \multicolumn{2}{l}{rel. efficiency cFit/weighted} & $0.298 \pm 0.006$ & $0.295 \pm 0.006$ & $0.290 \pm 0.006$ & $0.293 \pm 0.006$ & $0.296 \pm 0.006$ & $0.297 \pm 0.006$\\
\hline\end{tabular}
        }}
    \caption{Means and widths of the pull distribution for the different approaches to the uncertainty estimation, depending on the total yield $N_{\rm tot}$.
    In addition, the relative efficiencies (\ie\ the ratios of variances) of the cFit estimator and the weighted estimator defined by Eq.~\ref{eq:mlweighted} are given. 
    The different tables shown correspond to the different background models as specified in Sec.~\ref{sec:examples}.
    In this case, the mass range $[5\,267,5\,467]\mevcc$ is chosen, such that the nuisance parameter $\alpha_{\rm bkg}$ has negligible correlation with the event yields.\label{tab:pullsbnocorr}}  
\end{table}

\end{document}